\documentclass[aps,twocolumn,prb,amsmath,superscriptaddress,amssymb,floatfix,footinbib,longbibliography,bibnotes]{revtex4-2}

\usepackage[colorlinks=true,citecolor=blue,linkcolor=magenta]{hyperref}
\usepackage{amsmath}
\usepackage{graphicx}
\usepackage{soul}
\usepackage{changepage}
\usepackage{fancyhdr}
\usepackage{amsthm,amssymb}
\usepackage{floatflt}
\usepackage{float}
\usepackage{array}
\usepackage[usenames,dvipsnames]{color}
\usepackage{comment}

\usepackage{lipsum}
\usepackage{graphicx}
\usepackage{braket}
\usepackage{mathtools}
\usepackage{hyperref}
\usepackage[version=4]{mhchem}
\usepackage{siunitx}

\usepackage{color}
\usepackage[normalem]{ulem}
\usepackage{tikz-feynman}

\newcommand{\be}{\begin{align}}
\newcommand{\ee}{\end{align}}

\def \be{\begin{equation}}
\def \ee{\end{equation}}
\def \ba{\begin{array}}
\def \ea{\end{array}}
\def \bea{\begin{eqnarray}}
\def \eea{\end{eqnarray}}

\def \ba{\begin{align*}}
\def \ea{\end{align*}}

\newcommand {\apgt} {\ {\raise-.5ex\hbox{$\buildrel>\over\sim$}}\ }
\newcommand {\aplt} {\ {\raise-.5ex\hbox{$\buildrel<\over\sim$}}\ }

\def \dag{{\dagger}}

\def \ba{\begin{align}}
\def \ea{\end{align}}

\begin{document}

\title{Solvable Theory of a Strange Metal at the Breakdown of a Heavy Fermi Liquid}

\author{Erik E. Aldape}
\thanks{E.E.A. and T.C. contributed equally to this work.}

\affiliation{Department of Physics, University of California, Berkeley, CA 94720, USA}

\author{Tessa Cookmeyer}
\thanks{E.E.A. and T.C. contributed equally to this work.}

\affiliation{Department of Physics, University of California, Berkeley, CA 94720, USA}
\affiliation{Materials Science Division, Lawrence Berkeley National Laboratory, Berkeley, California 94720, USA}

\author{Aavishkar A. Patel}

\affiliation{Department of Physics, University of California, Berkeley, CA 94720, USA}

\author{Ehud Altman}

\affiliation{Department of Physics, University of California, Berkeley, CA 94720, USA}
\affiliation{Materials Science Division, Lawrence Berkeley National Laboratory, Berkeley, California 94720, USA}

\begin{abstract}

We introduce an effective theory for quantum critical points (QCPs) in heavy fermion systems, involving a change in carrier density without symmetry breaking. Our new theory captures a strongly coupled metallic QCP, leading to robust marginal Fermi liquid transport phenomenology, and associated linear in temperature ($T$) ``strange metal" resistivity, all within a controlled large $N$ limit. In the parameter regime of strong damping of emergent bosonic excitations, the QCP also displays a near-universal ``Planckian" transport lifetime, $\tau_{\mathrm{tr}}\sim\hbar/(k_BT)$. This is contrasted with the conventional so-called ``slave boson" theory of the Kondo breakdown, where the large $N$ limit describes a weak coupling fixed point and non-trivial transport behavior may only be obtained through uncontrolled $1/N$ corrections. We also compute the weak-field Hall coefficient within the effective model as the system is tuned across the transition. We further find that between the two plateaus, reflecting the different carrier densities in the two Fermi liquid phases, the Hall coefficient can develop a peak in the critical crossover regime, like in recent experimental findings, in the parameter regime of weak boson damping.

\end{abstract}

\maketitle

\section{Introduction}

The properties of heavy fermion materials (HFMs) have been a continued source of fascination, calling fundamental concepts of solid state physics into question \cite{si2010heavy}. An important ingredient in the physics of the HFMs is the coexistence and interplay of conduction electrons with a half filled localized valence electron band behaving as local spin-$1/2$ moments \cite{doniach1977kondo}. Early on, a mechanism was proposed whereby the valence levels (VLs) effectively hybridize with the conduction electrons through Kondo-like screening of their spin \cite{read1983solution,coleman1984new}. This mechanism explains the establishment of a heavy Fermi liquid (FL) with a large Fermi surface (FS) that includes both the conduction and valence electrons as required by Luttinger's theorem. However, many of these materials can be tuned through quantum critical points (QCPs) at which the large FS gives way to one with a small volume, equal to the filling of the conduction band alone \cite{shishido2005drastic,paschen2004hall,friedemann2010fermi,maksimovic2020}. 

Reconstruction of the FS can occur through two distinct routes. The first is through symmetry breaking, such as an antiferromagnetic transition, as seen in CeRhIn$_5$ \cite{shishido2005drastic}. In this case the emergent small FS satisfies Luttinger's theorem within the new reduced Brillouin zone. However recent experiments with a related material, CeCoIn$_5$ \cite{maksimovic2020} suggest a FS changing transition without symmetry breaking. 

Such a transition has a simple description within a scheme \cite{coleman1984new,senthil2004weak} in which the Kondo interaction is expressed as a coupling to a bosonic valence fluctuation, {\it i.e.} $c^\dagger_\sigma f_\sigma b$. Here, $c_\sigma$ represents the conduction electron with spin index $\sigma$, and $f_\sigma$ is a fermion operator carrying the spin of the singly occupied VLs. Hybridization between the conduction and valence bands emerges with condensation of the boson $b$, leading to the creation of a heavy FL phase.

As emphasized by Senthil et. al. \cite{senthil2004weak}, besides carrying a physical electron charge, this boson is also charged under an emergent $U(1)$ gauge field that fixes the local occupation of the VLs. Therefore, condensation of $b$ in the heavy FL phase leads to confinement through the Higgs mechanism. In the gapped (uncondensed) phase of the boson, on the other hand, the VLs effectively decouple from the Fermi sea and form a $U(1)$ spin liquid. This phase is referred to as a fractionalized Fermi liquid (FL$^\star$), and it was argued that it supports a small FS \cite{senthil2003fractionalized}, thereby obeying a generalized form of Luttinger's theorem \cite{oshikawa2000topological}. 

This so-called ``slave boson" theory \cite{coleman1984new,senthil2004weak} describes a route for a transition involving change in the FS volume without symmetry breaking. However, the standard large $N$ approach \cite{coleman1984new} used to approximate the theory fails to capture essential properties of QCPs seen in HFMs; it does not offer a robust explanation of the ubiquitous ``strange metal", with its linear in temperature ($T$) resistivity $\rho_{xx}$ at the QCP \cite{Stewart2001,maksimovic2020}. The essential problem in the theory is that the feedback of the single critical boson on a large number of $N$ fermion species is suppressed by $1/N$. The conduction electrons are therefore non-interacting at the large $N$ saddle point. Thus, the same feature that makes this theory solvable also prevents it from describing a fully strongly coupled QCP.

In this paper, we introduce a valence fluctuation theory, which captures a strongly coupled QCP showing marginal Fermi liquid (MFL) phenomenology \cite{Varma1996} and strange metal $T$-linear resistivity, in a solvable limit. We start from the same degrees of freedom as in the slave boson theory described above \cite{coleman1984new,Coleman2005}. However we introduce a different large $N$ limit, which allows controlled calculation of transport properties nonperturbatively. 

The new large $N$ limit is inspired by recent work on ``low rank" Sachdev-Ye-Kitaev (SYK) models, in which $N$ fermion flavors interact via random Yukawa couplings with $\alpha N$ boson flavors \cite{Bi2017,Patel2018gauge,Marcus2019,Wang2020,Esterlis2019,Kim2020}. 
Recently, this approach has been used to compute quantum critical properties and quantum chaos in the 2+1-dimensional Gross-Neveu-Yukawa model, namely, massless Dirac fermions coupled to a critical boson field \cite{Kim2021}. The critical exponents found at the saddle point level are in excellent agreement with those obtained from conformal bootstrap, even for moderate values of $N$. The key advantage compared to the standard large $N$ limit is that, because both the fermion and boson numbers scale with $N$, the saddle point equations include self-consistent feedback between them, allowing us to capture a strongly coupled QCP. 

We implement the large $N$ scheme in the Kondo lattice problem by introducing $N$ flavors of the spin-$1/2$ fermions $c_\sigma$ and $f_\sigma$, and of the valence fluctuation (spin-0) boson $b$, while retaining the global $su(2)$ spin symmetry. We consider two distinct models of the fermion-boson couplings $g^r_{ijk}$. In Model I, the couplings are spatially disordered, and in Model II they are flavor random but translationally invariant. Thus, the randomness in Model II is just a theoretical tool. Integrating over it may be viewed as averaging over an ensemble of translationally invariant models that all yield identical long wavelength behavior. 

In both models, we obtain a QCP showing linear in $T$ resistivity up to logarithmic corrections, however these critical points describe transitions between slightly different phases. In Model I, we obtain the linear in $T$ resistivity at the QCP only if the heavy FL transitions to a ``layered FL$^\star$" phase, where the spinons $f_\sigma$ and boson $b$ are deconfined only within two-dimensional (2D) planes. In Model II, on the other hand, the MFL is obtained at a transition to a fully three-dimensional (3D) FL$^\star$ phase. Moreover, not only is the resistivity linear in $T$ at the QCP, the transport lifetime always takes the universal ``Planckian" value $\tau_{tr}\approx \hbar/(k_B T)$. Model I by contrast can be tuned between a strongly damped "Planckian" regime, and a weakly damped MFL charaterized by a sub-Planckian linear in $T$ relaxation rate. Interestingly, in the weakly damped MFL regime, we find an enhancement of the Hall coefficient $R_H$ in the critical regime, similar to recent experimental findings in CeCoIn${}_5$ \cite{maksimovic2020}. 

The rest of the paper is organized as follows: in Section II, we review the standard large $N$ approach to Kondo lattice models and then introduce the new large $N$ limit. In Sections III and IV we solve two models, with and without translation invariance, in this large $N$ limit, and calculate transport quantities. We find strange metal behavior with $T$-linear resistivity at the QCP, and the evolution of the Hall resistivity across the QCP confirms a change of carrier density, with an additional enhancement of the Hall coefficent near criticality.

\section{Large $N$ Kondo Lattice Models}

In HFMs, rare earth or actinide ions contribute a lattice of localized valence spins $\vec{S}$ coupled to the mobile conduction electrons $c_{\sigma}$. The essential low energy physics of HFMs are generally believed to be captured by
the Kondo lattice model and variations of it \cite{doniach1977kondo}: 
\begin{equation}\label{eq:kondo_ham}
    H=\sum_{k,\alpha} \epsilon_{c,k}c^{\dag}_{k,\alpha}c_{k,\alpha}+J_K\sum_{r,\alpha,\beta}(\vec{S}_r\cdot c^{\dag}_{r,\alpha}\vec{\sigma}_{\alpha\beta}c_{r,\beta}),
\end{equation}
where $\epsilon_{c,k}$ is the momentum ($k$) space dispersion of the conduction electrons. The localized valence spin at lattice site $r$ can be expressed in terms of Abrikosov fermions or spinons: 
\begin{equation}
\label{eq:spin_fermion}
\vec{S}_r=\sum_{\alpha,\beta}f^{\dag}_{r,\alpha}\frac{\vec{\sigma}_{\alpha\beta}}{2}f_{r,\beta}, 
\quad n_{fr} = \sum_{\alpha}f^\dag_{r,\alpha}f_{r,\alpha} = 1,
\end{equation}
which can potentially hybridize with the conduction electrons $c$. The Kondo interaction can be written in a way that makes the hybridization manifest by substituting \eqref{eq:spin_fermion} into a simplified form of \eqref{eq:kondo_ham} \cite{Coqblin1969}, and decoupling the quartic interaction by the introduction of a Hubbard-Stratonovich boson $b$:
 \begin{align}
    \label{eq:kondo_hyb}
    J_K\sum_{r,\alpha,\beta}(\vec{S}_r\cdot c^{\dag}_{r,\alpha}\vec{\sigma}_{\alpha\beta}c_{r,\beta})&\rightarrow g\sum_{r,\alpha,\beta} (c^{\dag}_{r,\alpha}f_{r,\alpha}b_r+\mathrm{H.c.}), 
\end{align}
subject to the constraint
$n_{fr} - b^\dag_r b_r = 1$.
At the mean-field level the boson $b_r$ is equal to the hybridization $\langle \sum_\alpha c_{r,\alpha}f_{r,\alpha}^{\dag}\rangle$. Microscopically, one can view the boson as a bound singlet of a valence spin with a conduction electron. The addition of a boson to a site $r$ must be accompanied by removal of a spinon $f_{r\sigma}$. Thus, the constraint $n_{fr}=1$ required for description of valence spins in terms of the fermions, is generalized to the above constraint in presence of the bosons. The fixed occupancy constraint implies a $U(1)$ gauge structure in these degrees of freedom \cite{senthil2003fractionalized,senthil2004weak}. The $b$ and $f$ then carry charges $1$ and $-1$ respectively under a $U(1)$ gauge field that they are minimally coupled to, with the fixed occupancy constraint enforced by its time component acting as a Lagrange multiplier \cite{WenBook}. In the FL$^\star$ phase, characterized by a small FS, the boson is gapped and the gauge theory is in the deconfined phase. The heavy FL phase with a large Fermi surface is established at a QCP at which the boson $b$ condenses, thereby confining the gauge field through the Higgs mechanism.  

Condensation of the valence fluctuations provides a simple understanding for the main features of the heavy FL phase \cite{doniach1977kondo,Kondo1962,Anderson1969}. As evident from \eqref{eq:kondo_hyb}, the condensed boson hybridizes the $f$ fermions with the conduction electron. Thus, the Fermi surface must grow to encompass the full density of conduction and valence electrons. The coherent mixing between the mobile conduction electrons with the localized $f$ spinons also explains the large effective mass, which is the hallmark of the heavy FL phase. However, an exact description of the aforementioned Higgs transition within this model is in general hard, as it involves fluctuating gauge fields coupled to multiple matter particles. 

The standard approach to make analytic progress in the valence fluctuation theory has been to artificially enlarge the $su(2)$ spin symmetry to $su(N)$, and take the large $N$ limit. The large number $N$ of $c$ and $f$ fermion species, controls an exact saddle point solution equivalent to a static mean-field theory, where $b_r=\langle \sum_\alpha c_{r,\alpha}f_{r,\alpha}^{\dag}\rangle$ is obtained self-consistently \cite{Anderson1981, coleman1984new, auerbach1986kondo}. Because the critical fluctuations of the boson and the gauge field are suppressed, the conduction electrons remain non-interacting, or at least good quasiparticles. Hence, this large $N$ limit is not a good starting point for obtaining non-trivial critical transport properties, and in particular, the strange metal phenomenology that we want to describe. The main point of this paper is to introduce a new large $N$ limit that retains solubility of the problem yet describes non-quasiparticle physics already at the saddle point level. The most important difference between our approach and the previous large $N$ theories is that we keep the fermion spin indices $\sigma$ $su(2)$ instead of promoting them to $su(N)$, and instead endow all three species $c,f,b$ with a flavor index $i=1,\ldots ,N$. The large $N$ modification we make then is: 
\begin{equation}
    c_{\sigma}, f_{\sigma}, b \rightarrow c_{i,\sigma}, f_{i,\sigma}, b_{i}, \quad i \in 1,\ldots,N, \quad \sigma \in 1,2.
\end{equation}
One can also continuously vary the ratio of the numbers of flavors of each particle type, however here we fix the same $N$ for all particles. Because all the species have a comparable number of flavors, their self-energies all remain $\mathcal{O}(1)$ within the large $N$ limit.

The second new feature in our generalized large $N$ limit is the introduction of a random ensemble of interaction constants, similar to recently studied ``low-rank" SYK models, which involve fermions with random Yukawa coupling to bosons \cite{Bi2017,Patel2018gauge,Marcus2019,Wang2020,Esterlis2019,Kim2020,Kim2021}. The random interactions should be viewed as a mathematical construct implementing a particular type of controlled large $N$ limit. We therefore consider the following family of model Hamiltonians; 
\begin{equation}\label{eq:model_ham}
\begin{aligned}
    H&=\sum_{\lambda\in\{c_\sigma,f_\sigma,b\}} H_\lambda + H_\text{int}, \\
    H_\lambda &= \sum_{i=1}^N \sum_k (\epsilon_{\lambda,k}-\mu_\lambda) \lambda_{k,i}^\dagger \lambda_{k,i}, \\
    H_\text{int} &= \frac{1}{N} \sum_{i,j,l=1}^N\sum_{r,\sigma} (g^r_{ijl}c_{r,i,\sigma}^\dagger f_{r,j,\sigma}b_{r,l} + \text{H.c.}), \\
    \sum_{i=1}^N&\left(\sum_\sigma f^\dagger_{r,i,\sigma}f_{r,i,\sigma} - b^\dagger_{r,i}b_{r,i}\right)= -N\kappa.
\end{aligned}
\end{equation}
Here $g^{r}_{ijl}$ are complex Gaussian random variables. We have included emergent dispersions $\epsilon_{\lambda,k}$ for $\lambda=f_\sigma,b$, which are expected to be generated when integrating out higher energy modes. The last line of (\ref{eq:model_ham}) is the large $N$ generalization of the occupancy constraint in (\ref{eq:kondo_hyb}) ($\kappa$ is a free parameter).

We consider two models for the coupling tensors $g^{r}_{ijl}$. In Model I these are taken to be uncorrelated between different sites $r$, whereas they are identical on all sites in Model II:
\begin{align}
&\text{Model I:}~~~\overline{g^{r}_{ijl}\, g^{r'}_{i'j'l'}} = g^2\delta_{rr'}\delta_{ii'}\delta_{jj'}\delta_{ll'}, \nonumber \\
&\text{Model II:}~~~\overline{g^{r}_{ijl}\, g^{r'}_{i'j'l'}} = g^2\delta_{ii'}\delta_{jj'}\delta_{ll'}.
\label{eq:disorder_correlators}
\end{align}
Thus, Model I is spatially disordered, and should be viewed as a depiction of HFMs with spatially disordered Kondo couplings. Model II, on the other hand, is translationally invariant and should be viewed as a model for clean systems. The averaging over flavors in both models eliminates various intractable Feynman diagrams \cite{MS2016,Esterlis2019,Kim2021}, thus allowing controlled access to the QCP at strong coupling.

While the $f$ and $b$ are also additionally coupled to the emergent $U(1)$ gauge field $a$, the coupling constant scales as $1/\sqrt{N}$: $H_{\lambda a} \sim (a/\sqrt{N})\sum_{i=1}^N \lambda^\dag_i \lambda_i$, for $\lambda=f,b$. This ensures that the gauge field fluctuations do not contribute to the $f,b$ self-energies in the large $N$ limit. Nonetheless, integrating out the emergent gauge field propagator leads to exact Ioffe-Larkin constraints on the current correlators \cite{Ioffe1989,Lee1992}, tantamount to imposing series addition of the conductivities of $f,b$ (Appendix~\ref{app:Ioffe-Larkin}).

In both models we assume simple quadratic dispersions $\epsilon_\lambda=k^2/(2m_\lambda)$ for all three species $c,f,b$. We choose the masses to be appropriate for the creation of a heavy FL phase upon condensing the bosons, hybridizing a heavy $f$ with the freely mobile conduction electrons $c$, and $b$ arising from hybridization of $c$ with the heavy $f$. We therefore take the hierarchy $m_b > m_f \gg m_c$. This choice of masses implies the bandwidths of $c$ and $f$ are large relative to that of $b$. The $c,f$ chemical potentials are chosen such that the respective densities are close to equal, motivated by stoichiometric considerations, and by an observed near doubling in Hall coefficient across the transition from the heavy FL to FL$^{\star}$ in CeCoIn$_5$ \cite{maksimovic2020}.

The transition between the FL$^\star$ and heavy FL phases occurs, as in previous theories, through condensation of the boson $b$. A natural parameter that can control the transition across the QCP in experiments is the total physical charge density, $n_{el}=\langle n_b\rangle+\langle n_c\rangle$, while the VL occupation and hence $\langle n_f\rangle-\langle n_b\rangle = -\kappa$ is held fixed. However, for convenience of calculation we tune $\kappa$ instead. The two approaches are approximately equivalent in the regime we consider, where the bandwidths of $c$ and $f$ are much larger than that of $b$, with the difference between them amounting to small relative changes in the $c,f$ occupations, which only make small changes in the physical properties of $c,f$, and therefore will not significantly alter our results.

\begin{figure}
\centering
 	\includegraphics[width=0.45\textwidth]{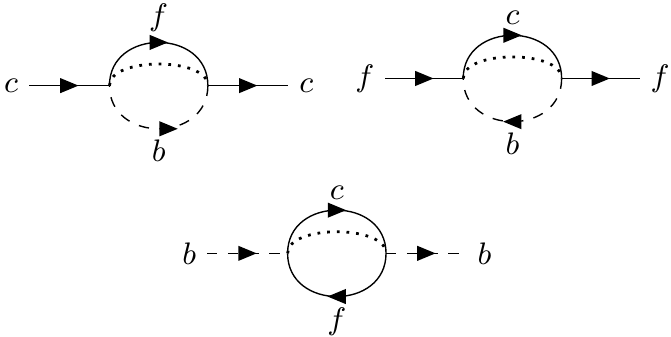}
\caption{The only contributing diagrams to the self-energies of the three particle species. All others are suppressed by the large $N$ limit or averaging over $g^r_{ijl}$. The averaging over the coupling tensors is indicated by the dotted lines, with its correlator given by (\ref{eq:disorder_correlators}). The dotted lines carry momentum for Model I but do not for Model II.}
\label{fig:SD_textfig}
\end{figure}

As in the case of previous work on SYK-like models \cite{MS2016,Esterlis2019,Kim2021}, the averaging over the coupling tensors in the large $N$ limit yields exact coupled Schwinger-Dyson (SD) equations for the Green's functions of the three species $c,f,b$, which we solve self-consistently throughout the phase diagram. The self-energies for these SD equations are shown in Fig.~\ref{fig:SD_textfig}. Using these, we compute non-perturbatively the $T$-dependent conductivity tensors in the two models, focusing in particular, on the critical regime.

In the analysis of Model I, we assume a special FL$^\star$ phase, in which the emergent gauge field, and thus also the $f$ and $b$ particles that are charged under it, are all deconfined only within individual 2D planes. The physical 3D system is a stack of these 2D layers. The behavior of the resistivity across the transition between the layered FL$^\star$ and the heavy FL phase is shown in Fig.~\ref{fig:phase_diagram}(a). In the quantum critical regime $\rho_{xx}$ shows a quasi-linear $T$ dependence (linear with a logarithmic correction). 

The nature of the critical MFL depends on a dimensionless coupling strength $\gamma$ between the bosons and fermions. For sufficiently strong coupling, the bosons are overdamped, and the QCP displays a near-universal ``Planckian" transport lifetime $\tau\sim \hbar/(k_B T)$, which is independent of all microscopic details of the model (up to logarithmic factors). In the opposite regime of weak damping ($\gamma\ll 1$), the critical behavior provides an example of a skewed MFL \cite{Georges2021}, in which the scattering rates of particle and hole excitations about the electron FS are different. The resistivity is linear in $T$ but sub-Planckian, and the fermion self-energies are asymmetric about $\omega=0$. On tuning across the QCP, the in-plane Hall coefficient $R_H$ computed for weak out-of-plane magnetic fields transitions between two plateau values that correspond to the different effective carrier densities of the FL$^{\star}$ and FL phase. In the weakly damped regime this change of $R_H$ is non-monotonic, developing a peak in the quantum critical region as a function of the tuning parameter $\kappa$ (Fig.~\ref{fig:phase_diagram}(b)). This enhancement of $R_H$ near criticality is reminiscent, yet much more modest than that observed in experiments on CeCoIn$_5$ \cite{maksimovic2020}. 

For Model II, we consider a fully 3D deconfined FL$^\star$ phase. We show that $\rho_{xx}$ is quasi-linear in $T$ in the critical region if the $f$ FS at the QCP matches that of the conduction electrons, and if the $f$ fermions and $b$ bosons additionally rapidly relax momentum via impurity scattering and/or self-interactions on the lattice. This is closely related to the work of Paul et. al. \cite{Paul2013}, who find a MFL for matching FS's coupled to a complex bosonic field under certain phenomenological assumptions. Within Model II however, this result is exact in the large $N$ limit. We further show that the two FS's may be naturally self-tuned to matching at the QCP, in order to maximize the free energy released when the bosons condense. Unlike in Model I, we find that the bosons in Model II are always overdamped, leading to Planckian transport lifetime at low temperatures independent of the coupling strength. Because of the overdamped nature of the bosons there is no enhancement of $R_H$ in Model II.

\section{Model I: Spatially Disordered Couplings}

In this section we solve for the Green's functions in Model I and calculate transport temperature dependence of transport quantities across the transition. We identify two regimes of the critical behavior, depending on the boson-fermion coupling strength. The calculation is exact in the large $N$ limit.

\subsection{Self-energies and phase diagram}

The starting point for obtaining the phase diagram and calculating the transport properties in this model at large $N$ are the coupled Schwinger Dyson equations for the Green's functions of the three species:
\begin{equation}
    \begin{aligned}
     G_c(i\omega) & = \frac{1}{V}\sum_k\frac{1}{i\omega-\epsilon_{c,k}+\mu_c-\Sigma_c(i\omega)},  \\
     G_f(i\omega) & = \frac{1}{V}\sum_k\frac{1}{i\omega-\epsilon_{f,k}+\mu_f-\Sigma_f(i\omega)},  \\
     G_b(i\omega) & = \frac{1}{V}\sum_k\frac{1}{-i\omega+\epsilon_{b,k}+\Delta_b-\Sigma_b(i\omega)}, \\
    \end{aligned}
    \label{eq:SDM1}
\end{equation}
where $V$ is the system volume. The self-energies $\Sigma_{c,f,b}$ in the large $N$ limit are given exactly by the diagrams in Fig.~\ref{fig:SD_textfig}, which read
\begin{equation}
    \begin{aligned}
        \Sigma_c(i\omega) &= g^2T\sum_{i\omega'} G_f(i\omega') G_b(i\omega-i\omega') , \\
     \Sigma_f(i\omega) &= g^2T\sum_{i\omega'} G_c(i\omega') G_b(i\omega'-i\omega) ,\\
      \Sigma_b(i\omega) &= -2g^2T\sum_{i\omega'} G_c(i\omega') G_f(i\omega'-i\omega). \\
    \end{aligned}
    \label{eq:SDmain}
\end{equation}
Here $V$ is the system volume, and the factor of $2$ in the equation for $\Sigma_b$ arises from the $su(2)$ spin degeneracy of $c$ and $f$. The self-energies only involve momentum-averaged Green's functions $G_{\lambda}(i\omega) = (1/V)\sum_k G_{\lambda}(i\omega,k)$ because the random interactions in Model I are uncorrelated between different sites. In the relevant regime where the fermion bandwidths are the largest scales, their momentum-averaged Green's functions take the simple form $G_{c,f}(i\omega) = -(i/2)\nu_{c,f}\,\text{sgn}(\omega)$ \cite{Patel2018}, where $\nu_{c,f}$ are the respective spinless densities of states at the Fermi energies. This allows to calculate the boson self-energy $\Sigma_b$
\begin{equation}
\Sigma_b(i\omega) = -2g^2T\sum_{i\omega'} G_c(i\omega') G_f(i\omega'-i\omega) = -\gamma|\omega|+C_b,
\end{equation}
here $\gamma=g^2\nu_c\nu_f/(2\pi)=g^2(3n_c)^{1/3}m_c m_f /(2\pi^{4/3})$ is a dimensionless coupling constant characterizing the strength of the boson damping. We will explain the effects of its magnitude on the physics of the system in the subsequent paragraphs. The $T$-independent constant $C_b$ can be absorbed by the $T=0$ chemical potential of the bosons.

\begin{figure}[b]
 	\centering
 	\includegraphics[width=0.48\textwidth]{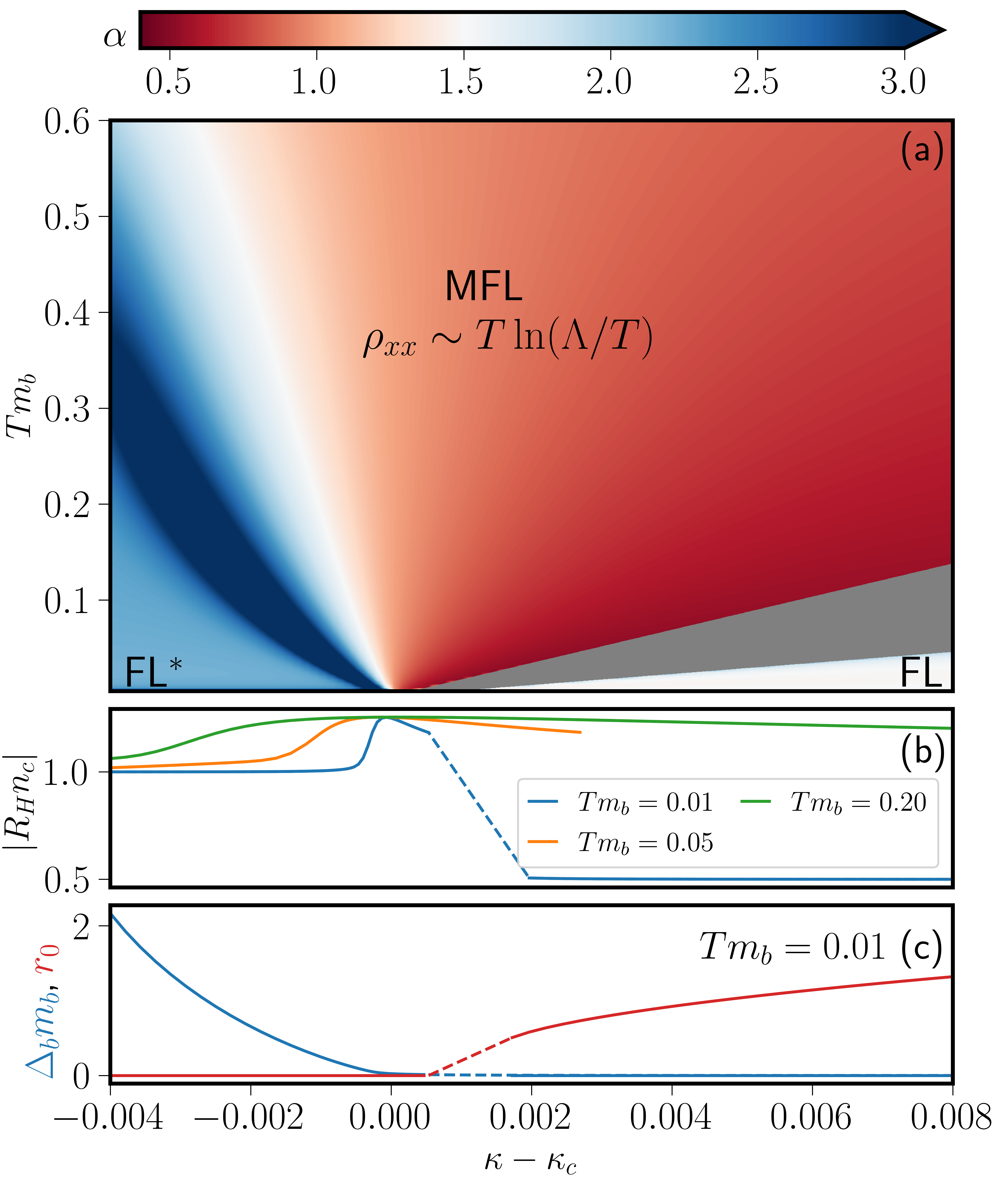}
     	\caption{(a) The phase diagram for Model I. We have $\rho_{xx}-\rho_{xx}(T=0) \sim T^\alpha \ln(\Lambda/T)$, and the color indicates the value of $\alpha=d\ln(\rho_{xx}/\ln(\Lambda/T))/d\ln(T)$. We exclude the (gray) crossover region where our approximate treatment of the 3D condensed phase breaks down. (b) The plot of $R_H$ at weak out-of-plane magnetic field vs. $\kappa-\kappa_c$. $R_H$ transitions non-monotonically between two plateau values controlled by the effective carrier densities in the FL$^\star$ and heavy FL phases respectively. The value of $R_H$ is enhanced in the quantum critical region. The dashed lines indicate points within the gray region that are omitted. Due to the different dimensionality of $f$ and $c$, the plateau value on the right is only roughly $1/(n_c+n_f)$. (c) The boson ``soft gap", $\Delta_b m_b$, and the strength of the boson condensate, $\langle b_{r,1} \rangle= r_0 \sqrt{N}$, are plotted vs. $\kappa-\kappa_c$ at low temperature. $\Delta_b$ is finite when $\kappa<\kappa_c$ and is exponentially suppressed when $\kappa>\kappa_c$, at which point inter-layer instabilities allow for a 3D boson condensate to form, forcing $\Delta_b=0$. Again, we omit the crossover between these two regimes (gray region). Here $\gamma=0.02\ll 1$, $n_c=n_f=1$, $\Lambda m_b=\pi^2/2$, $m_b=5 m_f=50 m_c$ (we set $\hbar=k_B=a_l=1$ everyhwere, where $a_l$ is the lattice constant).}
 	\label{fig:phase_diagram}
 \end{figure}

With the Green's functions in hand, the phase diagram is obtained by solving for the boson gap $\Delta_b(T)$ and the fermion chemical potential $\mu_f(T)$ that would satisfy the constraint $\langle n_f\rangle-\langle n_b\rangle = -\kappa$. In the relevant regime of large fermion bandwidth (or Fermi energy) compared to the temperature, the change in the fermion occupation with temperature is negligible. Therefore fixing $\kappa$ is essentially equivalent to fixing the boson occupation 
\begin{align}
\langle n_b\rangle=T\sum_\omega G_b(i\omega)=\kappa+\langle n_f\rangle,
\end{align}
where $\langle n_f\rangle$ is treated as a constant. The phase transition, associated with condensation of the boson, is then tuned by the parameter $\kappa$, analogous to the fixed length constraint in the $O(N)$ rotor model at large $N$ \cite{SachdevQPT}.  Similar to the rotor model, the boson occupation is fixed by solving for the variation of the ``soft gap'' $\Delta_b(T)$ in the boson Green's function \eqref{eq:SDmain} with temperature. 

The defining features of the zero temperature phases tuned by $\kappa$ are shown in  Fig.~\ref{fig:phase_diagram}(c).
In the FL$^*$ phase, obtained for $\kappa<\kappa_c$, the zero temperature gap $\Delta_b(0)$ is positive and vanishes continuously as $\kappa$ approaches the critical value $\kappa_c$. 
For $\kappa>\kappa_c$, on the other hand, one of the boson flavors is condensed at $T=0$ and acquires a condensate amplitude $|\langle b_{r,1}\rangle|=r_0\sqrt{N}$. This leads to the hybridization of the $f$ and $c$ fermion bands, which characterizes the heavy FL phase.  Details of the calculation are given in Appendix~\ref{app:gap}. 

The temperature dependence of the soft gap $\Delta_b(T)$ is crucial for determining the thermodynamic and transport properties. Solving the constraint equation at criticality we find that soft gap grows quasi-linearly with temperature as $\Delta_b(T)\sim T w_1(\gamma,T)$, where $w_1$ varies quasi-logarithmically with $T$ \footnote{The function $w_1(\gamma,T)$ vanishes quasilogarithmically as $T\rightarrow 0$, diverges logarithmically as $\gamma\rightarrow0$, and is quasi-linear in $\gamma$ for $\gamma\gg 1$}. Details of the calculation are given in appendix Appendix~\ref{app:limits}. In the $FL^\star$ phase ($\kappa<\kappa_c$) $\Delta_b(T)$ exhibits the critical behavior for $T\gg\Delta_b(0)$, while its temperature dependence is exponentially suppressed for $T\ll\Delta_b(0)$. 

In the heavy FL phase ($\kappa>\kappa_c$) the temperature dependence of $\Delta_b$ is more subtle because the $b$ and $f$ fermions are no longer confined to hop within planes in this phase. Once a condensate is established, the inter-layer interactions generate inter-layer hopping terms of the $b$ and $f$ partons of strength proportional to $r_0^2$, thus establishing a fully 3D Higgs phase (for full details see Appendix \ref{app:instabilities}). The approximate description of the Higgs phase in terms of a self-consistent 3D condensate remains valid in the heavy FL phase below a crossover temperature scale $T^\ast$ that vanishes at the QCP. Above the crossover scale $T^\ast$ the $b$ sector is dominated by 2D critical fluctuations \footnote{As is well known there is no phase transition between the low $T$ Higgs phase and high $T$ confined phase. Accordingly, there is no true finite $T$ Bose condensation transition, only a crossover}. In computing the transport properties for $\kappa>\kappa_c$ we will treat these two regimes separately, leaving out the more complicated crossover regime (gray region in Fig.~\ref{fig:phase_diagram}(a)).

We now turn to the fermion Green's functions, showing first that they accquire a MFL self-energy at the QCP. To calculate the fermionic self-energies $\Sigma_{c,f}$ we need the momentum-averaged $b$ Green's function: 
\begin{equation}
    \begin{aligned}
    G_b(i\omega)&=\int\frac{d^dk}{(2\pi)^d}\frac{1}{-i\omega+k^2/2m_b+\gamma|\omega|+\Delta_b} \\
    &\approx \frac{m_b}{2\pi} \ln \left(\frac{\Lambda}{-i\omega+\gamma|\omega|+\Delta_b}\right),~~d=2,
    \end{aligned}
    \label{eq:gbint}
\end{equation}
where $\Lambda = \pi^2/(2m_b)$ is the boson bandwidth. We always consider sufficiently low frequency and temperature such that $\mathrm{max}(|\omega|,\gamma|\omega|,\Delta_b)\ll \Lambda$. This ensures that self-energies remain smaller than the bandwidths of their respective species and thereby will keep our computations self-consistent. The logarithmic form in (\ref{eq:gbint}) is only obtained for 2D bosons. The QCP is defined by $\Delta_b=0$; when inserted into \eqref{eq:gbint} and \eqref{eq:SDmain}, we obtain MFL self-energies:
\begin{align}
\label{eq:sigmac}
\Sigma_{c}(i\omega,T=0) = g^2 \int \frac{d\omega'}{2\pi}  G_{f}(i\omega') G_b(i\omega-i\omega') \nonumber \\
=\frac{\gamma m_b}{2\pi \nu_c}\left[i\omega\ln\left(\frac{\sqrt{1+\gamma^2}}{e\Lambda/|\omega|}\right) + \cot ^{-1}(\gamma)|\omega|\right] + C_c. \nonumber \\
\Sigma_{f}(i\omega,T=0) = g^2 \int \frac{d\omega'}{2\pi}  G_{c}(i\omega') G_b(i\omega'-i\omega) \nonumber \\
=\frac{\gamma m_b}{2\pi \nu_f}\left[i\omega\ln\left(\frac{\sqrt{1+\gamma^2}}{e\Lambda/|\omega|}\right) - \cot ^{-1}(\gamma)|\omega|\right] + C_f.
\end{align}
The constants $C_{c,f}$ can be absorbed into $\mu_{c,f}$.

The parameter $\gamma$, related to the strength of damping of the $b$ bosons, allows us to tune between different physical regimes. In general we expect $\gamma$ to increase with the strength of the Kondo coupling $g$. In the limit of $\gamma \gg 1$, the analytic continuation of (\ref{eq:sigmac}) to real frequency gives
\begin{equation}
\mathrm{Im}[\Sigma_{c,f,R}](\omega,T = 0) = -\frac{\gamma m_b}{4\nu_{c,f}}|\omega|,    
\label{eq:sigmatrad}
\end{equation}
which is the traditional MFL form \cite{Varma1996}. On the other hand, when $\gamma \ll 1$, the fermion self-energies (\ref{eq:sigmac}) are {\it asymmetric} about $\omega = 0$:
\begin{align}\label{eq:sigmaskew}
&\mathrm{Im}[\Sigma_{c,R}](\omega,T = 0) = \frac{\gamma m_b}{2\pi\nu_{c,f}}|\omega|\left(-\frac{\pi}{2}-\cot^{-1}(\gamma)\mathrm{sgn}(\omega)\right), \\
&\mathrm{Im}[\Sigma_{f,R}](\omega,T = 0) = \frac{\gamma m_b}{2\pi\nu_{c,f}}|\omega| \left(-\frac{\pi}{2}+\cot^{-1}(\gamma)\mathrm{sgn}(\omega)\right). \nonumber
\end{align}
Thus, in this regime, our model provides a concrete example of a ``skewed" MFL \cite{Georges2021}. This skewed MFL is expected to have a nonvanishing Seebeck coefficient in the $T\rightarrow0$ limit  due to the asymmetric inelastic scattering rate in (\ref{eq:sigmaskew}) \cite{Georges2021,Arindam2020}. The nonvanishing Seebeck coefficient as $T\rightarrow0$, and the asymmetric frequency dependence of the electron spectral function, provide experimentally detectable signatures of the small $\gamma$ regime \footnote{The magnitude of the low-temperature Seebeck coefficient is $\sim k_B/e$ when $\gamma\ll1$, declining to zero as $\gamma$ is increased to $\gamma\gg1$.}. 

In the FL$^{\star}$ phase, where $\Delta_b(T = 0)>0$, we obtain, in a similar fashion to (\ref{eq:sigmac}), 
\begin{align}
\Sigma_{c,f}(i\omega,T = 0) &= -\frac{\gamma m_b\ln(\Lambda/\Delta_b(T = 0))}{\pi\nu_{c,f}}i\omega \nonumber \\
&+i\frac{\gamma^2 m_b}{2\pi\nu_{c,f}\Delta_b(T=0)}\omega^2.
\label{eq:flstarsigma}
\end{align}
The $\mathcal{O}(\omega^2)$ term leads to a Fermi liquid $\omega^2$ scattering rate on the real frequency axis, and hence a scattering rate $\propto \omega^2+\pi^2 T^2$ upon analytic continuation to the thermal circle for $T>0$. The $\mathcal{O}(\omega)$ term leads to a renormalization of the Fermi liquid quasiparticle weights, and hence an enhancement of the conduction electron effective mass, given by
\begin{equation}
m^\ast_c = m_c\left(1+\frac{\gamma m_b}{\pi \nu_c}\ln\left(\frac{\Lambda}{\Delta_b(T)}\right)\right).
\label{eq:effectivemass}
\end{equation}
Here, we extended the result to small nonvanishing temperatures by replacing $\Delta_b(0)\to \Delta_b(T)$. Since $\Delta_b(T = 0) \sim \kappa_c - \kappa$ vanishes on approach to the QCP, the zero temperature effective mass diverges, consistent with experimental findings in HFMs \cite{Stewart2001,Custers2003}. 
In the critical region $\Delta_b\propto T$ up to logarithmic corrections. Thus, the divergence of $m_c^\ast$ is cut-off logarithmically by the temperature at criticality.

We now calculate the imaginary part of the fermion self-energies at finite $T$, necessary for computing conductivities. The $c$ fermion self-energy in the Lehmann representation is given by: 
\begin{equation}
\begin{aligned}
    \Sigma_c(i\omega,T) 
    &= -g^2\int \frac{d\epsilon d\epsilon'}{(2\pi)^2} A_f(\epsilon)A_b(\epsilon') \frac{n_B(\epsilon')+n_F(-\epsilon)}{\epsilon'+\epsilon-i\omega},
\end{aligned}
\end{equation}
where ${n}_B,{n}_F$ are the Bose and Fermi functions at temperature $T$, $A_f(\epsilon) = -2 \text{Im}[ G_f^R(\epsilon)] = \nu_f$ is the fermion spectral function, and $A_b(\epsilon)$ the boson spectral function. 
We analytically continue $i\omega \to \omega + i \delta$ to obtain
\begin{equation}\label{eq:fermiselfenergy}
    \text{Im} [\Sigma_{c,R}(\omega,T)] = -g^2\nu_f \int \frac{d\epsilon }{4\pi} A_b(\epsilon) ({n}_B(\epsilon)+{n}_F(\epsilon-{\omega})).
\end{equation}
This expression also holds for $\text{Im}[\Sigma_{f,R}]$ with the change $\nu_f \to \nu_c$ and $\omega \to -\omega$.  The boson spectral function is derived in Appendix~\ref{app:gap} and is given by:
\begin{equation}
A_b(\omega) =\frac{m_b}{\pi}\left[\pi\Theta(\omega-\Delta_b) + \tan^{-1}\left(\frac{\gamma \omega}{\Delta_b-\omega}\right) \right],
\end{equation}
where $\Theta(x)$ is the Heaviside step function. Note that the temperature dependence of $A_b$ comes entirely from its dependence on $\Delta(T)$. We have shown that in the critical region $\Delta_b\propto T$ up to logarithmic corrections. Therefore, up to these corrections, the spectral function can be expressed as $A_b(\omega/T,z)$, with $z=\Delta_b/T$ a temperature independent constant. Using this expression in \eqref{eq:fermiselfenergy} and scaling the integration variable immediately gives a $T$-linear result up to the logarithmic corrections. We will show that this property implies near $T$-linearity of the resistivity. 

In the two limits $\gamma \gg \max(1,\Delta_b/T)$ and $\gamma\ll 1$ we obtain explicit expressions for the imaginary parts of the self-energy in the critical region (Appendix \ref{app:limits}). For large $\gamma$ we have
\begin{equation}
\begin{aligned}
    \mathrm{Im}[\Sigma_{c,R}(\omega,T)] &\approx -\frac{\gamma m_b}{2\pi\nu_c}T \Bigg[\frac{\Delta_b}{\gamma T}\ln\left(\frac{\Lambda e}{\Delta_b}\right)\nonumber \\
    &+\pi\ln\left(2\cosh\left(\frac{\omega}{2T}\right)\right)\Bigg]; \ \Delta_b/(\gamma T) < 1,
\end{aligned}
\label{eq:GammaLarge}
\end{equation}
\begin{equation}
    \Delta_b \approx \frac{\pi \gamma T}{\ln\left(\frac{\Lambda}{T \gamma e}\right)}W_0\left(\frac{2\sqrt{e}}{\pi^2}\ln\left(\frac{\Lambda}{T\gamma e}\right) \right),
\end{equation}
where $W_0(z)$ is the Lambert W function. For $\gamma \ll 1$, (\ref{eq:sigmac}) is well approximated by:
\begin{equation}
\mathrm{Im}[\Sigma_{c,R}(\omega,T)] \approx -\frac{\gamma^2 m_b}{2 \pi\nu_c} \, T\left(1+e^{\omega/T}\right),~|\omega|\lesssim T.
\label{eq:GammaSimple}
\end{equation}
Like at $T = 0$ (\ref{eq:sigmaskew}), this self-energy is asymmetric between positive and negative frequencies, and is therefore skewed.

\subsection{Conditions for Planckian dissipation}

It has been proposed that inelastic relaxation times, in most if not all situations, cannot be much smaller than the quantum mechanical ``Planckian'' time scale $\tau_P=\hbar/(k_B T)$ (see \cite{hartnoll2021planckian} and references therein). There is a growing list of materials, showing strange metal behavior at low temperatures, which seem to be close to this limit, namely they relax on the Planckian time scale up to a constant of order one \cite{Bruin804,Legros18,Paglione19,Cao2020,Patel2019}. Since the self-energies calculated above imply relaxation times proportional to $1/T$, it is interesting to ask how systems described by Model I line up with the proposed Planckian bound. 

Note however, that the correct quasiparticle relaxation time cannot be extracted directly as the inverse $\text{Im}\Sigma_R$. Rather it is renormalized by the same factor as the mass. To see this, we eliminate the prefactor of the $\omega$ term to obtain the standard Fermi liquid form of the Green's function 
\begin{equation}
    G_{c,R}(\omega,k)=\frac{Z}{\omega-Z\xi_k-i Z\text{Im}\left[\Sigma_{c,R}(\omega)\right]}
\end{equation}
with $Z=m/m^\ast_c$. From this we can immediately obtain   
$1/\tau_c=Z\,\text{Im}\left(\Sigma_{c,R}(\omega=0)\right)$.
This is the same timescale extracted from analysis of transport data pertaining to strange metal QCPs \cite{Bruin804,Legros18,Paglione19,Cao2020,Patel2019} using the Drude formula for quasi particle transport $\tau= m^\ast \sigma_{xx}/(n e^2)$. In the experiments the effective quasiparticle mass is measured slightly away from the critical point. 
Note that we focus here on the relaxation rates of the conduction electrons because, as shown in Sec. \ref{sec:transportI} below, they dominate the transport. 

In the strongly damped regime, where $\gamma\gg 1,\Delta_b/T$, equations  \eqref{eq:effectivemass} and \eqref{eq:GammaLarge} give
\begin{align}
    \tau_c &=\left(\frac{\pi\nu_c}{\gamma m_b}+\ln\left(\frac{\Lambda}{\gamma T}\right)\right)\frac{\hbar}{k_B T} \nonumber\\
    &\approx \ln\left(\frac{\Lambda}{\gamma T}\right)\frac{\hbar}{k_B T},
\end{align}
At realistic temperatures $\tau_c$ can be viewed as Planckian relaxation modified only by a slowly varying logarithmic function of temperature and nearly independent of the microscopic couplings. The result provides an appealing potential explanation for observation of near Planckian relaxation across different materials, with $\mathcal{O}(1)$ proportionality constants that vary only slightly between materials \cite{Bruin804}.

In the weakly damped regime $\gamma\ll 1$ equations \eqref{eq:effectivemass} and  \eqref{eq:GammaSimple} give
\begin{align}
    \tau_c ={1\over \gamma}\left(\frac{\pi\nu_c}{\gamma m_b}+\ln\left(\frac{\Lambda}{T\ln(\pi/\gamma)}\right)\right)\frac{\hbar}{k_B T},
    \end{align}
which is manifestly nonuniversal. The proposed Planckian lower bound is still obeyed, but exceeded by a large factor of at least $1/\gamma$. Thus we do not expect Planckian transport in the weak damping regime.

\begin{figure*}[ht]
    \includegraphics{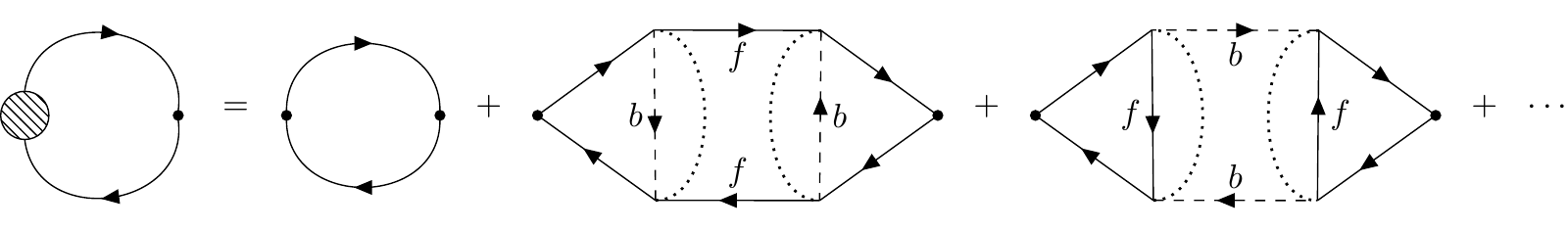}
    \caption{The diagrams that contribute to the $c$ conductivity. These diagrams are not suppressed by the large $N$ limit, but only the first (bubble) diagram is nonzero in Model I; in Model II, the corrections to the bubble do not identically vanish, but their effects are nevertheless suppressed (see main text). As in Fig.~\ref{fig:SD_textfig}, dotted lines indicate the averaging over the flavor random couplings $g^r_{ijl}$, which carry momentum in Model I (but not in Model II). Consequentially, the momentum integrals in the left and right loops of the correction diagrams are decoupled only in Model I. The diagrams that contribute to the $f$ and $b$ conductivities are analogous to the ones above. The diagrams that contribute to the cross-correlations of currents of different species are analogous to the vertex diagrams correcting the bubble diagram above, and also vanish in Model I (but not in Model II).}
    \label{fig:FD_polbubble}
\end{figure*}

\subsection{Transport}
\label{sec:transportI}

The computation of transport properties is greatly simplified in Model I due to the spatially disordered coupling $g^r_{ijl}$. To clarify this point, let us first ignore the effects of the emergent $U(1)$ gauge field. In this case the Kubo formula for Model I takes a particularly simple form involving only the bare bubble diagram for each of the three species (the first diagram in the series shown in Fig.~\ref{fig:FD_polbubble}). To see this, first note that only vertex corrections with non-crossing boson lines can potentially contribute in the large $N$ limit. However, in such diagrams, the momentum integral on the loop containing the bare current vertex is decoupled from the rest of the diagram due to averaging over the site-uncorrelated couplings $g^r_{ijl}$. Once decoupled, these loop integrals vanish because the current vertices and the propagators 
on the loop have opposite parities under spatial inversion. 
Note that all cross species current correlations must involve vertex corrections, which vanish by the same mechanism. Thus the conductivities associated with the different species can be separately calculated from their respective bubble diagrams. Physically, these diagrams describe current decay due to scattering of fermions on critical bosons, which is not momentum conserving due to the spatially disordered couplings.

The effects of the emergent $U(1)$ gauge field on transport can now be included by integrating it out exactly in the large $N$ limit. This leads to a Ioffe-Larkin composition rule for the in-plane conductivities of the three species, described by the respective bubble diagrams (see Appendix~\ref{app:Ioffe-Larkin}) \cite{Ioffe1989,Lee1992}: 
\begin{equation}
    \pmb{\sigma} = \begin{pmatrix}\sigma_{xx} & \sigma_{xy} \\ -\sigma_{xy} & \sigma_{yy} \end{pmatrix}=\pmb{\sigma}_c + (\pmb{\sigma}_b^{-1}+\pmb{\sigma}_f^{-1})^{-1} \equiv \pmb{\sigma}_c + \pmb{\sigma}_{bf}.
    \label{eq:IL}
\end{equation}
In other words, the conductivities of the $f$ fermions and the bosons, which carry a $U(1)$ gauge charge, are added in series and their combined current is added in parallel to that of the conduction electrons. 

The transport properties of the two phases can be easily understood from this composition rule. In the heavy fermi liquid phase, obtained for $\kappa>\kappa_c$, the boson is condensed and therefore contributes zero resistance to the in-series addition. The total conductivity is then a result of adding the $f$ and $c$ fermions currents in parallel, consistent with the expected increase of the carrier number associated with the large Fermi surface. In the FL$^\ast$ phase, obtained for $\kappa<\kappa_c$, the boson conductivity vanishes at zero temperature due to the soft gap. The combined conductivity of the bosons with the $f$ fermions also vanishes due to the series addition. Therefore the total conductivity is equal to just that of the conduction electrons $\pmb{\sigma}=\pmb{\sigma}_c$, compatible with a small fermi surface consisting of only those electrons.

\begin{figure}[htpb]
 	\centering
 	\includegraphics[width=0.48\textwidth]{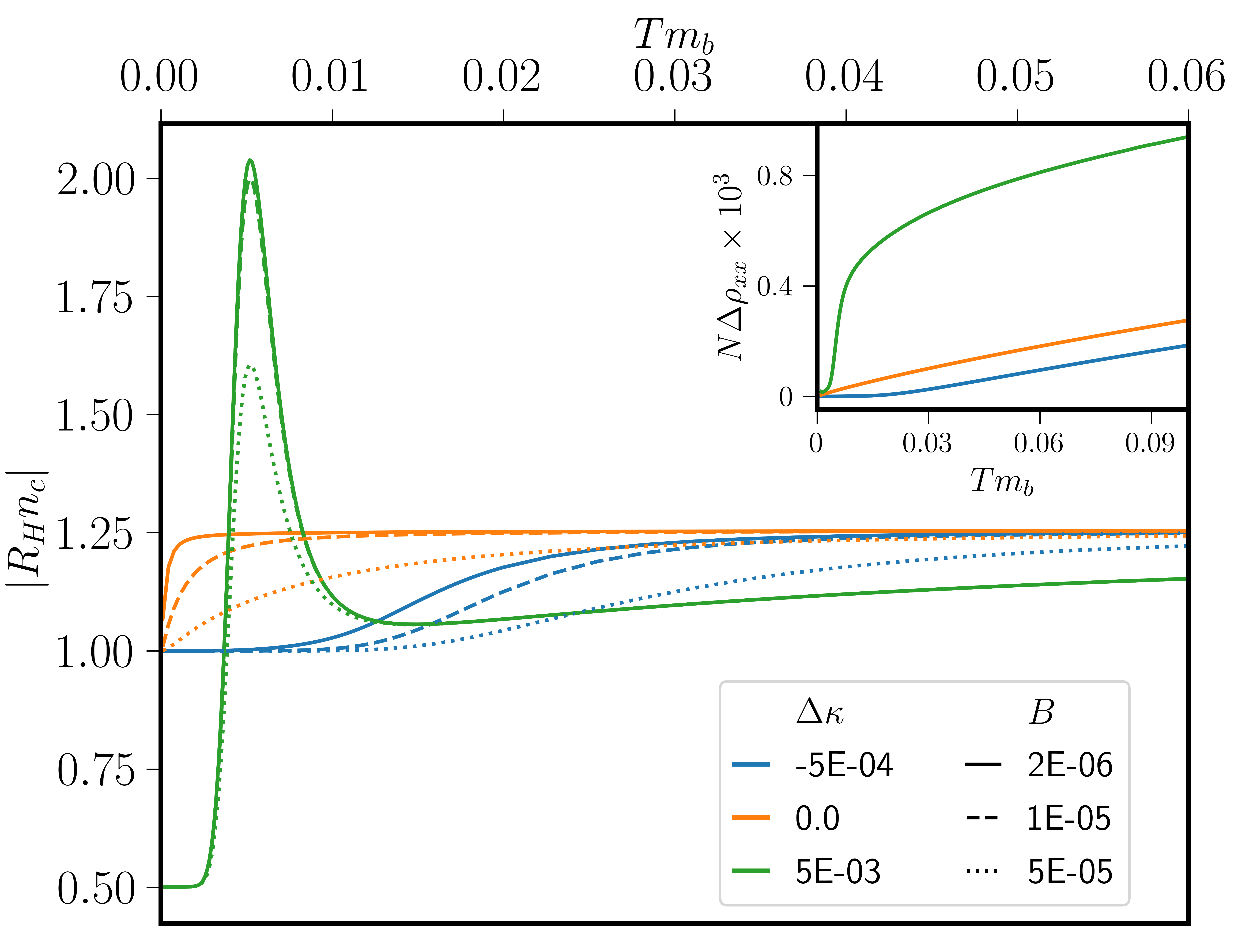}
 	 \caption{$R_H$ vs. $T$ in Model I for various $B$ and $\Delta \kappa=\kappa-\kappa_c$, computed numerically without any approximations. $R_H$ is roughly constant within the critical region and is higher than the expected $R_H\approx-1/n_c$ seen in the FL$^\star$ region (blue and orange curves). A larger $B$ suppresses $R_H$ slightly. There is a large enhancement in the crossover region between the condensed bosons and the quantum critical region, when we ignore inter-layer instabilities for $\kappa>\kappa_c$ (green curve). (Inset) $\Delta \rho_{xx}\equiv \rho_{xx}-\rho_{xx}(T=0)$ vs. $T$ for different values of $\Delta \kappa$. The other parameters are the same as in Fig.~\ref{fig:phase_diagram}.} 
 	\label{fig:Rhrhoxx_cuts}
\end{figure}

We now argue that in the quantum critical region at finite temperatures the transport is also dominated by the conduction electrons. To obtain the boson contribution $\sigma_b$, note that in the critical regime we have $\Delta_b(T) \sim T$ (up to logarithms), which retains the scaling of the Green's function as $1/\omega$. A simple scaling analysis of the bubble diagram then shows that $\pmb{\sigma}_b \sim T^0$, much smaller than $\sigma_f\sim 1/T$. Thus, the small boson conductivity bottlenecks the series addition with the spinons. Then the total conductivity is dominated by the much larger $\sigma_c\sim 1/T$ added in parallel. 
We confirm by exact numerical evaluation that indeed the total conductivity in the critical region is dominated by the conduction electrons (Fig.~\ref{fig:Rhrhoxx_cuts} inset). 

The longitudinal resistivity of the conduction electrons, derived from the bubble diagram in Fig.~\ref{fig:FD_polbubble}, takes the form  \cite{Patel2018} (see also Appendix \ref{app:Kubo}):
\begin{equation}
        \rho_{c,xx}=T\left(\frac{n_c}{8m_c} \int_{-\infty}^{\infty} d\omega \frac{\text{sech}^2(\omega/(2T))}{|\text{Im}[\Sigma_{c,R}(\omega,T)]|}\right)^{-1}.
        \label{eq:rxxc}
\end{equation}
In the critical region $\text{Im}[\Sigma_{c,R}(\omega,T)]\sim T$ for $|\omega|\lesssim T$, so that the integral in \eqref{eq:rxxc} is independent of $T$ at leading order. Thus we get nearly $T$-linear resistivity in the critical strange metal. 

In the FL$^\star$ phase we found in (\ref{eq:flstarsigma}) that  $|\text{Im}[\Sigma_{c,R}(\omega,T)]| \propto \omega^2+\pi^2 T^2$. Plugging this into (\ref{eq:flstarsigma}) gives $\rho_{xx}\propto T^2$ as in a normal Fermi liquid (Fig.~\ref{fig:phase_diagram}(a)).

In the heavy FL phase, the boson conductivity diverges due to the condensation of $\langle b_{r,1}\rangle \sim r_0\sqrt{N}$ and the Ioffe-Larkin composition rule therefore implies the parallel addition of the $c$ and $f$ conductivities. The condensate also generates inter-layer hopping of the bosons and spinons, which in return stabilize the condensate, within this mean-field treatment, at nonvanishing low temperatures. Details of this self-consistent model are described in Appendix~\ref{app:instabilities}.

Note that the $c$ and $f$ fermions continue to couple to the $N-1$ uncondensed gapless boson flavors $b_{2,...,N}$.  The 3D boson dispersion for $b_{2,...,N}$ implies that we must compute the equivaluent of (\ref{eq:gbint}) with an additional integral over the out-of-plane momentum, which leads to it having a $\omega^{1/2}$ frequency dependence (instead of $\ln(\omega)$), and subsequently to $\mathrm{Im}[\Sigma_{c,f,R}]\sim \mathrm{const}.+\mathrm{max}(T^{3/2},\omega^{3/2})$. This results in a resistivity that behaves as $\rho_{xx}\sim \mathrm{const}.+T^{3/2}$ at low $T$ as seen in Fig.~\ref{fig:phase_diagram}(a), where the constant contribution to $\mathrm{Im}[\Sigma_{c,f,R}]$ (and therefore $\rho_{xx}$) is generated by scattering off of the condensed $b_1$ mode. The $N-1$ uncondensed boson modes leading to the $T^{3/2}$ correction exist only as an artifact of the large $N$ limit, and they will not be present in the physical $N=1$ limit. Therefore, in the physical system we expect the finite $T$ corrections to the resistivity in the heavy FL phase to be weaker than $T^{3/2}$.

At nonzero out-of-plane magnetic fields, $B\neq 0$, $\pmb{\sigma}$ may be computed by expressing the Kubo formula in the basis of Landau levels, since the local self-energies are spatially independent. Vertex corrections to the current correlation functions continue to vanish even when $B\neq 0$ (Appendix \ref{app:Kubo}). As a result of integrating out the emergent $U(1)$ gauge field, the in-plane $\pmb{\sigma}_{b,f}$ are computed in presence of renormalized magnetic fields produced by the response of the emergent $U(1)$ gauge field to the (weak) external magnetic field $B$ (Appendix \ref{app:Ioffe-Larkin}):
\begin{equation}
    B_f = B \frac{\chi_b}{\chi_f+\chi_b},~B_b = B \frac{\chi_f}{\chi_f+\chi_b}.
    \label{eq:Beff}
\end{equation}
Here $\chi_y$ is the diamagnetic susceptibility for species $y$. We set $\chi_f = 1/(24 \pi m_f)$, {\it i.e.} the free fermion Landau diamagnetic susceptibility, corrections to which are suppressed by the large $f$ bandwidth (see Appendix~\ref{app:diamag}), and $\chi_b$ to its zero field value as we are only concerned with small $B$.

In the FL$^{\star}$ phase and the quantum critical region, since the transport is dominated by the conduction electrons as discussed above, we can express the weak-field Hall coefficient as 
\begin{equation}
\begin{gathered}
        R_H \approx R_H^c = \frac{\sigma_{c,xy}}{(\sigma_{c,xx})^2} \\ =-\frac{4T}{n_c}\frac{\int_{-\infty}^{\infty} d\omega\, \text{sech}^2\left(\omega/(2T)\right) \text{Im}[\Sigma_{c,R}(\omega,T)]^{-2}}{\left(\int_{-\infty}^{\infty}d\omega \,\text{sech}^2\left(\omega/(2T)\right) \text{Im}[\Sigma_{c,R}(\omega,T)]^{-1}\right)^2}. 
        \label{eq:RHc}
\end{gathered}
\end{equation}
When $\mathrm{Im}[\Sigma_{c,R}(\omega,T)]$ is independent of $\omega$, we get $R_H\approx -1/n_c$. Thus an enhancement of $R_H$ beyond this value requires a strong frequency dependence of $\text{Im}[\Sigma_{c,R}(\omega,T)]$ at $|\omega|\lesssim T$, as otherwise $R_H$ would be independent of the self-energy. In the FL$^\star$ phase, $\mathrm{Im}[\Sigma_{c,R}(\omega,T)]\propto \omega^2+\pi^2 T^2$, and $R_H\approx -1.05/n_c$. We find that in the quantum critical region, for weak damping $\gamma\ll 1$, $R_{H} \approx -4/(3n_c)$, which can be obtained by inserting (\ref{eq:GammaSimple}) into (\ref{eq:RHc}). Therefore, there is an enhancement of $R_H$ upon entering the quantum critical region from the FL$^\star$ phase. In the strongly damped $\gamma\gg 1$ regime, the frequency dependence of $\text{Im}[\Sigma_{c,R}(\omega,T)]$ in the quantum critical region (\ref{eq:GammaLarge}) is weaker than that in (\ref{eq:GammaSimple}), and consequently $R_H\approx -1.07/n_c$ in the quantum critical region, which is a negligible enhancement over the FL$^\star$ phase. In Fig.~\ref{fig:Rhrhoxx_cuts} we demonstrate the enhancement of $R_H$ for $\gamma\ll 1$, seen when crossing from the FL$^{\star}$ phase to the quantum critical region as a function of $T$, by computing the total conductivity numerically without any approximations. The enhancement of $R_H$ is suppressed by magnetic field and sharpened (as a function of $\kappa$) with decreasing temperature. \ref{fig:phase_diagram}(b) shows the enhancement of $R_H$ in the crossover between the two regimes as a function of the tuning the parameter $\kappa$ at constant temperature. This enhancement is more modest than that observed in experiments on CeCoIn$_5$ \cite{maksimovic2020}. 

We have noted already that upon tuning $\kappa$ into the heavy FL phase ($\kappa>\kappa_c$), the total conductivity tensor is simply $\mathbf{\sigma}_c+\mathbf{\sigma}_f$, because the boson is superconducting and connected in series to the $f$. Moreover, in the presence of an external magnetic field, the Meissner effect generated by the superconducting boson leads to a divergent susceptibility $\chi_b$ that screens the magnetic field seen by the boson, while the $f$ fermions see the full magnetic field up to the small Landau diamagnetism.  Consequently the Hall effect is just as it would be for a Fermi liquid composed of both the $c$ and $f$ fermions, $|R_H|n_c=n_c/(n_c+n_f)=1/2$. Thus, as seen in Fig.~\ref{fig:phase_diagram}(b), $|R_H| n_c$ changes from $\approx 1$ in the FL$^{\star}$ phase to $\approx 1/2$ in the heavy FL phase. 

Note, however that the calculation performed to obtain these plots is interrupted in the grayed out crossover region of Fig.~\ref{fig:phase_diagram}(a) between the critical and heavy FL regime. We can attempt to capture $R_H$ in this region by continuing the calculation from the critical regime, with the boson fluctuations decoupled between 2D layers, down to low temperatures. In this case we find a strong enhancement of the Hall coefficient over an intermediate temperature window (Fig.~\ref{fig:Rhrhoxx_cuts}) in the weakly damped $\gamma\ll 1$ regime. The enhancement is dominated by the contribution of the boson conductivity $\pmb{\sigma}_b$ to the total conductivity $\pmb{\sigma}$ . The strong non-monotonic behavior stems from a competition between two effects. On the one hand the boson gap decreases rapidly with decreasing temperature and becomes exponentially suppressed below the grayed out crossover regime,  $\Delta_b\sim T\exp\left[-\frac{2\pi(\kappa-\kappa_c)}{Tm_b}\right]$. This leads to a large ${\sigma}_{b,xy}$ due to bosons excited above the small gap. On the other hand, the susceptibility $\chi_b$ diverges rapidly ultimately leading to vanishing of $B_b$ and hence also of $\pmb{\sigma}_b$ at zero temperature. The interplay between these two effects leads to the sharp peak in $|R_H|$ versus temperature seen in Fig.~\ref{fig:Rhrhoxx_cuts}. This strong enhancement is more reminiscent of the experimental results on CeCoIn${}_5$ \cite{maksimovic2020}.

We note that when the boson is strongly damped, with $\gamma\gg 1$, this mechanism for enhancement of $R_H$ is not effective because the boson becomes nearly particle-hole symmetric with $G_b(i\omega)\approx G_b(-i\omega)$. 

\section{Model II: Translationally Invariant Couplings}

In this section we consider the model (\ref{eq:model_ham}) with random tensor couplings that are the {\it same} on all lattice sites, satisfying $\overline{g^{r}_{ijl}\, g^{r'}_{i'j'l'}} = g^2\delta_{ii'}\delta_{jj'}\delta_{ll'}$.
We also assume that, in the FL$^\star$ phase, the $U(1)$ gauge field is fully deconfined in three dimensions.
Due to the momentum conservation, the SD equations (with self-energies given by Fig.~\ref{fig:SD_textfig}) now involve momentum {\it dependent} (rather than momentum-averaged) Green's functions. We further specialize to the case where the $c$ and $f$ FS match \cite{Paul2013}, which we will demonstrate is a natural condition.
We will then continue to compute the transport quantities in parallel to the analysis of Model I. 

\subsection{Matched Fermi surfaces}

We argue that the matching of the FS's of $c$ and $f$ fermions is not as fine tuned a condition as it might appear. First, an equal site occupation $n\approx 1/2$ in both bands is in many cases a natural result of stoichiometry \cite{maksimovic2020}. But though having equal Fermi surface volumes is a necessary condition, it does not necessarily imply matching. 
A key point is that the $f$ fermions are emergent degrees of freedom (partons), whose dispersion is generated dynamically, unlike the dispersion of the $c$ which is fixed by microscopic material parameters. Below we argue that the dynamical variable that controls the dispersion of the $f$ fermions self tunes to match the FS of the $c$ fermions at the critical point as such matching maximizes the free energy relieved by condensation of the $b$.

To demonstrate the energetic mechanism behind the matching of the FS's, we assume that the $c$ and $f$ FS's are ellipsoidal, with the dispersions
\begin{equation}
\begin{aligned}
&\epsilon_{c,k} = \frac{k_x^2}{2m_{c,x}} + \frac{k_y^2}{2m_{c,y}} + \frac{k_z^2}{2m_{c,z}},  \\
&\epsilon_{f,k} = \frac{k_x^2}{2m_{f,x}} + \frac{k_y^2}{2m_{f,y}} + \frac{k_z^2}{2m_{f,z}},
\end{aligned}
\end{equation}
and that they have the same volume 
\begin{equation}
    \begin{aligned}
        V_{\mathrm{FS}}&=\mu_c^{3/2}\sqrt{2m_{c,x}m_{c,y}m_{c,z}}/(3\pi^2) \\ 
        &=\mu_f^{3/2}\sqrt{2m_{f,x}m_{f,y}m_{f,z}}/(3\pi^2).
    \end{aligned}
\end{equation}

The ratios $r_{\alpha=\{c,f\};\beta=\{y,z\}}=m_{\alpha,\beta}/m_{\alpha,x}$ control the shape of the Fermi surfaces. We will treat the parameters of the $f$ dispersion as variational parameters that minimize the ground state energy of the system upon boson condensation. When the boson is uncondensed, the grand free energy of the non-interacting fermion system at $T=0$ is $F_0 = -(2/5)V_{\mathrm{FS}}(\mu_c+\mu_f)$, not taking into account the fluctuations of the bosons. Upon condensing $b\rightarrow b_0$, and ignoring the remaining boson fluctuations, the mean-field Hamiltionian is
\begin{align}
H_0 &= \sum_{k,\sigma}\left[\left(\epsilon_{c,k}-\mu_c\right)c^\dagger_{k,\sigma}c_{k,\sigma}+\left(\epsilon_{f,k}-\mu_f\right)f^\dagger_{k,\sigma}f_{k,\sigma}\right] \nonumber \\
&+ b_0 \sum_{k,\sigma}\left[c^\dagger_{k,\sigma}f_{k,\sigma}+\mathrm{H.c}\right] + E(b_0),
\end{align}
where $E(b_0)\sim -b_0^2+b_0^4 < 0$ is the grand free energy arising from the purely bosonic part of the Hamiltonian. We then determine the change in grand free energy at $T=0$ of the two fermion bands produced by diagonalizing the $2\times 2$ $c,f$ Hamiltonian;
\begin{equation}
\begin{aligned}
F - F_0 &= \sum_{\pm}\int\frac{d^3k}{2\pi^3}\bigg((\epsilon_{c,k}-\mu_c)
 +(\epsilon_{f,k}-\mu_f) \\
&\pm\sqrt{(\epsilon_{c,k}-\epsilon_{f,k}+\mu_f-\mu_c)^2+4b_0^2}\bigg)  \\
&\times \theta\bigg(\mu_c+\mu_f-\epsilon_{c,k}-\epsilon_{f,k} \\
&\mp\sqrt{(\epsilon_{c,k}-\epsilon_{f,k}+\mu_f-\mu_c)^2+4b_0^2}\bigg) \\
&+\frac{2}{5}V_{\mathrm{FS}}(\mu_c+\mu_f) + E(b_0). 
\end{aligned}
\end{equation}

\begin{figure}[ht]
 	\centering
 	\includegraphics[width=0.5\textwidth]{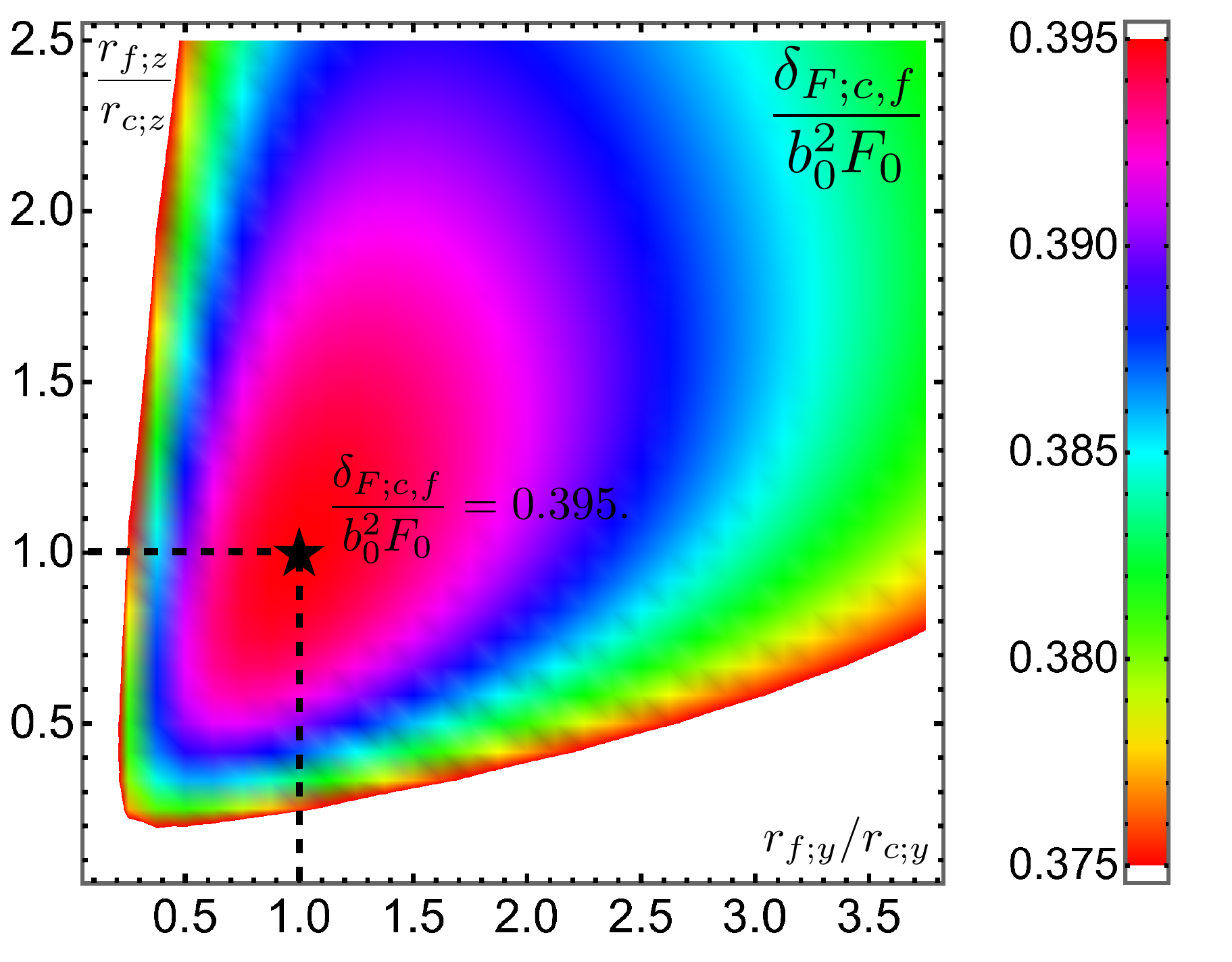}
     	\caption{The relative grand free energy relieved, $\delta_{F;c,f}/(b_0^2F_0)$, vs the eccentricity ratios controlling the shape of the ellipsoidal $f$ Fermi surface relative to that of the $c$ Fermi surface, for $\mu_f=\mu_c/10$. We use $V=20\sqrt{2}/(3\pi^2)$, $m_{c,x}=1.0$, $r_{c;y}=0.8$, $r_{c;z}=1.2$. It is maximized when the $f$ and $c$ Fermi surfaces are of the same shape, {\it i.e.} $r_{f;y} = r_{c;y}$ and $r_{f;z} = r_{c;z}$ (star).} 
 	\label{fig:Model2MatchedFS}
 \end{figure}

We can now consider the set of parameters for $f$ that maximize $\delta_{F;c,f}=F_0-F+E(b_0)$, which is the fermion contribution to the grand free energy relieved by boson condensation. The total grand free energy relieved, $F_0-F$, then is also maximized for fixed $b_0$. We know that, physically, the $f$ bandwidth is much smaller than the conduction electron bandwidth, so we fix $\mu_f$ at some value $\mu_f\ll \mu_c$. Eliminating $m_{f,x}$ through this and the constraint on $V$, we then vary the remaining parameters $r_{f;y}$ and $r_{f;z}$. We indeed find that $\delta_{F;c,f}$ is maximized when $r_{f;y,z}=r_{c;y,z}$ respectively (Fig.~\ref{fig:Model2MatchedFS}), implying the matching of the $c$ and $f$ Fermi surfaces in our toy mean-field calculation. We will study the renormalization of the $f$ dispersion at strong coupling beyond the mean-field level (which can be obtained by exact numerical solution of the SD equations and by minimizing the total interacting grand free energy at the large $N$ saddle point) in future work.

 \subsection{Self-energies and phase diagram}
 
The SD equations for Model II with the matched FS that we have motivated are given by:
\begin{equation}
    \begin{aligned}
     G_c(i\omega,k) & = \frac{1}{i\omega-\epsilon_{c,k}+\mu_c-\Sigma_c(i\omega,k)},  \\
     G_f(i\omega,k) & = \frac{1}{i\omega-\epsilon_{f,k}+\mu_f-\Sigma_f(i\omega,k)},  \\
     G_b(i\omega,k) & = \frac{1}{-i\omega+\epsilon_{b,k}+\Delta_b-\Sigma_b(i\omega,k)}. 
    \end{aligned}
\end{equation}
These equations are complemented by the expressions for the self-energies (diagrams in Fig.~\ref{fig:SD_textfig})
\begin{equation}
\begin{aligned}
\Sigma_{c}(i\omega,k) &= g^2T\sum_{i\nu}\int\frac{d^3q}{(2\pi)^3}G_{f}((i\omega+i\nu),k+q) \\
& \times G_b(-i\nu,q), \nonumber
\end{aligned}
\end{equation}
\begin{align}
\Sigma_{f}(i\omega,k) &= g^2T\sum_{i\nu}\int\frac{d^3q}{(2\pi)^3}G_{c}(-(i\omega+i\nu),k+q) \nonumber \\
& \times G_b(-i\nu,q), \\
\Sigma_b(i\omega,k) &= -2g^2T\sum_{i\nu}\int\frac{d^3q}{(2\pi)^3}G_c(i\omega+i\nu,k+q) \nonumber \\
& \times G_f(i\nu,q). \nonumber
\end{align}

Here again, the factor of $2$ in the equation for $\Sigma_b$ arises from the $su(2)$ spin degeneracy of $c$ and $f$. 
Although the self-energies here can have momentum dependence due to the translational invariance of Model II, let us assume to begin with that the fermionic ones are independent of momentum for $k$ near the FS, that is $\Sigma_{c,f}(i\omega,k)=\Sigma_{c,f}(i\omega)$. We will see below that this is a self-consistent assumption. 

With the assumption of momentum independent fermionic self-energies, we can average the contributions to the bosonic self-energy coming from small patches of the FS \cite{Metlitski2010}. The contribution from a given patch is 
\begin{align}
&\Sigma_b^p(i\omega,k) = \nonumber \\ &-2g^2T\sum_{i\nu}\int\frac{dq_\perp}{2\pi}\frac{d^2q_\parallel}{(2\pi)^2}
\Bigg(i\nu-v_{f,F} q_\perp-q_\parallel^2/(2m_f) \nonumber \\
&-\Sigma_f(i\omega)\Bigg)^{-1}\times\Bigg(i\nu+i\omega-v_{c,F} (q_\perp+k_\perp) \label{eq:bpse} \\
&-(q_\parallel+k_\parallel)^2/(2m_c)-\Sigma_c(i\omega+i\nu)\Bigg)^{-1}, \nonumber
\end{align}
where $\perp,\parallel$ define the directions relative to the patch of the matched FS, $v_{c,f,F}$ are the Fermi velocities, and $m_{c,f}$ are the fermion masses. After integrating over $q$ and averaging over patches (see Appendix~\ref{app:SDM2}), we obtain
\begin{equation}
\Sigma_b(i\omega,k) \approx -2 g^2 m_c m_f \frac{|\omega|}{k} \equiv -\gamma_2\frac{|\omega|}{k}.
\label{eq:bse}
\end{equation}
Here $\gamma_2=2g^2 m_c m_f$ is the natural dimensionless coupling for the boson damping, akin to $\gamma$ in Model I. The FS matching allows a small momentum boson ($k\to 0$) to decay into $c$-$f$ particle-hole pairs,  resulting in a low $k$ singularity of the boson self-energy. This self-energy is identical to the ``Landau damping" form \cite{Metlitski2010} of low momentum bosons coupled to low energy particle-hole excitations about the FS of an ordinary metal. The Landau damping we obtain implies a dynamical exponent $z=3$ for the critical bosonic fluctuations which has been shown to lead to MFL phenomenology in $d=3$ \cite{Paul2007,Paul2008}, as we will also explain in the following paragraphs.  

Having calculated the boson self-energy, we may now determine the boson gap $\Delta_b(T)$ using the occupancy constraint as done above for Model I. Similarly to Model I, we get a QCP separating the FL$^\star$ phase for $\kappa<\kappa_c$, where the boson is soft gapped at zero temperature and the heavy FL phase for $\kappa>\kappa_c$, where the boson is condensed $\langle b_1\rangle \sim r_0\sqrt{N}$. At the critical point we find $\Delta_b(T)\sim T^{5/4}$ at low temperatures (and $B=0$). The phase diagram of Model II is therefore qualitatively similar to Fig.~\ref{fig:phase_diagram}: the critical fan is flanked by a FL$^\star$ phase with a $T^2$ resistivity on the left, and a heavy FL phase with a large carrier density on the right.

With the boson Green's function determined we can compute the $c$ self-energy (the calculation for $f$ is almost identical):
\begin{equation}
\begin{aligned}
&\Sigma_{c}(i\omega,k) = \\ &g^2T\sum_{i\nu}\int\frac{dq_\perp}{2\pi}\frac{d^2q_\parallel}{(2\pi)^2}\frac{1}{i\nu+\gamma_2\frac{|\nu|}{q}+\frac{q^2}{2m_b}+\Delta_b} \\
&\times\Bigg(i\nu+i\omega-v_{f,F }(q_\perp+k_\perp) \\
&-(q_\parallel+k_\parallel)^2/(2m_{f})-\Sigma_{f}(i\omega+i\nu)\Bigg)^{-1}. 
\end{aligned}
\end{equation}
Since the fermion propagator (which depends on $q_\perp\sim q_\parallel^2$) is much more sensitive to $q_\perp$ at small frequencies and momenta than the boson propagator (which depends on $q_\perp^2 + q_\parallel^2 \sim q_\parallel^4+q_\parallel^2$), we can set $q_\perp=0$ in the boson propagator. As a result of this the self-energy takes a form similar to Model I, coupling momentum-averaged Green's functions ($q_\perp$ averaged for fermions and $q_\parallel$ for the bosons). Moreover, the self-energy we obtain resembles the behavior in Model I in that the momentum-averaged fermions couple to an effectively 2D boson; 
\begin{align}
\Sigma_{c}(i\omega,k) &\approx g^2T \sum_{i\nu}\left(
 \int\frac{dq_{\perp}}{2\pi}G_f(i(\nu+\omega,q+k)\right) \nonumber \\
 &\quad \times \left(\int \frac{d^2q_{\parallel}}{(2\pi)^2} G_b(-i\nu,q_{\parallel})\right) \\
&=-\frac{ig^2}{2v_{f,F}}T\sum_{i\nu}\int\frac{d^2q_\parallel}{(2\pi)^2}\frac{\mathrm{sgn}(\nu+\omega)}{i\nu+\gamma_2\frac{|\nu|}{q_\parallel}+\frac{q_\parallel^2}{2m_b}+\Delta_b}. \nonumber
\label{eq:scfm2}
\end{align}

This self-energy is indeed independent of momentum $k$, as promised earlier. We continue by noting that we can ignore the $i\nu$ term compared to the boson self-energy $\gamma_2|\nu|/q_\parallel$, which is much larger at low frequencies. Hence we obtain the self-energy in the low frequency limit and $T=0$:
\begin{align}
\Sigma_{c,f}(i\omega,T=0) &= -\frac{\gamma_2 m_b}{12\pi^2m_{c,f} k_F} \nonumber \\
&\times i\omega\ln\left(\frac{e\sqrt{2m_b\Lambda^3}}{\gamma_2|\omega|}\right)~~(\mathrm{QCP}), \nonumber \\
\Sigma_{c,f}(i\omega,T=0) &= -\frac{\gamma_2 m_b\ln(\Lambda/\Delta_b(T=0))}{8\pi^2 m_{c,f} k_F}i\omega \\ &+ i\frac{\gamma_2^2\sqrt{m_b/2}}{32\pi m_{c,f} k_F\Delta_b^{3/2}(T=0)}\omega^2~~(\mathrm{FL}^\star), \nonumber
\end{align}
where $\Lambda$ is the boson bandwidth and $k_F$ is the Fermi momentum of the matched FS's. Due to the strong Landau damping we obtain a non-skewed MFL for all values of the damping parameter $\gamma_2$. This should be contrasted with Model I, which leads to a skewed MFL for small damping parameter $\gamma$. However the renormalization of the effective fermion masses upon approaching the QCP from the FL$^\star$ phase are the same as in Model I.

At low but non-zero temperatures above the critical point, the Matsubara frequency sum in (\ref{eq:scfm2}) may be computed analytically upon ignoring the $i\nu$ term as before. Then, we can compute the $q_\parallel$ integral numerically with a UV cutoff $\sim\sqrt{2m_b\Lambda}$ to obtain
\begin{equation}
\Sigma_{c,f}(i\omega,T) = -\frac{i\gamma_2m_b}{m_{c,f}k_F}T\mathrm{sgn}(\omega)\varphi\left(\frac{|\omega|}{T},\frac{\Lambda}{T},\frac{\Lambda}{\Delta_b(T)}\right).
\label{eq:model2selfET}
\end{equation}
The dependencies on $\Lambda/T$ and $\Lambda/\Delta_b(T)$ are logarithmic, as in Model I. As we have seen in the calculations for Model I, this form of the self-energy leads to a universal Planckian scattering rate $\tau^{-1}=(m_c/m_c^\star(T))\mathrm{Im}[\Sigma_{c,R}(\omega=0,T)]$, up to slowly-varying logarithmic factors. Note that we obtain this Planckian scattering rate independent of the damping parameter $\gamma_2$ unlike in Model I, which resulted in Planckian scattering only in the strong damping regime $\gamma\gg 1$.

\subsection{Transport}
\label{sec:M2T}

An exact calculation of the transport properties in Model II is more complicated than in Model I, because the vertex correction diagrams in Fig.~\ref{fig:FD_polbubble} do not vanish. Due to momentum conservation, the momentum integrals in the left and right loops of these diagrams do not decouple as they do in Model I. Similarly, the cross-species current correlations do not vanish in Model II as they do in Model I and complicate the Ioffe-Larkin rule. Nonetheless, we will argue below that the effects of all of these corrections may be neglected, leading to transport properties that are dominated, as in Model I, by the self-energies obtained from the bubble diagram in the previous section (\ref{eq:model2selfET}) \footnote{While the transport vertex corrections can still be resummed exactly as a ladder series owing to our controlled large $N$ limit, unlike in previous work on fermions coupled to critical bosons \cite{KFWL}, this calculation is tedious, and we therefore defer it for future work}. We show that this results in strange metal phenomenology (nearly $T$-linear resistivity) in the critical fan for sufficiently low temperatures. In the subsequent paragraphs, we will explain explicitly how this comes about.

First, we note that the conductivity in the quantum critical and FL$^\star$ regimes is dominated by the conduction electrons $c$. The much heavier damped bosons, added in parallel, form an insulator in the FL$^\star$ phase, and a poor conductor at the QCP, and therefore contribute negligibly to the conductivity. In the heavy FL phase, condensation of $b$ leads to effective hybridization of the $c$ and $f$ bands, and the electrical transport is determined by the large hybridized Fermi surface.

Let us now turn to the question of vertex corrections. In conventional quantum critical systems, a single scattering of an electron off a low momentum critical boson ($\bf{q}\to 0$), included in the electron self-energy, leads to vanishing current relaxation. The transport time is therefore not set by the quasiparticle relaxation time, and is instead obtained from the Kubo formula only by summing over multiple scatterings, which are included in the current vertex corrections. In Model II, the situation is different because the decay process included in the electron self-energy $c_{\bf k} \rightarrow f_{{\bf k}-{\bf q}} + b_{\bf q}$ leads to significant current relaxation even at small momentum transfers ${\bf q}\to 0$. 
The final state current carried by the boson $\sim e q/m_b$, is much smaller than the initial state current $\sim e v_F$ carried by the conduction electron. Note that the $f$ fermion does not contribute any additional current to the final state: due to the local occupancy constraint enforced by the Ioffe-Larkin rules, the boson and $f$ fermion must carry the same current, which is also equal to the total current carried by them, as the $f$ fermion is uncharged.

Although single scattering events lead to current relaxation over short timescales, whose rate is set by the electron self-energy, as argued above (see also \cite{Paul2013}), we also need momentum relaxation in order to obtain a finite DC conductivity. The nonzero overlap of the total current and the conserved total momentum operators will prevent the current from fully relaxing over the long timescales relevant to DC transport, leading to an infinite DC conductivity \cite{HLS2018}. However, this problem is resolved in practice by the existence of an adequate amount of impurities that can scatter the heavy $f$ fermions and thus dissipate the momentum received from the $c$ fermions faster than the equilibration rate between the three species.
This eliminates the above ``momentum drag" phenomenon, and allows the self-energy to also set the current relaxation rate of the conduction electrons $c$ over long time scales. 

Our identification of the current relaxation rate with the rate set by the $c$ electron self-energy (\ref{eq:model2selfET}), just like in Model I, therefore allows for an identification of Planckian strange metal phenomenology in the critical regime of Model II at sufficiently low $T$. As in Model I, we can obtain the resistivity from Eq. (\ref{eq:rxxc}), which results in $\rho_{xx}\sim T \ln (\Lambda/T)$. 

Important differences from Model I, however, arise from the boson damping $\Sigma_b(i\omega,k) \sim |\omega|/k$, which is parametrically much larger at small $k$ than $\Sigma_b(i\omega,k) \sim \gamma|\omega|$ in Model I regardless of the value of $\gamma$. Because the momentum of occupied bosons is effectively cut off at $k\sim \sqrt{m_b T}$. We can identify an effective damping constant $\gamma(T)=\gamma_2/\sqrt{m_b T}$, which is always large at sufficiently low temperatures (see Appendix~\ref{app:SDM2}). Hence there is never any significant enhancement of $R_H$ in the critical regime at low $T$ in Model II, as there is no weak $b$ damping regime like the small $\gamma$ regime for Model I, that was required there to obtain an enhanced $R_H$. Furthermore, the strong damping ensures that Model II is always in the Planckian regime at low enough $T$, as opposed to Model I, which was Planckian only when $\gamma\gg 1$.

In Appendix~\ref{app:SDM2}, we consider a higher temperature regime, occuring for $T\gg \gamma_2^2/m_b$, in which the boson damping is weaker and an enhancement of $R_H$ is consequently obtained. However, the resistivity in this regime is no longer $T$-linear and instead scales as $\sim\sqrt{T}$.

\section{Discussion}

The new large $N$ approach formulated in this paper captures a strongly coupled QCP, showing linear in $T$ resistivity at a Kondo breakdown transition involving a change of the Fermi surface volume. Such MFL phenomenology, seen ubiquitously in experiments with heavy fermion materials, could not be obtained in a controlled way within previous large $N$ theories \cite{read1983solution,coleman1984new,auerbach1986kondo,senthil2004weak}. The essential new element in our formulation is that the number of fermions and critical boson species are both scaled with $N$. 

The MFL with linear in $T$ resistivity is obtained within two distinct models of the Kondo lattice. It is worth emphasizing the differences in the physical situations they describe, and in the predicted phenomena. Model I is disordered, and leads to a MFL only if the QCP and adjacent FL$^\star$ phase are deconfined in layers, that is deconfined inside 2D planes, yet confined between planes. 
This model can be tuned between two regimes by a coupling constant $\gamma$. In the strong damping limit $\gamma\gg 1$ the system exhibits Planckian dissipation, with a universal electron relaxation time $\tau_{tr}\approx \hbar/(k_B T)$. The strong damping also prevents any significant enhancement of the Hall coefficient $R_H$ in the critical regime. In the weak damping regime, $\gamma\ll 1$, the transport relaxation time is much larger than the Planckian time (by a factor $1/\gamma$) and the Hall coefficient $R_H$ is enhanced in the critical regime. Furthermore, the electron self-energy in this regime is ``skewed", with an asymmetry in the damping of particle vs. hole excitations (\ref{eq:GammaSimple}).

Model II, on the other hand, is translationally invariant, and describes a transition from a fully 3D FL$^\star$ with a small Fermi surface to a heavy Fermi liquid with a large Fermi surface. The critical boson is always strongly damped at low temperatures due to Landau damping, leading to Planckian dissipation with a universal electron transport lifetime $\tau_{\mathrm{tr}}\sim \hbar/(k_B T)$. The strong damping prevents enhancement of $R_H$ in the critical regime. 

A testable prediction, which follows from the analysis of the two models, is that Planckian dissipation at the QCP cannot be accompanied by enhancement of the Hall coefficient $R_H$. Enhancement of $R_H$ at the QCP, as has been observed in recent experiments with CeCoIn${}_5$ \cite{maksimovic2020}, can occur only in the weakly damped regime of Model I, where a set of additional features are predicted: first, the QCP and the nearby FL$^\star$ phase are deconfined only within 2D planes, which would have observable implications on transport. For example, the thermal conductivity is expected to be strongly anisotropic, because in this phase spinons contribute to the in-plane, but not to the out-of-plane thermal transport. The charge conductivity, on the other hand, is dominated by the conduction electrons, which can hop between planes, and would therefore be much more isotropic. Consequently, only the in-plane Lorenz ratio is expected to be significantly enhanced. Another unique property of the weakly damped ($\gamma\ll 1$) MFL, is a skewed fermion spectral function, which is expected to generate a low temperature Seebeck coefficient in the critical regime \cite{Georges2021,Patel2018}. Sizeable $T\rightarrow 0$ Seebeck coefficients have recently been reported experimentally in 2D strange metals \cite{Arindam2020,Taillefer2021Seebeck}, and it would be interesting to investigate whether these arise due to skewed electron self-energies.

The new large $N$ approach we have introduced to study the Kondo breakdown transition in HFM can also be useful in formulating a controlled theory of other quantum critical states. The high $T_c$ cuprate superconductors, for example, exhibit similar signatures of FS reconstruction near optimal doping \cite{TailleferPseudogap}, accompanied by $T$-linear resistivity \cite{TailleferSM}. While there are no local moments to be subsumed in the Fermi sea, a parton model describing a change in FS volume has recently been proposed \cite{Zhang2020}. Investigating this QCP using the new large $N$ scheme is an interesting problem for future work. Our approach can also be used to address the interplay of these critical fluctuations with superconductivity and magnetism, which appear to be crucial to cuprate phenomenology. 

Another interesting extension of this work would be to formulate a controlled treatment of gapless gauge field fluctuations coupled to matter fields. This is important, for example, for describing gapless $U(1)$ spin liquids or the Halperin-Lee-Read state in a half-filled Landau-Level \cite{Halperin1993,KFWL}. The standard large $N$ theory captures the gauge field fluctuations within a $1/N$ expansion, which is known to be uncontrolled \cite{sslee2009}. In the large $N$ models we introduced here, gauge field fluctuations are still suppressed by $1/N$, but the $1/N$ expansion could possibly be better controlled. Furthermore, it is interesting to explore generalizations of the scheme to include $N$ flavors of $U(1)$ gauge fields with flavor-random gauge couplings, and thereby capture the feedback of the gauge field fluctuations self-consistently at the saddle point level itself. 

\noindent{\it Acknowledgements --} We thank James Analytis for helpful discussions. T.C. was supported by the NSF Graduate Research Fellowship Program, NSF DGE No. 1752814. A.A.P. was supported by the Miller Institute for Basic Research in Science. E. A. acknowledges support from a Department of Energy grant DE-SC0019380.

\bibliography{FermiRecon.bib}

\begin{thebibliography}{63}%
\makeatletter
\providecommand \@ifxundefined [1]{%
 \@ifx{#1\undefined}
}%
\providecommand \@ifnum [1]{%
 \ifnum #1\expandafter \@firstoftwo
 \else \expandafter \@secondoftwo
 \fi
}%
\providecommand \@ifx [1]{%
 \ifx #1\expandafter \@firstoftwo
 \else \expandafter \@secondoftwo
 \fi
}%
\providecommand \natexlab [1]{#1}%
\providecommand \enquote  [1]{``#1''}%
\providecommand \bibnamefont  [1]{#1}%
\providecommand \bibfnamefont [1]{#1}%
\providecommand \citenamefont [1]{#1}%
\providecommand \href@noop [0]{\@secondoftwo}%
\providecommand \href [0]{\begingroup \@sanitize@url \@href}%
\providecommand \@href[1]{\@@startlink{#1}\@@href}%
\providecommand \@@href[1]{\endgroup#1\@@endlink}%
\providecommand \@sanitize@url [0]{\catcode `\\12\catcode `\$12\catcode
  `\&12\catcode `\#12\catcode `\^12\catcode `\_12\catcode `\%12\relax}%
\providecommand \@@startlink[1]{}%
\providecommand \@@endlink[0]{}%
\providecommand \url  [0]{\begingroup\@sanitize@url \@url }%
\providecommand \@url [1]{\endgroup\@href {#1}{\urlprefix }}%
\providecommand \urlprefix  [0]{URL }%
\providecommand \Eprint [0]{\href }%
\providecommand \doibase [0]{https://doi.org/}%
\providecommand \selectlanguage [0]{\@gobble}%
\providecommand \bibinfo  [0]{\@secondoftwo}%
\providecommand \bibfield  [0]{\@secondoftwo}%
\providecommand \translation [1]{[#1]}%
\providecommand \BibitemOpen [0]{}%
\providecommand \bibitemStop [0]{}%
\providecommand \bibitemNoStop [0]{.\EOS\space}%
\providecommand \EOS [0]{\spacefactor3000\relax}%
\providecommand \BibitemShut  [1]{\csname bibitem#1\endcsname}%
\let\auto@bib@innerbib\@empty
\bibitem [{\citenamefont {Si}\ and\ \citenamefont
  {Steglich}(2010)}]{si2010heavy}%
  \BibitemOpen
  \bibfield  {author} {\bibinfo {author} {\bibfnamefont {Q.}~\bibnamefont
  {Si}}\ and\ \bibinfo {author} {\bibfnamefont {F.}~\bibnamefont {Steglich}},\
  }\bibfield  {title} {\bibinfo {title} {Heavy fermions and quantum phase
  transitions},\ }\href {https://science.sciencemag.org/content/329/5996/1161}
  {\bibfield  {journal} {\bibinfo  {journal} {Science}\ }\textbf {\bibinfo
  {volume} {329}},\ \bibinfo {pages} {1161} (\bibinfo {year}
  {2010})}\BibitemShut {NoStop}%
\bibitem [{\citenamefont {Doniach}(1977)}]{doniach1977kondo}%
  \BibitemOpen
  \bibfield  {author} {\bibinfo {author} {\bibfnamefont {S.}~\bibnamefont
  {Doniach}},\ }\bibfield  {title} {\bibinfo {title} {The {K}ondo lattice and
  weak antiferromagnetism},\ }\href
  {https://www.sciencedirect.com/science/article/abs/pii/0378436377901905}
  {\bibfield  {journal} {\bibinfo  {journal} {Physica B+C}\ }\textbf {\bibinfo
  {volume} {91}},\ \bibinfo {pages} {231} (\bibinfo {year} {1977})}\BibitemShut
  {NoStop}%
\bibitem [{\citenamefont {Read}\ and\ \citenamefont
  {Newns}(1983)}]{read1983solution}%
  \BibitemOpen
  \bibfield  {author} {\bibinfo {author} {\bibfnamefont {N.}~\bibnamefont
  {Read}}\ and\ \bibinfo {author} {\bibfnamefont {D.}~\bibnamefont {Newns}},\
  }\bibfield  {title} {\bibinfo {title} {On the solution of the
  {C}oqblin-{S}chreiffer {H}amiltonian by the large-${N}$ expansion
  technique},\ }\href
  {https://iopscience.iop.org/article/10.1088/0022-3719/16/17/014} {\bibfield
  {journal} {\bibinfo  {journal} {Journal of Physics C: Solid State Physics}\
  }\textbf {\bibinfo {volume} {16}},\ \bibinfo {pages} {3273} (\bibinfo {year}
  {1983})}\BibitemShut {NoStop}%
\bibitem [{\citenamefont {Coleman}(1984)}]{coleman1984new}%
  \BibitemOpen
  \bibfield  {author} {\bibinfo {author} {\bibfnamefont {P.}~\bibnamefont
  {Coleman}},\ }\bibfield  {title} {\bibinfo {title} {New approach to the
  mixed-valence problem},\ }\href
  {https://journals.aps.org/prb/abstract/10.1103/PhysRevB.29.3035} {\bibfield
  {journal} {\bibinfo  {journal} {Physical Review B}\ }\textbf {\bibinfo
  {volume} {29}},\ \bibinfo {pages} {3035} (\bibinfo {year}
  {1984})}\BibitemShut {NoStop}%
\bibitem [{\citenamefont {Shishido}\ \emph {et~al.}(2005)\citenamefont
  {Shishido}, \citenamefont {Settai}, \citenamefont {Harima},\ and\
  \citenamefont {{\=O}nuki}}]{shishido2005drastic}%
  \BibitemOpen
  \bibfield  {author} {\bibinfo {author} {\bibfnamefont {H.}~\bibnamefont
  {Shishido}}, \bibinfo {author} {\bibfnamefont {R.}~\bibnamefont {Settai}},
  \bibinfo {author} {\bibfnamefont {H.}~\bibnamefont {Harima}},\ and\ \bibinfo
  {author} {\bibfnamefont {Y.}~\bibnamefont {{\=O}nuki}},\ }\bibfield  {title}
  {\bibinfo {title} {A drastic change of the {F}ermi surface at a critical
  pressure in {C}e{R}h{I}n$_5$: d{H}v{A} study under pressure},\ }\href
  {https://journals.jps.jp/doi/10.1143/JPSJ.74.1103} {\bibfield  {journal}
  {\bibinfo  {journal} {Journal of the Physical Society of Japan}\ }\textbf
  {\bibinfo {volume} {74}},\ \bibinfo {pages} {1103} (\bibinfo {year}
  {2005})}\BibitemShut {NoStop}%
\bibitem [{\citenamefont {Paschen}\ \emph {et~al.}(2004)\citenamefont
  {Paschen}, \citenamefont {L{\"u}hmann}, \citenamefont {Wirth}, \citenamefont
  {Gegenwart}, \citenamefont {Trovarelli}, \citenamefont {Geibel},
  \citenamefont {Steglich}, \citenamefont {Coleman},\ and\ \citenamefont
  {Si}}]{paschen2004hall}%
  \BibitemOpen
  \bibfield  {author} {\bibinfo {author} {\bibfnamefont {S.}~\bibnamefont
  {Paschen}}, \bibinfo {author} {\bibfnamefont {T.}~\bibnamefont
  {L{\"u}hmann}}, \bibinfo {author} {\bibfnamefont {S.}~\bibnamefont {Wirth}},
  \bibinfo {author} {\bibfnamefont {P.}~\bibnamefont {Gegenwart}}, \bibinfo
  {author} {\bibfnamefont {O.}~\bibnamefont {Trovarelli}}, \bibinfo {author}
  {\bibfnamefont {C.}~\bibnamefont {Geibel}}, \bibinfo {author} {\bibfnamefont
  {F.}~\bibnamefont {Steglich}}, \bibinfo {author} {\bibfnamefont
  {P.}~\bibnamefont {Coleman}},\ and\ \bibinfo {author} {\bibfnamefont
  {Q.}~\bibnamefont {Si}},\ }\bibfield  {title} {\bibinfo {title} {Hall-effect
  evolution across a heavy-fermion quantum critical point},\ }\href
  {https://www.nature.com/articles/nature03129} {\bibfield  {journal} {\bibinfo
   {journal} {Nature}\ }\textbf {\bibinfo {volume} {432}},\ \bibinfo {pages}
  {881} (\bibinfo {year} {2004})}\BibitemShut {NoStop}%
\bibitem [{\citenamefont {Friedemann}\ \emph {et~al.}(2010)\citenamefont
  {Friedemann}, \citenamefont {Oeschler}, \citenamefont {Wirth}, \citenamefont
  {Krellner}, \citenamefont {Geibel}, \citenamefont {Steglich}, \citenamefont
  {Paschen}, \citenamefont {Kirchner},\ and\ \citenamefont
  {Si}}]{friedemann2010fermi}%
  \BibitemOpen
  \bibfield  {author} {\bibinfo {author} {\bibfnamefont {S.}~\bibnamefont
  {Friedemann}}, \bibinfo {author} {\bibfnamefont {N.}~\bibnamefont
  {Oeschler}}, \bibinfo {author} {\bibfnamefont {S.}~\bibnamefont {Wirth}},
  \bibinfo {author} {\bibfnamefont {C.}~\bibnamefont {Krellner}}, \bibinfo
  {author} {\bibfnamefont {C.}~\bibnamefont {Geibel}}, \bibinfo {author}
  {\bibfnamefont {F.}~\bibnamefont {Steglich}}, \bibinfo {author}
  {\bibfnamefont {S.}~\bibnamefont {Paschen}}, \bibinfo {author} {\bibfnamefont
  {S.}~\bibnamefont {Kirchner}},\ and\ \bibinfo {author} {\bibfnamefont
  {Q.}~\bibnamefont {Si}},\ }\bibfield  {title} {\bibinfo {title}
  {Fermi-surface collapse and dynamical scaling near a quantum-critical
  point},\ }\href {https://www.pnas.org/content/107/33/14547} {\bibfield
  {journal} {\bibinfo  {journal} {Proceedings of the National Academy of
  Sciences}\ }\textbf {\bibinfo {volume} {107}},\ \bibinfo {pages} {14547}
  (\bibinfo {year} {2010})}\BibitemShut {NoStop}%
\bibitem [{\citenamefont {Maksimovic}\ \emph {et~al.}(2020)\citenamefont
  {Maksimovic}, \citenamefont {Cookmeyer}, \citenamefont {Rusz}, \citenamefont
  {Nagarajan}, \citenamefont {Gong}, \citenamefont {Wan}, \citenamefont
  {Faubel}, \citenamefont {Hayes}, \citenamefont {Jang}, \citenamefont
  {Werman}, \citenamefont {Oppeneer}, \citenamefont {Altman},\ and\
  \citenamefont {Analytis}}]{maksimovic2020}%
  \BibitemOpen
  \bibfield  {author} {\bibinfo {author} {\bibfnamefont {N.}~\bibnamefont
  {Maksimovic}}, \bibinfo {author} {\bibfnamefont {T.}~\bibnamefont
  {Cookmeyer}}, \bibinfo {author} {\bibfnamefont {J.}~\bibnamefont {Rusz}},
  \bibinfo {author} {\bibfnamefont {V.}~\bibnamefont {Nagarajan}}, \bibinfo
  {author} {\bibfnamefont {A.}~\bibnamefont {Gong}}, \bibinfo {author}
  {\bibfnamefont {F.}~\bibnamefont {Wan}}, \bibinfo {author} {\bibfnamefont
  {S.}~\bibnamefont {Faubel}}, \bibinfo {author} {\bibfnamefont {I.~M.}\
  \bibnamefont {Hayes}}, \bibinfo {author} {\bibfnamefont {S.}~\bibnamefont
  {Jang}}, \bibinfo {author} {\bibfnamefont {Y.}~\bibnamefont {Werman}},
  \bibinfo {author} {\bibfnamefont {P.~M.}\ \bibnamefont {Oppeneer}}, \bibinfo
  {author} {\bibfnamefont {E.}~\bibnamefont {Altman}},\ and\ \bibinfo {author}
  {\bibfnamefont {J.~G.}\ \bibnamefont {Analytis}},\ }\href@noop {} {\bibinfo
  {title} {Evidence for freezing of charge degrees of freedom across a critical
  point in {C}e{C}o{I}n$_5$}} (\bibinfo {year} {2020}),\ \Eprint
  {https://arxiv.org/abs/2011.12951} {arXiv:2011.12951 [cond-mat.str-el]}
  \BibitemShut {NoStop}%
\bibitem [{\citenamefont {Senthil}\ \emph {et~al.}(2004)\citenamefont
  {Senthil}, \citenamefont {Vojta},\ and\ \citenamefont
  {Sachdev}}]{senthil2004weak}%
  \BibitemOpen
  \bibfield  {author} {\bibinfo {author} {\bibfnamefont {T.}~\bibnamefont
  {Senthil}}, \bibinfo {author} {\bibfnamefont {M.}~\bibnamefont {Vojta}},\
  and\ \bibinfo {author} {\bibfnamefont {S.}~\bibnamefont {Sachdev}},\
  }\bibfield  {title} {\bibinfo {title} {Weak magnetism and non-{F}ermi liquids
  near heavy-fermion critical points},\ }\href
  {https://journals.aps.org/prb/abstract/10.1103/PhysRevB.69.035111} {\bibfield
   {journal} {\bibinfo  {journal} {Physical Review B}\ }\textbf {\bibinfo
  {volume} {69}},\ \bibinfo {pages} {035111} (\bibinfo {year}
  {2004})}\BibitemShut {NoStop}%
\bibitem [{\citenamefont {Senthil}\ \emph {et~al.}(2003)\citenamefont
  {Senthil}, \citenamefont {Sachdev},\ and\ \citenamefont
  {Vojta}}]{senthil2003fractionalized}%
  \BibitemOpen
  \bibfield  {author} {\bibinfo {author} {\bibfnamefont {T.}~\bibnamefont
  {Senthil}}, \bibinfo {author} {\bibfnamefont {S.}~\bibnamefont {Sachdev}},\
  and\ \bibinfo {author} {\bibfnamefont {M.}~\bibnamefont {Vojta}},\ }\bibfield
   {title} {\bibinfo {title} {Fractionalized {F}ermi liquids},\ }\href
  {https://journals.aps.org/prl/abstract/10.1103/PhysRevLett.90.216403}
  {\bibfield  {journal} {\bibinfo  {journal} {Physical review letters}\
  }\textbf {\bibinfo {volume} {90}},\ \bibinfo {pages} {216403} (\bibinfo
  {year} {2003})}\BibitemShut {NoStop}%
\bibitem [{\citenamefont {Oshikawa}(2000)}]{oshikawa2000topological}%
  \BibitemOpen
  \bibfield  {author} {\bibinfo {author} {\bibfnamefont {M.}~\bibnamefont
  {Oshikawa}},\ }\bibfield  {title} {\bibinfo {title} {Topological approach to
  {L}uttinger's theorem and the {F}ermi surface of a {K}ondo lattice},\ }\href
  {https://journals.aps.org/prl/abstract/10.1103/PhysRevLett.84.3370}
  {\bibfield  {journal} {\bibinfo  {journal} {Physical Review Letters}\
  }\textbf {\bibinfo {volume} {84}},\ \bibinfo {pages} {3370} (\bibinfo {year}
  {2000})}\BibitemShut {NoStop}%
\bibitem [{\citenamefont {Stewart}(2001)}]{Stewart2001}%
  \BibitemOpen
  \bibfield  {author} {\bibinfo {author} {\bibfnamefont {G.~R.}\ \bibnamefont
  {Stewart}},\ }\bibfield  {title} {\bibinfo {title} {Non-{F}ermi-liquid
  behavior in $d$- and $f$-electron metals},\ }\href
  {https://doi.org/10.1103/RevModPhys.73.797} {\bibfield  {journal} {\bibinfo
  {journal} {Rev. Mod. Phys.}\ }\textbf {\bibinfo {volume} {73}},\ \bibinfo
  {pages} {797} (\bibinfo {year} {2001})}\BibitemShut {NoStop}%
\bibitem [{\citenamefont {Varma}\ \emph {et~al.}(1989)\citenamefont {Varma},
  \citenamefont {Littlewood}, \citenamefont {Schmitt-Rink}, \citenamefont
  {Abrahams},\ and\ \citenamefont {Ruckenstein}}]{Varma1996}%
  \BibitemOpen
  \bibfield  {author} {\bibinfo {author} {\bibfnamefont {C.~M.}\ \bibnamefont
  {Varma}}, \bibinfo {author} {\bibfnamefont {P.~B.}\ \bibnamefont
  {Littlewood}}, \bibinfo {author} {\bibfnamefont {S.}~\bibnamefont
  {Schmitt-Rink}}, \bibinfo {author} {\bibfnamefont {E.}~\bibnamefont
  {Abrahams}},\ and\ \bibinfo {author} {\bibfnamefont {A.~E.}\ \bibnamefont
  {Ruckenstein}},\ }\bibfield  {title} {\bibinfo {title} {Phenomenology of the
  normal state of {C}u-{O} high-temperature superconductors},\ }\href
  {https://doi.org/10.1103/PhysRevLett.63.1996} {\bibfield  {journal} {\bibinfo
   {journal} {Phys. Rev. Lett.}\ }\textbf {\bibinfo {volume} {63}},\ \bibinfo
  {pages} {1996} (\bibinfo {year} {1989})}\BibitemShut {NoStop}%
\bibitem [{\citenamefont {Coleman}\ \emph {et~al.}(2005)\citenamefont
  {Coleman}, \citenamefont {Paul},\ and\ \citenamefont {Rech}}]{Coleman2005}%
  \BibitemOpen
  \bibfield  {author} {\bibinfo {author} {\bibfnamefont {P.}~\bibnamefont
  {Coleman}}, \bibinfo {author} {\bibfnamefont {I.}~\bibnamefont {Paul}},\ and\
  \bibinfo {author} {\bibfnamefont {J.}~\bibnamefont {Rech}},\ }\bibfield
  {title} {\bibinfo {title} {Sum rules and {W}ard identities in the {K}ondo
  lattice},\ }\href {https://doi.org/10.1103/PhysRevB.72.094430} {\bibfield
  {journal} {\bibinfo  {journal} {Phys. Rev. B}\ }\textbf {\bibinfo {volume}
  {72}},\ \bibinfo {pages} {094430} (\bibinfo {year} {2005})}\BibitemShut
  {NoStop}%
\bibitem [{\citenamefont {Bi}\ \emph {et~al.}(2017)\citenamefont {Bi},
  \citenamefont {Jian}, \citenamefont {You}, \citenamefont {Pawlak},\ and\
  \citenamefont {Xu}}]{Bi2017}%
  \BibitemOpen
  \bibfield  {author} {\bibinfo {author} {\bibfnamefont {Z.}~\bibnamefont
  {Bi}}, \bibinfo {author} {\bibfnamefont {C.-M.}\ \bibnamefont {Jian}},
  \bibinfo {author} {\bibfnamefont {Y.-Z.}\ \bibnamefont {You}}, \bibinfo
  {author} {\bibfnamefont {K.~A.}\ \bibnamefont {Pawlak}},\ and\ \bibinfo
  {author} {\bibfnamefont {C.}~\bibnamefont {Xu}},\ }\bibfield  {title}
  {\bibinfo {title} {Instability of the non-{F}ermi-liquid state of the
  {S}achdev-{Y}e-{K}itaev model},\ }\href
  {https://doi.org/10.1103/PhysRevB.95.205105} {\bibfield  {journal} {\bibinfo
  {journal} {Phys. Rev. B}\ }\textbf {\bibinfo {volume} {95}},\ \bibinfo
  {pages} {205105} (\bibinfo {year} {2017})}\BibitemShut {NoStop}%
\bibitem [{\citenamefont {Patel}\ and\ \citenamefont
  {Sachdev}(2018)}]{Patel2018gauge}%
  \BibitemOpen
  \bibfield  {author} {\bibinfo {author} {\bibfnamefont {A.~A.}\ \bibnamefont
  {Patel}}\ and\ \bibinfo {author} {\bibfnamefont {S.}~\bibnamefont
  {Sachdev}},\ }\bibfield  {title} {\bibinfo {title} {Critical strange metal
  from fluctuating gauge fields in a solvable random model},\ }\href
  {https://doi.org/10.1103/PhysRevB.98.125134} {\bibfield  {journal} {\bibinfo
  {journal} {Phys. Rev. B}\ }\textbf {\bibinfo {volume} {98}},\ \bibinfo
  {pages} {125134} (\bibinfo {year} {2018})}\BibitemShut {NoStop}%
\bibitem [{\citenamefont {Marcus}\ and\ \citenamefont
  {Vandoren}(2019)}]{Marcus2019}%
  \BibitemOpen
  \bibfield  {author} {\bibinfo {author} {\bibfnamefont {E.}~\bibnamefont
  {Marcus}}\ and\ \bibinfo {author} {\bibfnamefont {S.}~\bibnamefont
  {Vandoren}},\ }\bibfield  {title} {\bibinfo {title} {A new class of
  {S}{Y}{K}-like models with maximal chaos},\ }\href
  {https://doi.org/10.1007/JHEP01(2019)166} {\bibfield  {journal} {\bibinfo
  {journal} {Journal of High Energy Physics}\ }\textbf {\bibinfo {volume}
  {2019}},\ \bibinfo {pages} {166} (\bibinfo {year} {2019})}\BibitemShut
  {NoStop}%
\bibitem [{\citenamefont {Wang}(2020)}]{Wang2020}%
  \BibitemOpen
  \bibfield  {author} {\bibinfo {author} {\bibfnamefont {Y.}~\bibnamefont
  {Wang}},\ }\bibfield  {title} {\bibinfo {title} {Solvable strong-coupling
  quantum-dot model with a non-{F}ermi-liquid pairing transition},\ }\href
  {https://doi.org/10.1103/PhysRevLett.124.017002} {\bibfield  {journal}
  {\bibinfo  {journal} {Phys. Rev. Lett.}\ }\textbf {\bibinfo {volume} {124}},\
  \bibinfo {pages} {017002} (\bibinfo {year} {2020})}\BibitemShut {NoStop}%
\bibitem [{\citenamefont {Esterlis}\ and\ \citenamefont
  {Schmalian}(2019)}]{Esterlis2019}%
  \BibitemOpen
  \bibfield  {author} {\bibinfo {author} {\bibfnamefont {I.}~\bibnamefont
  {Esterlis}}\ and\ \bibinfo {author} {\bibfnamefont {J.}~\bibnamefont
  {Schmalian}},\ }\bibfield  {title} {\bibinfo {title} {Cooper pairing of
  incoherent electrons: An electron-phonon version of the
  {S}achdev-{Y}e-{K}itaev model},\ }\href
  {https://doi.org/10.1103/PhysRevB.100.115132} {\bibfield  {journal} {\bibinfo
   {journal} {Phys. Rev. B}\ }\textbf {\bibinfo {volume} {100}},\ \bibinfo
  {pages} {115132} (\bibinfo {year} {2019})}\BibitemShut {NoStop}%
\bibitem [{\citenamefont {Kim}\ \emph {et~al.}(2020)\citenamefont {Kim},
  \citenamefont {Cao},\ and\ \citenamefont {Altman}}]{Kim2020}%
  \BibitemOpen
  \bibfield  {author} {\bibinfo {author} {\bibfnamefont {J.}~\bibnamefont
  {Kim}}, \bibinfo {author} {\bibfnamefont {X.}~\bibnamefont {Cao}},\ and\
  \bibinfo {author} {\bibfnamefont {E.}~\bibnamefont {Altman}},\ }\bibfield
  {title} {\bibinfo {title} {Low-rank {S}achdev-{Y}e-{K}itaev models},\ }\href
  {https://doi.org/10.1103/PhysRevB.101.125112} {\bibfield  {journal} {\bibinfo
   {journal} {Phys. Rev. B}\ }\textbf {\bibinfo {volume} {101}},\ \bibinfo
  {pages} {125112} (\bibinfo {year} {2020})}\BibitemShut {NoStop}%
\bibitem [{\citenamefont {Kim}\ \emph {et~al.}(2021)\citenamefont {Kim},
  \citenamefont {Altman},\ and\ \citenamefont {Cao}}]{Kim2021}%
  \BibitemOpen
  \bibfield  {author} {\bibinfo {author} {\bibfnamefont {J.}~\bibnamefont
  {Kim}}, \bibinfo {author} {\bibfnamefont {E.}~\bibnamefont {Altman}},\ and\
  \bibinfo {author} {\bibfnamefont {X.}~\bibnamefont {Cao}},\ }\bibfield
  {title} {\bibinfo {title} {Dirac fast scramblers},\ }\href
  {https://doi.org/10.1103/PhysRevB.103.L081113} {\bibfield  {journal}
  {\bibinfo  {journal} {Phys. Rev. B}\ }\textbf {\bibinfo {volume} {103}},\
  \bibinfo {pages} {L081113} (\bibinfo {year} {2021})}\BibitemShut {NoStop}%
\bibitem [{\citenamefont {Coqblin}\ and\ \citenamefont
  {Schrieffer}(1969)}]{Coqblin1969}%
  \BibitemOpen
  \bibfield  {author} {\bibinfo {author} {\bibfnamefont {B.}~\bibnamefont
  {Coqblin}}\ and\ \bibinfo {author} {\bibfnamefont {J.~R.}\ \bibnamefont
  {Schrieffer}},\ }\bibfield  {title} {\bibinfo {title} {Exchange interaction
  in alloys with cerium impurities},\ }\href
  {https://doi.org/10.1103/PhysRev.185.847} {\bibfield  {journal} {\bibinfo
  {journal} {Phys. Rev.}\ }\textbf {\bibinfo {volume} {185}},\ \bibinfo {pages}
  {847} (\bibinfo {year} {1969})}\BibitemShut {NoStop}%
\bibitem [{\citenamefont {Gang}(2007)}]{WenBook}%
  \BibitemOpen
  \bibfield  {author} {\bibinfo {author} {\bibfnamefont {W.~X.}\ \bibnamefont
  {Gang}},\ }\href {https://doi.org/10.1093/acprof:oso/9780199227259.001.0001}
  {\emph {\bibinfo {title} {{Quantum Field Theory of Many-Body Systems: From
  the Origin of Sound to an Origin of Light and Electrons}}}}\ (\bibinfo
  {publisher} {Oxford {U}niversity {P}ress},\ \bibinfo {year}
  {2007})\BibitemShut {NoStop}%
\bibitem [{\citenamefont {Kondo}(1962)}]{Kondo1962}%
  \BibitemOpen
  \bibfield  {author} {\bibinfo {author} {\bibfnamefont {J.}~\bibnamefont
  {Kondo}},\ }\bibfield  {title} {\bibinfo {title} {{g-Shift and Anomalous Hall
  Effect in Gadolinium Metals}},\ }\href {https://doi.org/10.1143/PTP.28.846}
  {\bibfield  {journal} {\bibinfo  {journal} {Progress of Theoretical Physics}\
  }\textbf {\bibinfo {volume} {28}},\ \bibinfo {pages} {846} (\bibinfo {year}
  {1962})},\ \Eprint
  {https://arxiv.org/abs/https://academic.oup.com/ptp/article-pdf/28/5/846/5258393/28-5-846.pdf}
  {https://academic.oup.com/ptp/article-pdf/28/5/846/5258393/28-5-846.pdf}
  \BibitemShut {NoStop}%
\bibitem [{\citenamefont {Anderson}\ and\ \citenamefont
  {Yuval}(1969)}]{Anderson1969}%
  \BibitemOpen
  \bibfield  {author} {\bibinfo {author} {\bibfnamefont {P.~W.}\ \bibnamefont
  {Anderson}}\ and\ \bibinfo {author} {\bibfnamefont {G.}~\bibnamefont
  {Yuval}},\ }\bibfield  {title} {\bibinfo {title} {Exact results in the
  {K}ondo problem: Equivalence to a classical one-dimensional {C}oulomb gas},\
  }\href {https://doi.org/10.1103/PhysRevLett.23.89} {\bibfield  {journal}
  {\bibinfo  {journal} {Phys. Rev. Lett.}\ }\textbf {\bibinfo {volume} {23}},\
  \bibinfo {pages} {89} (\bibinfo {year} {1969})}\BibitemShut {NoStop}%
\bibitem [{\citenamefont {Anderson}(1981)}]{Anderson1981}%
  \BibitemOpen
  \bibfield  {author} {\bibinfo {author} {\bibfnamefont {P.}~\bibnamefont
  {Anderson}},\ }\href@noop {} {\emph {\bibinfo {title} {Valence fluctuations
  in solids : Santa Barbara Institute for Theoretical Physics Conferences,
  Santa Barbara, California, January 27-30, 1981 / edited by L.M. Falicov, W.
  Hanke, M.B. Maple}}}\ (\bibinfo  {publisher} {North-Holland ; Sole
  distributors for the U.S.A. and Canada, Elsevier North-Holland Amsterdam ;
  New York : New York},\ \bibinfo {year} {1981})\ p.\ \bibinfo {pages} {p.
  451}\BibitemShut {NoStop}%
\bibitem [{\citenamefont {Auerbach}\ and\ \citenamefont
  {Levin}(1986)}]{auerbach1986kondo}%
  \BibitemOpen
  \bibfield  {author} {\bibinfo {author} {\bibfnamefont {A.}~\bibnamefont
  {Auerbach}}\ and\ \bibinfo {author} {\bibfnamefont {K.}~\bibnamefont
  {Levin}},\ }\bibfield  {title} {\bibinfo {title} {{K}ondo bosons and the
  {K}ondo lattice: Microscopic basis for the heavy {F}ermi liquid},\ }\href
  {https://journals.aps.org/prl/abstract/10.1103/PhysRevLett.57.877} {\bibfield
   {journal} {\bibinfo  {journal} {Physical review letters}\ }\textbf {\bibinfo
  {volume} {57}},\ \bibinfo {pages} {877} (\bibinfo {year} {1986})}\BibitemShut
  {NoStop}%
\bibitem [{\citenamefont {Maldacena}\ and\ \citenamefont
  {Stanford}(2016)}]{MS2016}%
  \BibitemOpen
  \bibfield  {author} {\bibinfo {author} {\bibfnamefont {J.}~\bibnamefont
  {Maldacena}}\ and\ \bibinfo {author} {\bibfnamefont {D.}~\bibnamefont
  {Stanford}},\ }\bibfield  {title} {\bibinfo {title} {Remarks on the
  {S}achdev-{Y}e-{K}itaev model},\ }\href
  {https://doi.org/10.1103/PhysRevD.94.106002} {\bibfield  {journal} {\bibinfo
  {journal} {Phys. Rev. D}\ }\textbf {\bibinfo {volume} {94}},\ \bibinfo
  {pages} {106002} (\bibinfo {year} {2016})}\BibitemShut {NoStop}%
\bibitem [{\citenamefont {Ioffe}\ and\ \citenamefont
  {Larkin}(1989)}]{Ioffe1989}%
  \BibitemOpen
  \bibfield  {author} {\bibinfo {author} {\bibfnamefont {L.~B.}\ \bibnamefont
  {Ioffe}}\ and\ \bibinfo {author} {\bibfnamefont {A.~I.}\ \bibnamefont
  {Larkin}},\ }\bibfield  {title} {\bibinfo {title} {Gapless fermions and gauge
  fields in dielectrics},\ }\href {https://doi.org/10.1103/PhysRevB.39.8988}
  {\bibfield  {journal} {\bibinfo  {journal} {Phys. Rev. B}\ }\textbf {\bibinfo
  {volume} {39}},\ \bibinfo {pages} {8988} (\bibinfo {year}
  {1989})}\BibitemShut {NoStop}%
\bibitem [{\citenamefont {Lee}\ and\ \citenamefont {Nagaosa}(1992)}]{Lee1992}%
  \BibitemOpen
  \bibfield  {author} {\bibinfo {author} {\bibfnamefont {P.~A.}\ \bibnamefont
  {Lee}}\ and\ \bibinfo {author} {\bibfnamefont {N.}~\bibnamefont {Nagaosa}},\
  }\bibfield  {title} {\bibinfo {title} {Gauge theory of the normal state of
  high-${T}_{\mathit{c}}$ superconductors},\ }\href
  {https://doi.org/10.1103/PhysRevB.46.5621} {\bibfield  {journal} {\bibinfo
  {journal} {Phys. Rev. B}\ }\textbf {\bibinfo {volume} {46}},\ \bibinfo
  {pages} {5621} (\bibinfo {year} {1992})}\BibitemShut {NoStop}%
\bibitem [{\citenamefont {{Georges}}\ and\ \citenamefont
  {{Mravlje}}(2021)}]{Georges2021}%
  \BibitemOpen
  \bibfield  {author} {\bibinfo {author} {\bibfnamefont {A.}~\bibnamefont
  {{Georges}}}\ and\ \bibinfo {author} {\bibfnamefont {J.}~\bibnamefont
  {{Mravlje}}},\ }\bibfield  {title} {\bibinfo {title} {{Skewed Non-Fermi
  Liquids and the Seebeck Effect}},\ }\href@noop {} {\bibfield  {journal}
  {\bibinfo  {journal} {arXiv e-prints}\ ,\ \bibinfo {eid} {arXiv:2102.13224}}
  (\bibinfo {year} {2021})},\ \Eprint {https://arxiv.org/abs/2102.13224}
  {arXiv:2102.13224 [cond-mat.str-el]} \BibitemShut {NoStop}%
\bibitem [{\citenamefont {Paul}\ \emph {et~al.}(2013)\citenamefont {Paul},
  \citenamefont {P\'epin},\ and\ \citenamefont {Norman}}]{Paul2013}%
  \BibitemOpen
  \bibfield  {author} {\bibinfo {author} {\bibfnamefont {I.}~\bibnamefont
  {Paul}}, \bibinfo {author} {\bibfnamefont {C.}~\bibnamefont {P\'epin}},\ and\
  \bibinfo {author} {\bibfnamefont {M.~R.}\ \bibnamefont {Norman}},\ }\bibfield
   {title} {\bibinfo {title} {Equivalence of single-particle and transport
  lifetimes from hybridization fluctuations},\ }\href
  {https://doi.org/10.1103/PhysRevLett.110.066402} {\bibfield  {journal}
  {\bibinfo  {journal} {Phys. Rev. Lett.}\ }\textbf {\bibinfo {volume} {110}},\
  \bibinfo {pages} {066402} (\bibinfo {year} {2013})}\BibitemShut {NoStop}%
\bibitem [{\citenamefont {Patel}\ \emph
  {et~al.}(2018{\natexlab{a}})\citenamefont {Patel}, \citenamefont {McGreevy},
  \citenamefont {Arovas},\ and\ \citenamefont {Sachdev}}]{Patel2018}%
  \BibitemOpen
  \bibfield  {author} {\bibinfo {author} {\bibfnamefont {A.~A.}\ \bibnamefont
  {Patel}}, \bibinfo {author} {\bibfnamefont {J.}~\bibnamefont {McGreevy}},
  \bibinfo {author} {\bibfnamefont {D.~P.}\ \bibnamefont {Arovas}},\ and\
  \bibinfo {author} {\bibfnamefont {S.}~\bibnamefont {Sachdev}},\ }\bibfield
  {title} {\bibinfo {title} {Magnetotransport in a model of a disordered
  strange metal},\ }\href {https://doi.org/10.1103/PhysRevX.8.021049}
  {\bibfield  {journal} {\bibinfo  {journal} {Phys. Rev. X}\ }\textbf {\bibinfo
  {volume} {8}},\ \bibinfo {pages} {021049} (\bibinfo {year}
  {2018}{\natexlab{a}})}\BibitemShut {NoStop}%
\bibitem [{\citenamefont {Sachdev}(2011)}]{SachdevQPT}%
  \BibitemOpen
  \bibfield  {author} {\bibinfo {author} {\bibfnamefont {S.}~\bibnamefont
  {Sachdev}},\ }\href
  {https://www.cambridge.org/core/books/quantum-phase-transitions/33C1C81500346005E54C1DE4223E5562}
  {\emph {\bibinfo {title} {Quantum phase transitions}}}\ (\bibinfo
  {publisher} {Cambridge University Press},\ \bibinfo {year}
  {2011})\BibitemShut {NoStop}%
\bibitem [{Note1()}]{Note1}%
  \BibitemOpen
  \bibinfo {note} {The function $w_1(\gamma ,T)$ vanishes quasilogarithmically
  as $T\rightarrow 0$, diverges logarithmically as $\gamma \rightarrow 0$, and
  is quasi-linear in $\gamma $ for $\gamma \gg 1$}\BibitemShut {NoStop}%
\bibitem [{Note2()}]{Note2}%
  \BibitemOpen
  \bibinfo {note} {As is well known there is no phase transition between the
  low $T$ Higgs phase and high $T$ confined phase. Accordingly, there is no
  true finite $T$ Bose condensation transition, only a crossover}\BibitemShut
  {NoStop}%
\bibitem [{\citenamefont {{Ghawri}}\ \emph {et~al.}(2020)\citenamefont
  {{Ghawri}}, \citenamefont {{Mahapatra}}, \citenamefont {{Mandal}},
  \citenamefont {{Jayaraman}}, \citenamefont {{Garg}}, \citenamefont
  {{Watanabe}}, \citenamefont {{Taniguchi}}, \citenamefont {{Krishnamurthy}},
  \citenamefont {{Jain}}, \citenamefont {{Banerjee}}, \citenamefont
  {{Chandni}},\ and\ \citenamefont {{Ghosh}}}]{Arindam2020}%
  \BibitemOpen
  \bibfield  {author} {\bibinfo {author} {\bibfnamefont {B.}~\bibnamefont
  {{Ghawri}}}, \bibinfo {author} {\bibfnamefont {P.~S.}\ \bibnamefont
  {{Mahapatra}}}, \bibinfo {author} {\bibfnamefont {S.}~\bibnamefont
  {{Mandal}}}, \bibinfo {author} {\bibfnamefont {A.}~\bibnamefont
  {{Jayaraman}}}, \bibinfo {author} {\bibfnamefont {M.}~\bibnamefont {{Garg}}},
  \bibinfo {author} {\bibfnamefont {K.}~\bibnamefont {{Watanabe}}}, \bibinfo
  {author} {\bibfnamefont {T.}~\bibnamefont {{Taniguchi}}}, \bibinfo {author}
  {\bibfnamefont {H.~R.}\ \bibnamefont {{Krishnamurthy}}}, \bibinfo {author}
  {\bibfnamefont {M.}~\bibnamefont {{Jain}}}, \bibinfo {author} {\bibfnamefont
  {S.}~\bibnamefont {{Banerjee}}}, \bibinfo {author} {\bibfnamefont
  {U.}~\bibnamefont {{Chandni}}},\ and\ \bibinfo {author} {\bibfnamefont
  {A.}~\bibnamefont {{Ghosh}}},\ }\bibfield  {title} {\bibinfo {title} {{Excess
  entropy and breakdown of semiclassical description of thermoelectricity in
  twisted bilayer graphene close to half filling}},\ }\href@noop {} {\bibfield
  {journal} {\bibinfo  {journal} {arXiv e-prints}\ ,\ \bibinfo {eid}
  {arXiv:2004.12356}} (\bibinfo {year} {2020})},\ \Eprint
  {https://arxiv.org/abs/2004.12356} {arXiv:2004.12356 [cond-mat.mes-hall]}
  \BibitemShut {NoStop}%
\bibitem [{Note3()}]{Note3}%
  \BibitemOpen
  \bibinfo {note} {The magnitude of the low-temperature Seebeck coefficient is
  $\sim k_B/e$ when $\gamma \ll 1$, declining to zero as $\gamma $ is increased
  to $\gamma \gg 1$.}\BibitemShut {Stop}%
\bibitem [{\citenamefont {Custers}\ \emph {et~al.}(2003)\citenamefont
  {Custers}, \citenamefont {Gegenwart}, \citenamefont {Wilhelm}, \citenamefont
  {Neumaier}, \citenamefont {Tokiwa}, \citenamefont {Trovarelli}, \citenamefont
  {Geibel}, \citenamefont {Steglich}, \citenamefont {P{\'e}pin},\ and\
  \citenamefont {Coleman}}]{Custers2003}%
  \BibitemOpen
  \bibfield  {author} {\bibinfo {author} {\bibfnamefont {J.}~\bibnamefont
  {Custers}}, \bibinfo {author} {\bibfnamefont {P.}~\bibnamefont {Gegenwart}},
  \bibinfo {author} {\bibfnamefont {H.}~\bibnamefont {Wilhelm}}, \bibinfo
  {author} {\bibfnamefont {K.}~\bibnamefont {Neumaier}}, \bibinfo {author}
  {\bibfnamefont {Y.}~\bibnamefont {Tokiwa}}, \bibinfo {author} {\bibfnamefont
  {O.}~\bibnamefont {Trovarelli}}, \bibinfo {author} {\bibfnamefont
  {C.}~\bibnamefont {Geibel}}, \bibinfo {author} {\bibfnamefont
  {F.}~\bibnamefont {Steglich}}, \bibinfo {author} {\bibfnamefont
  {C.}~\bibnamefont {P{\'e}pin}},\ and\ \bibinfo {author} {\bibfnamefont
  {P.}~\bibnamefont {Coleman}},\ }\bibfield  {title} {\bibinfo {title} {The
  break-up of heavy electrons at a quantum critical point},\ }\href
  {https://doi.org/10.1038/nature01774} {\bibfield  {journal} {\bibinfo
  {journal} {Nature}\ }\textbf {\bibinfo {volume} {424}},\ \bibinfo {pages}
  {524} (\bibinfo {year} {2003})}\BibitemShut {NoStop}%
\bibitem [{\citenamefont {Hartnoll}\ and\ \citenamefont
  {Mackenzie}(2021)}]{hartnoll2021planckian}%
  \BibitemOpen
  \bibfield  {author} {\bibinfo {author} {\bibfnamefont {S.~A.}\ \bibnamefont
  {Hartnoll}}\ and\ \bibinfo {author} {\bibfnamefont {A.~P.}\ \bibnamefont
  {Mackenzie}},\ }\href@noop {} {\bibinfo {title} {Planckian dissipation in
  metals}} (\bibinfo {year} {2021}),\ \Eprint
  {https://arxiv.org/abs/2107.07802} {arXiv:2107.07802 [cond-mat.str-el]}
  \BibitemShut {NoStop}%
\bibitem [{\citenamefont {Bruin}\ \emph {et~al.}(2013)\citenamefont {Bruin},
  \citenamefont {Sakai}, \citenamefont {Perry},\ and\ \citenamefont
  {Mackenzie}}]{Bruin804}%
  \BibitemOpen
  \bibfield  {author} {\bibinfo {author} {\bibfnamefont {J.~A.~N.}\
  \bibnamefont {Bruin}}, \bibinfo {author} {\bibfnamefont {H.}~\bibnamefont
  {Sakai}}, \bibinfo {author} {\bibfnamefont {R.~S.}\ \bibnamefont {Perry}},\
  and\ \bibinfo {author} {\bibfnamefont {A.~P.}\ \bibnamefont {Mackenzie}},\
  }\bibfield  {title} {\bibinfo {title} {Similarity of scattering rates in
  metals showing ${T}$-linear resistivity},\ }\href
  {https://doi.org/10.1126/science.1227612} {\bibfield  {journal} {\bibinfo
  {journal} {Science}\ }\textbf {\bibinfo {volume} {339}},\ \bibinfo {pages}
  {804} (\bibinfo {year} {2013})},\ \Eprint
  {https://arxiv.org/abs/https://science.sciencemag.org/content/339/6121/804.full.pdf}
  {https://science.sciencemag.org/content/339/6121/804.full.pdf} \BibitemShut
  {NoStop}%
\bibitem [{\citenamefont {{Legros}}\ \emph {et~al.}(2018)\citenamefont
  {{Legros}}, \citenamefont {{Benhabib}}, \citenamefont {{Tabis}},
  \citenamefont {{Lalibert{\'e}}}, \citenamefont {{Dion}}, \citenamefont
  {{Lizaire}}, \citenamefont {{Vignolle}}, \citenamefont {{Vignolles}},
  \citenamefont {{Raffy}}, \citenamefont {{Li}}, \citenamefont
  {{Auban-Senzier}}, \citenamefont {{Doiron-Leyraud}}, \citenamefont
  {{Fournier}}, \citenamefont {{Colson}}, \citenamefont {{Taillefer}},\ and\
  \citenamefont {{Proust}}}]{Legros18}%
  \BibitemOpen
  \bibfield  {author} {\bibinfo {author} {\bibfnamefont {A.}~\bibnamefont
  {{Legros}}}, \bibinfo {author} {\bibfnamefont {S.}~\bibnamefont
  {{Benhabib}}}, \bibinfo {author} {\bibfnamefont {W.}~\bibnamefont {{Tabis}}},
  \bibinfo {author} {\bibfnamefont {F.}~\bibnamefont {{Lalibert{\'e}}}},
  \bibinfo {author} {\bibfnamefont {M.}~\bibnamefont {{Dion}}}, \bibinfo
  {author} {\bibfnamefont {M.}~\bibnamefont {{Lizaire}}}, \bibinfo {author}
  {\bibfnamefont {B.}~\bibnamefont {{Vignolle}}}, \bibinfo {author}
  {\bibfnamefont {D.}~\bibnamefont {{Vignolles}}}, \bibinfo {author}
  {\bibfnamefont {H.}~\bibnamefont {{Raffy}}}, \bibinfo {author} {\bibfnamefont
  {Z.~Z.}\ \bibnamefont {{Li}}}, \bibinfo {author} {\bibfnamefont
  {P.}~\bibnamefont {{Auban-Senzier}}}, \bibinfo {author} {\bibfnamefont
  {N.}~\bibnamefont {{Doiron-Leyraud}}}, \bibinfo {author} {\bibfnamefont
  {P.}~\bibnamefont {{Fournier}}}, \bibinfo {author} {\bibfnamefont
  {D.}~\bibnamefont {{Colson}}}, \bibinfo {author} {\bibfnamefont
  {L.}~\bibnamefont {{Taillefer}}},\ and\ \bibinfo {author} {\bibfnamefont
  {C.}~\bibnamefont {{Proust}}},\ }\bibfield  {title} {\bibinfo {title}
  {{Universal $T$-linear resistivity and {P}lanckian dissipation in overdoped
  cuprates}},\ }\href {https://doi.org/10.1038/s41567-018-0334-2} {\bibfield
  {journal} {\bibinfo  {journal} {Nature Physics}\ }\textbf {\bibinfo {volume}
  {15}},\ \bibinfo {pages} {142} (\bibinfo {year} {2018})},\ \Eprint
  {https://arxiv.org/abs/1805.02512} {arXiv:1805.02512 [cond-mat.supr-con]}
  \BibitemShut {NoStop}%
\bibitem [{\citenamefont {{Nakajima}}\ \emph {et~al.}(2019)\citenamefont
  {{Nakajima}}, \citenamefont {{Metz}}, \citenamefont {{Eckberg}},
  \citenamefont {{Kirshenbaum}}, \citenamefont {{Hughes}}, \citenamefont
  {{Wang}}, \citenamefont {{Wang}}, \citenamefont {{Saha}}, \citenamefont
  {{Liu}}, \citenamefont {{Butch}}, \citenamefont {{Campbell}}, \citenamefont
  {{Eo}}, \citenamefont {{Graf}}, \citenamefont {{Liu}}, \citenamefont
  {{Borisenko}}, \citenamefont {{Zavalij}},\ and\ \citenamefont
  {{Paglione}}}]{Paglione19}%
  \BibitemOpen
  \bibfield  {author} {\bibinfo {author} {\bibfnamefont {Y.}~\bibnamefont
  {{Nakajima}}}, \bibinfo {author} {\bibfnamefont {T.}~\bibnamefont {{Metz}}},
  \bibinfo {author} {\bibfnamefont {C.}~\bibnamefont {{Eckberg}}}, \bibinfo
  {author} {\bibfnamefont {K.}~\bibnamefont {{Kirshenbaum}}}, \bibinfo {author}
  {\bibfnamefont {A.}~\bibnamefont {{Hughes}}}, \bibinfo {author}
  {\bibfnamefont {R.}~\bibnamefont {{Wang}}}, \bibinfo {author} {\bibfnamefont
  {L.}~\bibnamefont {{Wang}}}, \bibinfo {author} {\bibfnamefont {S.~R.}\
  \bibnamefont {{Saha}}}, \bibinfo {author} {\bibfnamefont {I.-L.}\
  \bibnamefont {{Liu}}}, \bibinfo {author} {\bibfnamefont {N.~P.}\ \bibnamefont
  {{Butch}}}, \bibinfo {author} {\bibfnamefont {D.}~\bibnamefont {{Campbell}}},
  \bibinfo {author} {\bibfnamefont {Y.~S.}\ \bibnamefont {{Eo}}}, \bibinfo
  {author} {\bibfnamefont {D.}~\bibnamefont {{Graf}}}, \bibinfo {author}
  {\bibfnamefont {Z.}~\bibnamefont {{Liu}}}, \bibinfo {author} {\bibfnamefont
  {S.~V.}\ \bibnamefont {{Borisenko}}}, \bibinfo {author} {\bibfnamefont
  {P.~Y.}\ \bibnamefont {{Zavalij}}},\ and\ \bibinfo {author} {\bibfnamefont
  {J.}~\bibnamefont {{Paglione}}},\ }\bibfield  {title} {\bibinfo {title}
  {{{P}lanckian dissipation and scale invariance in a quantum-critical
  disordered pnictide}},\ }\href@noop {} {\bibfield  {journal} {\bibinfo
  {journal} {arXiv e-prints}\ } (\bibinfo {year} {2019})},\ \Eprint
  {https://arxiv.org/abs/1902.01034} {arXiv:1902.01034 [cond-mat.str-el]}
  \BibitemShut {NoStop}%
\bibitem [{\citenamefont {Cao}\ \emph {et~al.}(2020)\citenamefont {Cao},
  \citenamefont {Chowdhury}, \citenamefont {Rodan-Legrain}, \citenamefont
  {Rubies-Bigorda}, \citenamefont {Watanabe}, \citenamefont {Taniguchi},
  \citenamefont {Senthil},\ and\ \citenamefont {Jarillo-Herrero}}]{Cao2020}%
  \BibitemOpen
  \bibfield  {author} {\bibinfo {author} {\bibfnamefont {Y.}~\bibnamefont
  {Cao}}, \bibinfo {author} {\bibfnamefont {D.}~\bibnamefont {Chowdhury}},
  \bibinfo {author} {\bibfnamefont {D.}~\bibnamefont {Rodan-Legrain}}, \bibinfo
  {author} {\bibfnamefont {O.}~\bibnamefont {Rubies-Bigorda}}, \bibinfo
  {author} {\bibfnamefont {K.}~\bibnamefont {Watanabe}}, \bibinfo {author}
  {\bibfnamefont {T.}~\bibnamefont {Taniguchi}}, \bibinfo {author}
  {\bibfnamefont {T.}~\bibnamefont {Senthil}},\ and\ \bibinfo {author}
  {\bibfnamefont {P.}~\bibnamefont {Jarillo-Herrero}},\ }\bibfield  {title}
  {\bibinfo {title} {Strange metal in magic-angle graphene with near
  {P}lanckian dissipation},\ }\href
  {https://doi.org/10.1103/PhysRevLett.124.076801} {\bibfield  {journal}
  {\bibinfo  {journal} {Phys. Rev. Lett.}\ }\textbf {\bibinfo {volume} {124}},\
  \bibinfo {pages} {076801} (\bibinfo {year} {2020})}\BibitemShut {NoStop}%
\bibitem [{\citenamefont {Patel}\ and\ \citenamefont
  {Sachdev}(2019)}]{Patel2019}%
  \BibitemOpen
  \bibfield  {author} {\bibinfo {author} {\bibfnamefont {A.~A.}\ \bibnamefont
  {Patel}}\ and\ \bibinfo {author} {\bibfnamefont {S.}~\bibnamefont
  {Sachdev}},\ }\bibfield  {title} {\bibinfo {title} {Theory of a {P}lanckian
  metal},\ }\href {https://doi.org/10.1103/PhysRevLett.123.066601} {\bibfield
  {journal} {\bibinfo  {journal} {Phys. Rev. Lett.}\ }\textbf {\bibinfo
  {volume} {123}},\ \bibinfo {pages} {066601} (\bibinfo {year}
  {2019})}\BibitemShut {NoStop}%
\bibitem [{\citenamefont {Metlitski}\ and\ \citenamefont
  {Sachdev}(2010)}]{Metlitski2010}%
  \BibitemOpen
  \bibfield  {author} {\bibinfo {author} {\bibfnamefont {M.~A.}\ \bibnamefont
  {Metlitski}}\ and\ \bibinfo {author} {\bibfnamefont {S.}~\bibnamefont
  {Sachdev}},\ }\bibfield  {title} {\bibinfo {title} {Quantum phase transitions
  of metals in two spatial dimensions. {I}. {I}sing-nematic order},\ }\href
  {https://doi.org/10.1103/PhysRevB.82.075127} {\bibfield  {journal} {\bibinfo
  {journal} {Phys. Rev. B}\ }\textbf {\bibinfo {volume} {82}},\ \bibinfo
  {pages} {075127} (\bibinfo {year} {2010})}\BibitemShut {NoStop}%
\bibitem [{\citenamefont {Paul}\ \emph {et~al.}(2007)\citenamefont {Paul},
  \citenamefont {P\'epin},\ and\ \citenamefont {Norman}}]{Paul2007}%
  \BibitemOpen
  \bibfield  {author} {\bibinfo {author} {\bibfnamefont {I.}~\bibnamefont
  {Paul}}, \bibinfo {author} {\bibfnamefont {C.}~\bibnamefont {P\'epin}},\ and\
  \bibinfo {author} {\bibfnamefont {M.~R.}\ \bibnamefont {Norman}},\ }\bibfield
   {title} {\bibinfo {title} {Kondo breakdown and hybridization fluctuations in
  the kondo-heisenberg lattice},\ }\href
  {https://doi.org/10.1103/PhysRevLett.98.026402} {\bibfield  {journal}
  {\bibinfo  {journal} {Phys. Rev. Lett.}\ }\textbf {\bibinfo {volume} {98}},\
  \bibinfo {pages} {026402} (\bibinfo {year} {2007})}\BibitemShut {NoStop}%
\bibitem [{\citenamefont {Paul}\ \emph {et~al.}(2008)\citenamefont {Paul},
  \citenamefont {P\'epin},\ and\ \citenamefont {Norman}}]{Paul2008}%
  \BibitemOpen
  \bibfield  {author} {\bibinfo {author} {\bibfnamefont {I.}~\bibnamefont
  {Paul}}, \bibinfo {author} {\bibfnamefont {C.}~\bibnamefont {P\'epin}},\ and\
  \bibinfo {author} {\bibfnamefont {M.~R.}\ \bibnamefont {Norman}},\ }\bibfield
   {title} {\bibinfo {title} {Multiscale fluctuations near a {K}ondo breakdown
  quantum critical point},\ }\href {https://doi.org/10.1103/PhysRevB.78.035109}
  {\bibfield  {journal} {\bibinfo  {journal} {Phys. Rev. B}\ }\textbf {\bibinfo
  {volume} {78}},\ \bibinfo {pages} {035109} (\bibinfo {year}
  {2008})}\BibitemShut {NoStop}%
\bibitem [{Note4()}]{Note4}%
  \BibitemOpen
  \bibinfo {note} {While the transport vertex corrections can still be resummed
  exactly as a ladder series owing to our controlled large $N$ limit, unlike in
  previous work on fermions coupled to critical bosons \cite {KFWL}, this
  calculation is tedious, and we therefore defer it for future
  work}\BibitemShut {NoStop}%
\bibitem [{\citenamefont {Hartnoll}\ \emph {et~al.}(2018)\citenamefont
  {Hartnoll}, \citenamefont {Lucas},\ and\ \citenamefont {Sachdev}}]{HLS2018}%
  \BibitemOpen
  \bibfield  {author} {\bibinfo {author} {\bibfnamefont {S.~A.}\ \bibnamefont
  {Hartnoll}}, \bibinfo {author} {\bibfnamefont {A.}~\bibnamefont {Lucas}},\
  and\ \bibinfo {author} {\bibfnamefont {S.}~\bibnamefont {Sachdev}},\ }\href
  {https://mitpress.mit.edu/books/holographic-quantum-matter} {\emph {\bibinfo
  {title} {Holographic quantum matter}}}\ (\bibinfo  {publisher} {MIT press},\
  \bibinfo {year} {2018})\BibitemShut {NoStop}%
\bibitem [{\citenamefont {Collignon}\ \emph {et~al.}(2021)\citenamefont
  {Collignon}, \citenamefont {Ataei}, \citenamefont {Gourgout}, \citenamefont
  {Badoux}, \citenamefont {Lizaire}, \citenamefont {Legros}, \citenamefont
  {Licciardello}, \citenamefont {Wiedmann}, \citenamefont {Yan}, \citenamefont
  {Zhou}, \citenamefont {Ma}, \citenamefont {Gaulin}, \citenamefont
  {Doiron-Leyraud},\ and\ \citenamefont {Taillefer}}]{Taillefer2021Seebeck}%
  \BibitemOpen
  \bibfield  {author} {\bibinfo {author} {\bibfnamefont {C.}~\bibnamefont
  {Collignon}}, \bibinfo {author} {\bibfnamefont {A.}~\bibnamefont {Ataei}},
  \bibinfo {author} {\bibfnamefont {A.}~\bibnamefont {Gourgout}}, \bibinfo
  {author} {\bibfnamefont {S.}~\bibnamefont {Badoux}}, \bibinfo {author}
  {\bibfnamefont {M.}~\bibnamefont {Lizaire}}, \bibinfo {author} {\bibfnamefont
  {A.}~\bibnamefont {Legros}}, \bibinfo {author} {\bibfnamefont
  {S.}~\bibnamefont {Licciardello}}, \bibinfo {author} {\bibfnamefont
  {S.}~\bibnamefont {Wiedmann}}, \bibinfo {author} {\bibfnamefont {J.-Q.}\
  \bibnamefont {Yan}}, \bibinfo {author} {\bibfnamefont {J.-S.}\ \bibnamefont
  {Zhou}}, \bibinfo {author} {\bibfnamefont {Q.}~\bibnamefont {Ma}}, \bibinfo
  {author} {\bibfnamefont {B.~D.}\ \bibnamefont {Gaulin}}, \bibinfo {author}
  {\bibfnamefont {N.}~\bibnamefont {Doiron-Leyraud}},\ and\ \bibinfo {author}
  {\bibfnamefont {L.}~\bibnamefont {Taillefer}},\ }\bibfield  {title} {\bibinfo
  {title} {Thermopower across the phase diagram of the cuprate
  ${\mathrm{la}}_{1.6\ensuremath{-}x}{\mathrm{nd}}_{0.4}{\mathrm{sr}}_{x}{\mathrm{cuo}}_{4}$:
  Signatures of the pseudogap and charge density wave phases},\ }\href
  {https://doi.org/10.1103/PhysRevB.103.155102} {\bibfield  {journal} {\bibinfo
   {journal} {Phys. Rev. B}\ }\textbf {\bibinfo {volume} {103}},\ \bibinfo
  {pages} {155102} (\bibinfo {year} {2021})}\BibitemShut {NoStop}%
\bibitem [{\citenamefont {Proust}\ and\ \citenamefont
  {Taillefer}(2019)}]{TailleferPseudogap}%
  \BibitemOpen
  \bibfield  {author} {\bibinfo {author} {\bibfnamefont {C.}~\bibnamefont
  {Proust}}\ and\ \bibinfo {author} {\bibfnamefont {L.}~\bibnamefont
  {Taillefer}},\ }\bibfield  {title} {\bibinfo {title} {The remarkable
  underlying ground states of cuprate superconductors},\ }\href
  {https://doi.org/10.1146/annurev-conmatphys-031218-013210} {\bibfield
  {journal} {\bibinfo  {journal} {Annual Review of Condensed Matter Physics}\
  }\textbf {\bibinfo {volume} {10}},\ \bibinfo {pages} {409} (\bibinfo {year}
  {2019})},\ \Eprint
  {https://arxiv.org/abs/https://doi.org/10.1146/annurev-conmatphys-031218-013210}
  {https://doi.org/10.1146/annurev-conmatphys-031218-013210} \BibitemShut
  {NoStop}%
\bibitem [{\citenamefont {Taillefer}(2010)}]{TailleferSM}%
  \BibitemOpen
  \bibfield  {author} {\bibinfo {author} {\bibfnamefont {L.}~\bibnamefont
  {Taillefer}},\ }\bibfield  {title} {\bibinfo {title} {Scattering and pairing
  in cuprate superconductors},\ }\href
  {https://doi.org/10.1146/annurev-conmatphys-070909-104117} {\bibfield
  {journal} {\bibinfo  {journal} {Annual Review of Condensed Matter Physics}\
  }\textbf {\bibinfo {volume} {1}},\ \bibinfo {pages} {51} (\bibinfo {year}
  {2010})},\ \Eprint
  {https://arxiv.org/abs/https://doi.org/10.1146/annurev-conmatphys-070909-104117}
  {https://doi.org/10.1146/annurev-conmatphys-070909-104117} \BibitemShut
  {NoStop}%
\bibitem [{\citenamefont {Zhang}\ and\ \citenamefont
  {Sachdev}(2020)}]{Zhang2020}%
  \BibitemOpen
  \bibfield  {author} {\bibinfo {author} {\bibfnamefont {Y.-H.}\ \bibnamefont
  {Zhang}}\ and\ \bibinfo {author} {\bibfnamefont {S.}~\bibnamefont
  {Sachdev}},\ }\bibfield  {title} {\bibinfo {title} {From the pseudogap metal
  to the {F}ermi liquid using ancilla qubits},\ }\href
  {https://doi.org/10.1103/PhysRevResearch.2.023172} {\bibfield  {journal}
  {\bibinfo  {journal} {Phys. Rev. Research}\ }\textbf {\bibinfo {volume}
  {2}},\ \bibinfo {pages} {023172} (\bibinfo {year} {2020})}\BibitemShut
  {NoStop}%
\bibitem [{\citenamefont {Halperin}\ \emph {et~al.}(1993)\citenamefont
  {Halperin}, \citenamefont {Lee},\ and\ \citenamefont {Read}}]{Halperin1993}%
  \BibitemOpen
  \bibfield  {author} {\bibinfo {author} {\bibfnamefont {B.~I.}\ \bibnamefont
  {Halperin}}, \bibinfo {author} {\bibfnamefont {P.~A.}\ \bibnamefont {Lee}},\
  and\ \bibinfo {author} {\bibfnamefont {N.}~\bibnamefont {Read}},\ }\bibfield
  {title} {\bibinfo {title} {Theory of the half-filled {L}andau level},\ }\href
  {https://doi.org/10.1103/PhysRevB.47.7312} {\bibfield  {journal} {\bibinfo
  {journal} {Phys. Rev. B}\ }\textbf {\bibinfo {volume} {47}},\ \bibinfo
  {pages} {7312} (\bibinfo {year} {1993})}\BibitemShut {NoStop}%
\bibitem [{\citenamefont {Kim}\ \emph {et~al.}(1994)\citenamefont {Kim},
  \citenamefont {Furusaki}, \citenamefont {Wen},\ and\ \citenamefont
  {Lee}}]{KFWL}%
  \BibitemOpen
  \bibfield  {author} {\bibinfo {author} {\bibfnamefont {Y.~B.}\ \bibnamefont
  {Kim}}, \bibinfo {author} {\bibfnamefont {A.}~\bibnamefont {Furusaki}},
  \bibinfo {author} {\bibfnamefont {X.-G.}\ \bibnamefont {Wen}},\ and\ \bibinfo
  {author} {\bibfnamefont {P.~A.}\ \bibnamefont {Lee}},\ }\bibfield  {title}
  {\bibinfo {title} {Gauge-invariant response functions of fermions coupled to
  a gauge field},\ }\href {https://doi.org/10.1103/PhysRevB.50.17917}
  {\bibfield  {journal} {\bibinfo  {journal} {Phys. Rev. B}\ }\textbf {\bibinfo
  {volume} {50}},\ \bibinfo {pages} {17917} (\bibinfo {year}
  {1994})}\BibitemShut {NoStop}%
\bibitem [{\citenamefont {Lee}(2009)}]{sslee2009}%
  \BibitemOpen
  \bibfield  {author} {\bibinfo {author} {\bibfnamefont {S.-S.}\ \bibnamefont
  {Lee}},\ }\bibfield  {title} {\bibinfo {title} {Low-energy effective theory
  of {F}ermi surface coupled with {U}(1) gauge field in $2+1$ dimensions},\
  }\href {https://doi.org/10.1103/PhysRevB.80.165102} {\bibfield  {journal}
  {\bibinfo  {journal} {Phys. Rev. B}\ }\textbf {\bibinfo {volume} {80}},\
  \bibinfo {pages} {165102} (\bibinfo {year} {2009})}\BibitemShut {NoStop}%
\bibitem [{\citenamefont {Hugenholtz}\ and\ \citenamefont
  {Pines}(1959)}]{HP1959}%
  \BibitemOpen
  \bibfield  {author} {\bibinfo {author} {\bibfnamefont {N.~M.}\ \bibnamefont
  {Hugenholtz}}\ and\ \bibinfo {author} {\bibfnamefont {D.}~\bibnamefont
  {Pines}},\ }\bibfield  {title} {\bibinfo {title} {Ground-state energy and
  excitation spectrum of a system of interacting bosons},\ }\href
  {https://doi.org/10.1103/PhysRev.116.489} {\bibfield  {journal} {\bibinfo
  {journal} {Phys. Rev.}\ }\textbf {\bibinfo {volume} {116}},\ \bibinfo {pages}
  {489} (\bibinfo {year} {1959})}\BibitemShut {NoStop}%
\bibitem [{\citenamefont {Mahan}(2013)}]{Mahan}%
  \BibitemOpen
  \bibfield  {author} {\bibinfo {author} {\bibfnamefont {G.~D.}\ \bibnamefont
  {Mahan}},\ }\href {https://www.springer.com/gp/book/9780306434235} {\emph
  {\bibinfo {title} {Many-particle physics}}}\ (\bibinfo  {publisher} {Springer
  Science \& Business Media},\ \bibinfo {year} {2013})\BibitemShut {NoStop}%
\bibitem [{\citenamefont {Patel}\ and\ \citenamefont
  {Sachdev}(2014)}]{Patel2014}%
  \BibitemOpen
  \bibfield  {author} {\bibinfo {author} {\bibfnamefont {A.~A.}\ \bibnamefont
  {Patel}}\ and\ \bibinfo {author} {\bibfnamefont {S.}~\bibnamefont
  {Sachdev}},\ }\bibfield  {title} {\bibinfo {title} {{DC} resistivity at the
  onset of spin density wave order in two-dimensional metals},\ }\href
  {https://doi.org/10.1103/PhysRevB.90.165146} {\bibfield  {journal} {\bibinfo
  {journal} {Phys. Rev. B}\ }\textbf {\bibinfo {volume} {90}},\ \bibinfo
  {pages} {165146} (\bibinfo {year} {2014})}\BibitemShut {NoStop}%
\bibitem [{Note5()}]{Note5}%
  \BibitemOpen
  \bibinfo {note} {The polarization bubbles $\protect \pmb {\Pi }_{f,b}$
  involve the subtraction of diamagnetic terms not explicitly shown in
  Fig.~\ref {fig:diag_IL}, which render $\protect \pmb {\Pi }_{f,b}(\omega
  ,q)=\protect \pmb {\Pi }_{f,b}(\omega ,q)-\protect \pmb {\Pi }_{f,b}(\omega
  =0,q=0)$}\BibitemShut {NoStop}%
\bibitem [{\citenamefont {Patel}\ \emph
  {et~al.}(2018{\natexlab{b}})\citenamefont {Patel}, \citenamefont {Lawler},\
  and\ \citenamefont {Kim}}]{PatelKim2018}%
  \BibitemOpen
  \bibfield  {author} {\bibinfo {author} {\bibfnamefont {A.~A.}\ \bibnamefont
  {Patel}}, \bibinfo {author} {\bibfnamefont {M.~J.}\ \bibnamefont {Lawler}},\
  and\ \bibinfo {author} {\bibfnamefont {E.-A.}\ \bibnamefont {Kim}},\
  }\bibfield  {title} {\bibinfo {title} {Coherent superconductivity with a
  large gap ratio from incoherent metals},\ }\href
  {https://doi.org/10.1103/PhysRevLett.121.187001} {\bibfield  {journal}
  {\bibinfo  {journal} {Phys. Rev. Lett.}\ }\textbf {\bibinfo {volume} {121}},\
  \bibinfo {pages} {187001} (\bibinfo {year} {2018}{\natexlab{b}})}\BibitemShut
  {NoStop}%
\bibitem [{Note6()}]{Note6}%
  \BibitemOpen
  \bibinfo {note} {$H'_{bb}$ will also generate inter-layer boson pairing terms
  $\sim c'_b b^\dagger _l b^\dagger _{l'}$, but the Hugenholtz-Pines theorem
  \cite {HP1959} nevertheless ensures a 3D gapless boson phase, with the same
  effects on the fermions.}\BibitemShut {Stop}%
\end{thebibliography}%

\appendix

\section{Kubo formula in Landau Level basis for Model I}\label{app:Kubo}

In this Appendix, we obtain expressions for the conductivities of the different species in Model I via the Kubo formula, which are given by their respective bubble diagrams of Fig.~\ref{fig:FD_polbubble}, as described in the main text. We compute these generally at nonzero values of the out-of-plane magnetic field $B$ by working in the Landau Level basis in the $x-y$ plane with
wavefunctions
\begin{align}
    \psi_{n,k}(x,y) &= \frac{1}{\sqrt{L_x \ell}} e^{ikx}\phi_{n,k}(y/\ell); \\
    \phi_{n,k}(z) &= \frac{\pi^{-1/4}}{\sqrt{2^n n!}} H_n(z+k\ell) \exp\left(-\frac{(z+k\ell)^2}2\right), \nonumber
\end{align}
where $\ell = 1/\sqrt{eB}$ and $H_n(x)$ are the (physicist's) Hermite polynomials satisfying the recursion relation $H_{n+1}(x)=2xH_{n}(x)-H_{n}'(x)$. The energy of the states is $\omega_{c\lambda}(n+1/2)$ where $\omega_{c\lambda} = |e|B/m_\lambda$, where $\lambda\in\{c,f,b\}$. The use of the Landau level basis is possible because the self-energies of all three species are independent of momentum and therefore proportional to the identity matrix in real space, which implies that they are also proportional to the identity matrix in the Landau level basis, greatly simplifying the computation. Results such as (\ref{eq:rxxc}) and (\ref{eq:RHc}) in the weak magnetic field limit can be obtained by taking the $B\rightarrow0$ limit of our expressions here.

\begin{widetext}
It is important to recall the following identities:
\begin{equation} \label{eq:ll_ortho}
\begin{aligned}
    \int dz \phi_{n,k}(z) \partial_z \phi_{m,k}(z) &= \sqrt{\frac{m}2}\delta_{n,m-1} - \sqrt{\frac{m+1}2} \delta_{n,m+1}, \\
    \int dz \phi_{n,k}(z) \partial_z (z+k\ell) \phi_{m,k}(z) &= \sqrt{\frac{m}2}\delta_{n,m-1} + \sqrt{\frac{m+1}2} \delta_{n,m+1}.
\end{aligned}
\end{equation}
 
Now, our starting point is the Kubo formula in momentum space, which we will transform to the Landau Level basis. Recall that \cite{Mahan} $\sigma_{\lambda,\alpha\beta}(\omega,q) = -\text{Im} \Pi_{\lambda,\alpha\beta}^R(\omega,q)/\omega$ where
\begin{equation}
\begin{aligned}
    \Pi_{\lambda,\alpha\beta}=-\frac{1}{V} \int dx dx' dy dy' e^{iq_x (x-x')}e^{iq_y(y-y')} \int_0^{1/T} d\tau e^{i\omega\tau}\langle T_\tau J_{\lambda,\alpha}^\dagger(r,\tau) J_{\lambda,\beta}(r',0)\rangle, 
\end{aligned}
\end{equation}
where $\tau$ is imaginary time. With the above identities, a straightforward calculation will yield the spatially-integrated current as
\begin{equation}
\begin{aligned}
   \frac{2 m_\lambda i}e \int dx dy J_\lambda(r, \tau) &\equiv \int dx dy \left(\lambda^\dagger_r(\tau) (\nabla -ie A)\lambda_r(\tau) - (\nabla + i e A)\lambda_r^\dagger(\tau) \lambda_r(\tau)\right) \\
&=  \frac{2}\ell\sum_{k,n} \left(\begin{pmatrix} i \\ 1 \end{pmatrix}\sqrt{\frac{n+1}2} \lambda_{nk}^\dagger(\tau)\lambda_{n+1,k}(\tau) +  \begin{pmatrix} i \\ -1 \end{pmatrix}\sqrt{\frac{n}{2}} \lambda_{nk}^\dagger(\tau)\lambda_{n-1,k}(\tau)\right).
\end{aligned}
\end{equation}

We now evaluate $\Pi_{\lambda,xx}$ and $\Pi_{\lambda,xy}$ at $q=0$ using this expression. Using $G_{\lambda nk}(\tau) = \langle \lambda_{nk}(\tau) \lambda_{nk}^\dagger(0)\rangle$, we get
\begin{align}
    \begin{pmatrix}\Pi_{\lambda,xx} \\ \Pi_{\lambda,xy}\end{pmatrix}&= -\eta \frac{e^2}{V \ell^2m_\lambda^2}  \int_0^{1/T} d\tau e^{i\omega\tau}\sum_{nk} \begin{pmatrix}\frac{n+1}2 G_{\lambda nk}(\tau)G_{\lambda,n+1,k}(-\tau) +\frac{n}{2} G_{\lambda nk}(\tau)G_{\lambda,n-1,k}(-\tau) \\
 -i\frac{n+1}2 G_{\lambda nk}(\tau)G_{\lambda,n+1,k}(-\tau) +i\frac{n}{2} G_{\lambda nk}(\tau)G_{\lambda,n-1,k}(-\tau)
    \end{pmatrix} \\
    &= -\eta \frac{e^2\omega_{c\lambda}^2}{2\pi} T \sum_{i\nu_n}\left(\sum_{n}\frac{n+1}2 \begin{pmatrix}1 \\ -i \end{pmatrix} G_{\lambda n}(i\nu_n)G_{\lambda,n+1}(i\nu_n-i\omega)+\frac{n}{2}\begin{pmatrix}1 \\ i \end{pmatrix} G_{\lambda n}(i\nu_n)G_{\lambda,n-1}(i\nu_n-i\omega)\right). \nonumber
\end{align}
where $\eta=\pm$ for bosons and fermions, respectively, because of time-ordering. In the second step, we switched to Matsubara frequencies, used the fact that $G_{nk}(\tau)\equiv G_n(\tau)$ is independent of $k$, and there are $L_x L_y/\ell^2/(2\pi)$ terms in the $k$ sum. 

We have neglected the vertex corrections to the conductivity in Fig.~\ref{fig:FD_polbubble} here, which can be shown to vanish even at $B\neq0$. Since the disordered interactions $g_{ijk}^r$ are uncorrelated between different sites in Model I, such corrections can be written as
\begin{equation}
\delta\Pi_\lambda = \Bigg\langle\int dx dy J_\lambda(r,\tau)\int dx_{1,2} dy_{1,2} d\tau_{1,2,3,4} \lambda^\dagger_{r_1}(\tau_1) \lambda_{r_1}(\tau_2) K(\tau_1,\tau_2,\tau_3,\tau_4) \lambda^\dagger_{r_2}(\tau_3) \lambda_{r_2}(\tau_4) \int dx'dy' J_\lambda(r',\tau')\Bigg\rangle. 
\end{equation}
Since $\lambda_r(\tau)=\sum_{n,k}\psi_{n,k}(r)\lambda_{nk}(\tau)$, and $G_{\lambda nk}(\tau)$ are independent of $k$, the identity
\begin{equation}
\int dk H_n(z+kl) H_{n\pm 1}(z+kl) \mathrm{exp}\left(-(z+kl)^2\right) = 0,
\end{equation}
ensures that these corrections vanish.

Proceeding similarly as to \cite{Mahan}, we next switch to the Lehmann representation, analytically continue, take the imaginary part, and expand for small $\omega$. We find
\begin{equation}
\begin{aligned}
    \sigma_{\lambda,xx} &= -s_{\lambda} \lim_{\omega \to 0}\frac{\text{Im}[\Pi_{\lambda,xx}(\omega)]}{\omega} = -\frac{s_{\lambda} e^2 \omega_{c\lambda}^2}{4 \pi} \sum_n (n+1) \int \frac{d\epsilon}{(2\pi)}A_{\lambda n}(\epsilon) A_{\lambda,n+1}(\epsilon)\left(\frac{\partial n_\eta(\epsilon)}{\partial \epsilon}\right) \\
    &=-\frac{s_{\lambda} e^2}{4\pi}\int \frac{d\epsilon}{2\pi}\frac{4 \Sigma_{\lambda}''(\epsilon)\frac{\partial n_\eta(\epsilon)}{\partial \epsilon}}{4 [\Sigma_{\lambda}''(\epsilon)]^2+\omega_{c\lambda}^2}\left(2\Sigma_{\lambda}''(\epsilon)+2 (\epsilon+\tilde \mu_{\lambda}) \text{Im}\left[ \psi_0\left(\frac{1}{2} + \frac{-\epsilon + i \Sigma_{\lambda}''(\epsilon)-\tilde \mu_{\lambda}}{\omega_{c\lambda}}\right) \right] \right),
\end{aligned}
\end{equation}
where $s_{\lambda}$ is the spin degeneracy of the species $\lambda$. We performed the Landau level sum in terms of the digamma function, $\psi_0$, and we used $\psi_0(z)=\psi_0(1+z)-1/z$ and 
\begin{equation}
    A_{\lambda n}(\epsilon) = \frac{2\eta\Sigma_{\lambda}''(\epsilon)}{(\epsilon + \tilde \mu_\lambda -(n+1/2)\omega_{c\lambda})^2 + [\Sigma_{\lambda}''(\epsilon)]^2},
\end{equation}
so that $\Sigma_{\lambda}''(\epsilon) = \text{Im}[\Sigma_{\lambda,R}(\epsilon)]$ and $\tilde{\mu}_\lambda = \mu_\lambda - \text{Re}[\Sigma_{\lambda,R}(\epsilon)]$. 

For $\sigma_{\lambda,xy}$, 
we convert to relative and center of mass coordinates $\epsilon_c =(\epsilon+\epsilon')/2$ and $\epsilon_r=\epsilon-\epsilon'$. We then symmetrize with respect to $\epsilon_r$ in order to get an integral from $0$ to $\infty$. We find
\begin{equation}
\begin{aligned}
\Pi_{\lambda,xy}(\omega \to 0) &= - i \frac{e^2\omega_{c\lambda}^2}{4\pi}\sum_n(n+1) \int \frac{d\epsilon d\epsilon'}{(2\pi)^2} A_{\lambda n}(\epsilon) A_{\lambda,n+1}(\epsilon') (n_\eta(\epsilon)-  n_\eta(\epsilon') )\left[\frac{2(\omega+i\delta)}{(\epsilon-\epsilon')^2}\right],\\
\sigma_{\lambda,xy}&= -\frac{ s_{\lambda} e^2\omega_{c\lambda}^2}{2 \pi}\sum_n(n+1) \int_0^\infty \frac{d\epsilon_r}{2\pi} \int_{-\infty}^\infty \frac{d\epsilon_c }{2\pi}  \frac{\sinh\left(\frac{\epsilon_r}{2T}\right)}{\cosh\left(\frac{\epsilon_c}{T}\right)-\eta\cosh\left(\frac{\epsilon_r}{2T}\right)} \frac{1}{\epsilon_r^2} \\
& \times \left[A_{\lambda n}\left(\epsilon_c + \frac{\epsilon_r}2\right) A_{\lambda,n+1}\left(\epsilon_c - \frac{\epsilon_r}2\right)-A_{\lambda n}\left(\epsilon_c - \frac{\epsilon_r}2\right) A_{\lambda,n+1}\left(\epsilon_c + \frac{\epsilon_r}2\right)\right]\\
&=-\frac{s_{\lambda} e^2}{(2\pi)^3}\int_0^\infty d\epsilon_r \int_{-\infty}^\infty d \epsilon_c \left(F_{\lambda}(\epsilon_c,\epsilon_r)-F_{\lambda}(\epsilon_c,-\epsilon_r)\right)\frac{\sinh\left(\frac{\epsilon_r}{2T}\right)}{\cosh\left(\frac{\epsilon_c}{T}\right)-\eta\cosh\left(\frac{\epsilon_r}{2T}\right)}\frac{1}{\epsilon_r^2},
\end{aligned}
\end{equation}
The sum can be done to give an explicit expression for $F_{\lambda}(\epsilon_c,\epsilon_r)$ as
\begin{align}
    &\frac{F_{\lambda}\left(\frac{\epsilon+\epsilon'}2,\epsilon-\epsilon'\right)}{2\Sigma_{\lambda}''(\epsilon)\Sigma_{\lambda}''(\epsilon')} = \text{Im}\left[\frac{\psi_0\left(\frac{2\epsilon-2i\Sigma_{\lambda}''(\epsilon)-2\tilde \mu_{\lambda}(\epsilon)}{2\omega_{c\lambda}}-\frac12\right)\left(2\epsilon -\omega_{c\lambda}-2 i \Sigma_{\lambda}''(\epsilon)-2\tilde \mu_{\lambda}(\epsilon)\right)}{\Sigma_{\lambda}''(\epsilon)(\Sigma_{\lambda}''(\epsilon')^2+(\epsilon'-\epsilon+\omega_{c\lambda}+i\Sigma_{\lambda}''(\epsilon)-\tilde \mu_{\lambda}(\epsilon')+\tilde \mu_{\lambda}(\epsilon))^2)}\right]\\
    &+\text{Im}\left[\frac{\psi_0\left(\frac{2\epsilon'+2i\Sigma_{\lambda}''(\epsilon')-2\tilde \mu_{\lambda}(\epsilon')}{2\omega_{c\lambda}}+\frac12\right)\left(2\epsilon' +\omega_{c\lambda}+2 i \Sigma_{\lambda}''(\epsilon')-2\tilde \mu_{\lambda}(\epsilon')\right)}{\Sigma_{\lambda}''(\epsilon')(\Sigma_{\lambda}''(\epsilon')^2-\Sigma_{\lambda}''(\epsilon)^2+2i \Sigma_{\lambda}''(\epsilon')(\epsilon-\epsilon'-\omega_{c\lambda}+\tilde \mu_{\lambda}(\epsilon')-\tilde \mu_{\lambda}(\epsilon))-(\epsilon-\epsilon'-\omega_{c\lambda}+\mu_{\lambda}(\epsilon')-\mu_{\lambda}(\epsilon))^2)}\right]. \nonumber
\end{align}

For the fermions, for small magnetic fields, these expressions give the same result as the expressions derived from the Boltzmann equations in \cite{Patel2018} with the identification $v_F^2 \nu/(4\pi) \to n/m$ where $v_F$ is the Fermi velocity, $n$ is the density, and $m$ is the mass. However, for large magnetic fields, our expressions will have quantum oscillations that are absent in the Boltzmann treatment.

\end{widetext}

\section{Boson spectral function and $\Delta_b$ in Model I}\label{app:gap}

In this appendix, we derive the boson spectral function and the soft gap $\Delta_b$ generally for a nonzero out-of-plane magnetic field. As in the derivation of the Kubo formula, we use the Landau level basis, which is made possible by the spatial locality and site-invariance of the occupancy constraint in the last line of (\ref{eq:model_ham}). The values of $\Delta_b$ at small magnetic fields can be obtained by taking the $B\rightarrow0$ limit in our expressions. 
 
Because $\mu_c, \mu_f \gg \omega_{c,c/f}$, we still have the original result for the fermion Green's function that $G_{c,f}(i\omega) = -i(\nu_{c,f}/2)\text{sgn}(\omega)$. That is, the fermions are less affected by the Landau level quantization than the bosons, and, consequently, the boson self-energy calculation in the main text is unaffected.

However, the boson spectral function must be calculated by summing over the spectral functions in each Landau level instead of integrating over momentum. The result is
\begin{align}\label{eq:bos_spec}
 A_b(\omega) &= \frac{1}{\ell^2 2\pi}\sum_{m} \frac{2\gamma\omega}{(\omega-(m+1/2)\omega_{cb}-\Delta_b)^2+\gamma^2\omega^2} \nonumber \\
 &= -\frac{m_b}{\pi} \text{Im}\left[ \psi_0\left(\frac{1}{2}-\frac{-\Delta_b+\omega + i \gamma \omega}{\omega_{bc}}\right) \right]
\end{align}
\begin{equation}
\quad\xrightarrow{B\to0} \frac{m_b}{\pi}\left[\pi\Theta(\omega-\Delta_b) + \tan^{-1}\left(\frac{\gamma \omega}{\Delta_b-\omega}\right) \right], \nonumber
\end{equation}
where $\Theta(x)$ is the Heaviside step function, $\psi_0(z)$ is the digamma function, $\ell = 1/\sqrt{e_bB}$, and $\omega_{cb}= e_bB/m_b$ with $e_b, m_b$ the charge and mass of the boson respectively.

Now, recall from the main text that the scaling of the fermion self-energy expressions above depends crucially on $\Delta_b(T)$. It can be easily checked that the change in the number of $f$ fermions in response to a shifting chemical potential is suppressed by $\Delta \mu_f/\Lambda_f$ where $\Lambda_f$, the $f$ fermion bandwidth, is assumed to be large. Therefore, the constraint can be written as
\begin{equation}
    \kappa-\kappa_c = (G_b(\tau=0^{-},\Delta_b(T))-G_b(\tau=0^{-},\Delta_{b,c}(0)),
\end{equation}
and $\Delta_b$ depends on both temperature and $\kappa$, but we suppress the $\kappa$ dependence generally. When $\kappa=\kappa_c$, $\Delta_b=\Delta_{b,c}$ and $\Delta_{b,c}(T=0)=0$. This is reminiscent of the $O(N)$ rotor model \cite{SachdevQPT} and the calculation of the thermal mass in \cite{Patel2014}.

Although we can do this calculation carefully in multiple ways, we will recall that $G_b(\tau=0^{-}) = \sum_{i} \langle b^\dagger_i(\tau=0^{-}) b_i(\tau=0^{-}) \rangle \equiv n_b$, which is the number density of bosons. For this number to converge, we choose to regulate it in the usual way (see \cite{Mahan})
\begin{equation}
    n_b = \frac{1}{V} \sum_{nk} \int_{-\infty}^\infty \frac{d\omega}{2\pi} n_B(\omega) A_{bn}(\omega,\Delta_b),
\end{equation}
where $A_{bn}$ is the summand seen in \eqref{eq:bos_spec}. 

Fig.~\ref{fig:boson_Mass} summarizes the behavior of $\Delta_b(T)$ in the three phases at zero and finite applied field. The important feature is the $T$-linear (up to logarithmic corrections) growth in the critical region. Low $T$ transport is dictated by the limit of $z=\Delta _b/T$ which shifts from $\infty$ to zero across the transition.

\begin{widetext}

\begin{figure}
	\centering
 	\includegraphics[width=0.48\textwidth]{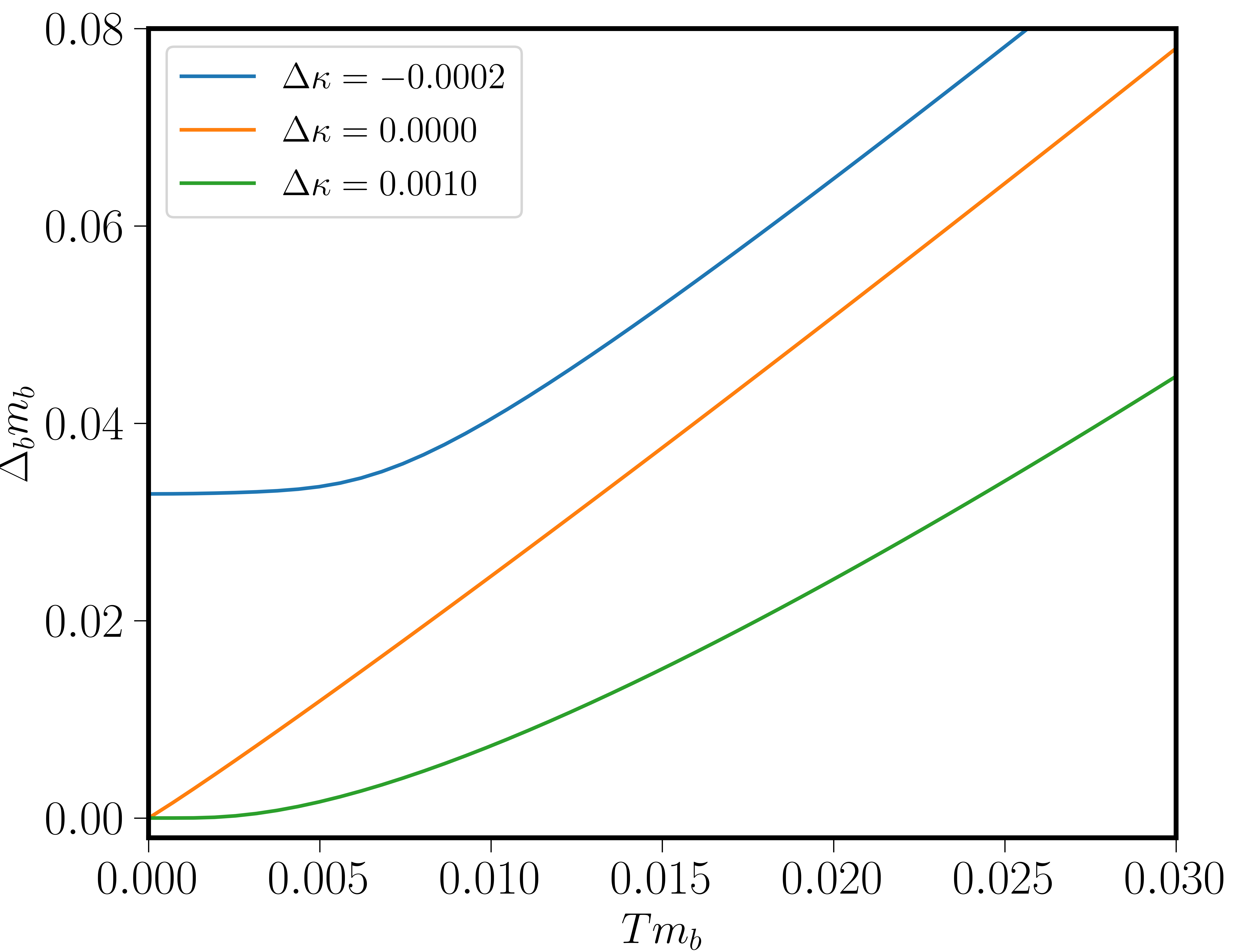}
 	\caption{We plot $\Delta_b$ vs. $T$ for various $\Delta \kappa=\kappa-\kappa_c$ with the color indicating $\Delta \kappa$. All curves become $T$-linear upon entering the critical region, but are either exponentially suppressed or approach a constant as $T\rightarrow 0$ if $\Delta_{\kappa}> 0$ or $\Delta_{\kappa} < 0$ respectively. All other parameters are the same as in Fig.~\ref{fig:phase_diagram} in the main text.} 
 	\label{fig:boson_Mass}
\end{figure}

\begin{figure}
    \centering
    \includegraphics[width=0.48\textwidth]{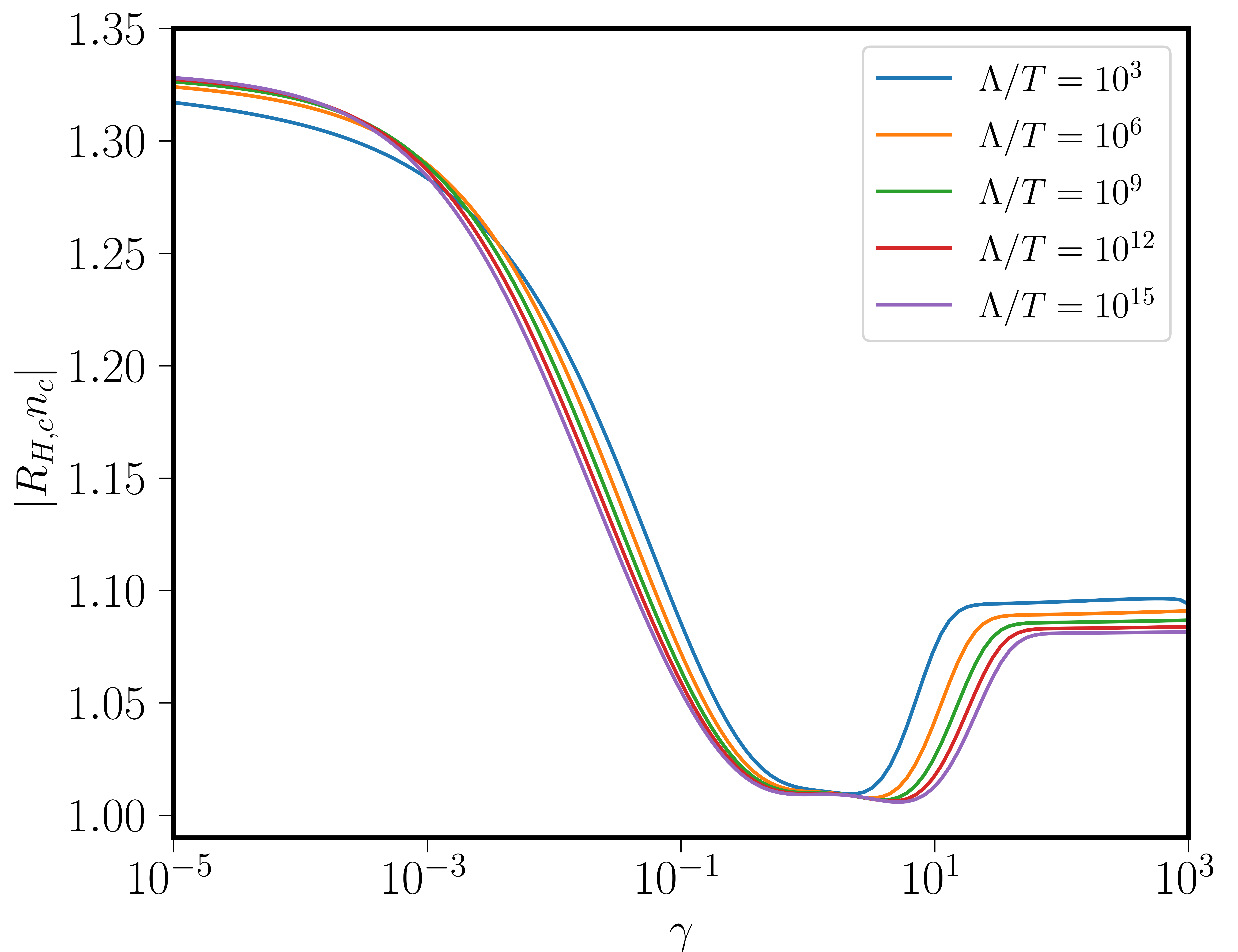}
    \caption{We plot $R_{H,c}n_c$, the Hall coefficient for the $c$ electrons when $\kappa=\kappa_c$, which approximates the total $R_H$ at low temperatures. In this regime, it depends on only two parameters: $\gamma$ and $\Lambda/T$. We find in this supplement that the peak at low $\gamma$ is exactly at $4/3$.}
    \label{fig:RHc_ALamoT}
\end{figure}

Note that
\begin{equation}
    \int_{-\infty}^\infty d\omega n_B(\omega) A_{bn}(\omega,\Delta_b) = \int_{0}^\infty d\omega n_B(\omega) (A_{bn}(\omega,\Delta_b)-A_{bn}(-\omega,\Delta_b))-\int_0^\infty d\omega A_{bn}(-\omega,\Delta_b),
\end{equation}
and that the first integral on the right-hand side is 0 when $T=0$. Recalling the form of the boson's spectral function from \eqref{eq:bos_spec}, we will find
\begin{equation}\label{eq:boson_tm_B}
\begin{aligned}
    2\pi (\kappa-\kappa_c) 
    &=\frac{\omega_{cb} m_b}{2\pi} \left[\int_0^\infty n_B(\omega)\frac{(A_{b}(\omega,\Delta_b) - A_{b}(-\omega,\Delta_b))}{1/(\ell^22\pi)}  +\frac{2\gamma}{\gamma^2+1}\ln\left(\frac{\Gamma(\mathcal{N}+3/2)\Gamma(1/2+\Delta_b/\omega_{cb})}{\Gamma(\mathcal{N}+3/2+\Delta_b/\omega_{cb})\Gamma(1/2)}\right)\right],
\end{aligned}
\end{equation}
where we have cut off the Landau level sum at $\mathcal{N}=\Lambda / \omega_{bc}$ and $\Gamma(n)$ is the Gamma function.

Taking the $B\to 0$ limit of \eqref{eq:boson_tm_B}, we can scale out $\Delta_b=zT$ and $x=\omega/T$ to find 
\begin{equation}\label{eq:B0bosonmassexp}
   2\pi^2(\kappa-\kappa_c)\frac{1}{Tm_b} = \int_0^\infty \frac{dx}{e^x-1}\left[ \tan^{-1}\left(\frac{\gamma x}{z-x}\right)+\tan^{-1}\left(\frac{\gamma x}{z+x}\right)\right]-\pi \ln\left(1-e^{-z}\right) -\frac{\gamma}{\gamma^2+1}z\ln\left(\frac{\Lambda e}{z T}\right).
\end{equation}
As $z\to 0$, the first two terms of the left-hand side dominate and as $z \to \infty$, the rightmost term dominates, so we see that there is a solution with $z$, whose value will change logarithmically, as $T\to \infty$. As expected, there is always a solution, so the bosons are not truly condensed as long as their dispersion is strictly 2D. Instead, for $\kappa>\kappa_c$ the gap becomes exponentially small in $
(\kappa-\kappa_c)/T$, i.e. $\Delta_b\sim T\exp\left[-\frac{2\pi(\kappa-\kappa_c)}{Tm_b}\right]$. In reality, however, there is a stable condensate solution at low temperature, facilitated by the 3D boson dispersion self-consistently generated by the presence of the condensate. For this reason, we have treated this low-temperature regime of the large Fermi surface phase ($\kappa>\kappa_c$) separately (see Appendix~\ref{app:instabilities}).
\end{widetext}

\section{Limiting self-energy calculations in Model I}\label{app:limits}

At low temperatures over the critical region, $\Delta_b/T$ is order one, so $\sigma_{xx}^b$ and $\sigma_{xy}^b$ are suppressed relative to the fermions, which are gapless. Therefore, by the Ioffe-Larkin composition rules (see Appendix~\ref{app:Ioffe-Larkin}), $\sigma^{bf}_{xx} \approx \sigma^{b}_{xx}$ and $R_H\approx R_{H,c}$ at low temperatures, which we confirm numerically. $R_{H,c}$, in turn, is determined by (\ref{eq:RHc}), and depends on the dimensionless parameters $(\kappa-\kappa_c)/(Tm_b)$, $\Lambda/T$, and $\gamma$. In Fig.~\ref{fig:RHc_ALamoT} we plot the dependence of $R_{H,c} n_c$ at criticality ($\kappa=\kappa_c$) on the latter two parameters.

To understand this behavior, we now derive simple limiting forms for the low-temperature $\Delta_b$ and fermion self-energy at criticality at low $B$. We'll consider three limits $\gamma \to 0$ with $T$ small but finite, $\gamma\to \infty$ with $T$ small but finite, and $T\to 0$ with $\gamma$ fixed. These expressions are used to obtain an estimate of the enhancement of the Hall coefficient given in the main text.

We first wish to solve \eqref{eq:B0bosonmassexp} when $\kappa=\kappa_c$ and $\gamma \to 0$ at fixed $T$. The integral on the right-hand side (RHS) is smaller than the other two terms, in this limit. We make the guess that $e^{-z} \ll 1$, so we arrive at
\begin{equation}
\begin{aligned}
\label{eq:zexact}
    z &= \frac{\Delta_b}{T} =\ln\left[ \frac{\pi}{\gamma z \ln\left(\frac{\Lambda e}{z T}\right)}\right]=\ln\left(\frac{\pi}{\gamma}\right) - \ln\left[ z \ln\left(\frac{\Lambda e}{z T}\right) \right]\\
    &\approx \ln\left(\frac{\pi}{\gamma}\right) - \ln\left[ \ln\left(\frac{\pi}{\gamma}\right) \ln\left(\frac{\Lambda e}{\ln(\pi/\gamma) T}\right) \right]; \qquad \gamma \to 0,
\end{aligned}
\end{equation}
which justifies our assumption.  In the last step, we used the fact that the second term is smaller than the first as $\gamma \to 0$, so we obtained an approximate expression for $z$ by simply substituting $z=\ln(\pi/\gamma)$ on the RHS. Better approximations are obtained by iteration, by substituting the improved expression for $z$. 

By inserting \eqref{eq:bos_spec} into \eqref{eq:fermiselfenergy}, we can evaluate the self-energy at leading order in $\gamma$ at criticality: 
\begin{equation}
\begin{aligned}
\label{eq:fermiselfenergyapprox}
    \Gamma_{\omega,T} &\equiv \text{Im}[\Sigma_{c,R}(\omega,T)] \\
    &=\lim_{\gamma \to 0} -\frac{\gamma m_b}{2\pi \nu_c} T \Bigg[\ln\left(\frac{1+e^{(\omega-\Delta_b)/T}}{1-e^{-\Delta_b/T}}\right)
    \\&+ \gamma\left(-\frac{\omega}{T}+\frac{\Delta_b}{T}\ln\left(\frac{\Delta_b}{|\Delta_b-\omega|}\right)\right)\Bigg].
\end{aligned}
\end{equation}
The $\mathcal O(1)$ term arises from approximating the spectral function as a step function. In the limit that $T$ is fixed and $\gamma\to0$, $\Delta_b / T \sim \ln(1/\gamma)$. Corrections to the spectral function, therefore, need only be integrated against $n_F(\epsilon'-\omega)-\Theta(-\epsilon')$, which we evaluate with the Sommerfeld approximation. The first term in \eqref{eq:fermiselfenergyapprox} goes as $Te^{-\Delta_b/T} \sim \gamma \Delta_b$, but, in this limit, the second term goes as $\gamma \omega^2/\Delta_b$ and is therefore higher order. Computing $R_{H,c} n_c$ using (\ref{eq:RHc}), and using just the first term in (\ref{eq:fermiselfenergyapprox}), gives exactly $4/3$ when $\gamma \ll 1$.

Turning to the $\gamma \to \infty$ limit, we see that the integral in \eqref{eq:B0bosonmassexp} is well approximated by taking the integrand as $\pi$ from $z \pi / (2 \gamma)$ to $z$, and as $0$ everywhere else. The error from this approximation is roughly a constant close to $\pi/2$ as $\gamma \to \infty$, so we end up needing to solve
\begin{equation}
    -\pi \ln\left(\frac{1-e^{-z\pi/(2\gamma)}}{\sqrt{e}}\right)=\frac{z}{\gamma}\ln\left(\frac{\Lambda e}{z T}\right).
\end{equation}
If $T$ is small enough, $z/\gamma$ will be small, which allows us to approximate the left-hand side as $-\pi \ln[(z \pi)/(2\sqrt{e}\gamma)]$. Finally, since $z/\gamma$ is small, we neglect the term $(z/\gamma)\ln(z/\gamma)$ that appears on the right-hand side. These approximations altogether yield
\begin{equation}
    z \approx \frac{\pi \gamma }{\ln\left(\frac{\Lambda}{T \gamma e}\right)}W_0\left(\frac{2\sqrt{e}}{\pi^2}\ln\left(\frac{\Lambda}{T\gamma e}\right) \right); \qquad \gamma \to \infty,
\label{eq:Sigmac1smallgamma}
\end{equation}
where $W_0(z)$ is the Lambert W function.

The self-energy in the large $\gamma$ limit is well approximated by the following:
\begin{align}
    \Gamma_{\omega,T} &= -\frac{\gamma m_b}{2\pi\nu_c}T \Big[\frac{z}{\gamma}\ln\left(\frac{\Lambda e}{z T}\right) + \pi \ln(1+e^{\omega/T}) \\
    &-\tan^{-1}(\gamma)\ln\left(\frac{1+e^{\omega/T}}{1+e^{-\omega/T}}\right)\Big]; \qquad \gamma \to \infty, \ z/\gamma < 1, \nonumber
\end{align}
where the integrals over the fermion occupation functions are done by setting $\Delta_b\to 0$ in the spectral function, which is accurate so long as $\Delta_b/(T \gamma) \ll 1$. When $z\to 0$ limit of that expression is plugged into (\ref{eq:RHc}), one finds $R_{H,c} \approx -1.07/n_c$ in good agreement with the numerics. Numerical studies confirm $R_{H,c}n_c$ increases near $\gamma=0,\infty$ with a single minimum near $\gamma=1$, the maximum being $4/3$. 

To understand the temperature dependence of the resistivity at criticality and small $\gamma$ we use the formula \cite{Patel2018}.
\begin{equation}
\begin{aligned}
        \rho_{c,xx}&=\left( \frac{n_c}{8 m_c T} \int_{-\infty}^{\infty} d\epsilon \frac{\text{sech}^2(\epsilon/2T)}{\Gamma_{\epsilon,T}}\right)^{-1}
        \\&=T\left(\frac{n_c}{8m_c} \int_{-\infty}^{\infty} d x \frac{\text{sech}^2(x/2)}{(\Gamma_{xT,T})/T}\right)^{-1}.
\end{aligned}
\end{equation}
Plugging the value of $\Delta_b$ \eqref{eq:zexact} into \eqref{eq:fermiselfenergyapprox} or the exact result we get that $\Gamma_{xT,T}/T $ depends on $T$ only through logarithmic corrections. 

To calculate the self-energy in the low-temperature limit at fixed field--as we do in our numerical calculations--we must use the finite field expression \eqref{eq:bos_spec}. For temperatures sufficiently lower than $B$ the self-energy takes the form $\Gamma_c\approx (T^2/B)g(\omega/T)$ and will be dominated by the cyclotron frequency. This will invalidate the small field approximation. In this case we must include the quadratic terms in $B$ for the Hall coefficient \cite{Patel2018}. The Hall coefficient then goes to one as $\Gamma/B \to 0$.

\section{Derivation of the Ioffe-Larkin condition for Model I}\label{app:Ioffe-Larkin}

The Kubo formula allows us to evaluate the conductivity tensors for the three species. To find the total conductivity, however, we must combine the contribution from the three species. Although the $c$ fermions are a separate species and will be added in parallel to the $b$ and $f$ contribution, the latter two species add together in series instead of in parallel due to the Ioffe-Larkin composition rule. In this section, we will derive the Ioffe-Larkin composition rule closely following Lee and Nagaosa \cite{Lee1992}. Our derivation is exact in the large $N$ limit.

Due to the emergent gauge field, the charge of the $b$ bosons and $f$ fermions is renormalized. The physical condition is that $e_b+e_f=-1$ as the $bfc^\dagger$ term in the Lagrangian must conserve charge. How the charge is distributed is a gauge choice, with the emergent gauge field ensuring the physical results are independent of this choice. 

\begin{figure*}
 \includegraphics{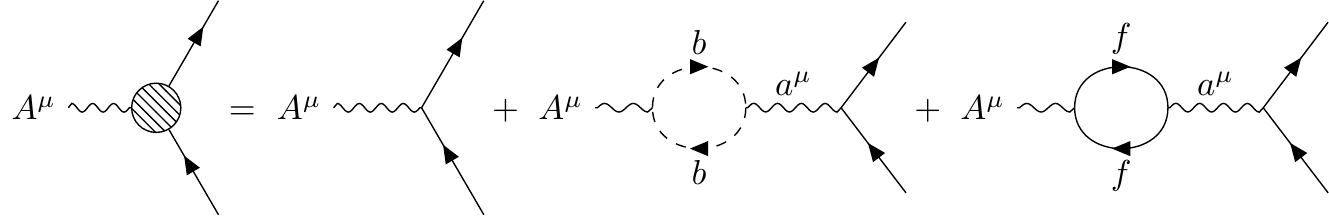}
    \caption{The diagrams that contribute to the renormalized charge. The propagators and polarization bubbles are all fully dressed. $A^\mu$ is the external gauge field, $a^\mu$ is the emergent gauge field, and the lines to the right of the diagrams are either $b$ or $f$ propagators depending on whether the renormalized $b$ charge or renormalized $f$ charge is being computed.}
    \label{fig:diag_IL}
\end{figure*}

We see in Fig.~\ref{fig:diag_IL} that there are three diagrams that contribute to the renormalization of the charge. In the diagrams, the polarization bubbles, $\pmb{\Pi}$, \footnote{The polarization bubbles $\pmb{\Pi}_{f,b}$ involve the subtraction of diamagnetic terms not explicitly shown in Fig.~\ref{fig:diag_IL}, which render $\pmb{\Pi}_{f,b}(\omega,q)=\pmb{\Pi}_{f,b}(\omega,q)-\pmb{\Pi}_{f,b}(\omega=0,q=0)$} and propagators are fully renormalized (with the fermionic spin degeneracy included). Any other diagram is either zero because of the locality of the SYK-type interaction or suppressed by $1/N$. We note that the propagator for the emergent $U(1)$ gauge field is \cite{Lee1992} 
\begin{equation}
    \pmb{D}(\tau-\tau') = - \langle a(\tau)a(\tau')\rangle = -(\pmb{\Pi}_f+\pmb{\Pi}_b)^{-1},
\end{equation}
and the boldface is indicating tensors, which follows if the inverse bare propagator is taken to be infinitesimal. 

Summing these diagrams for, e.g. the $f$ fermions gives
\begin{align}
    \pmb{ e}_f^r &= e_f - e_f \pmb{\Pi}_f(\pmb{\Pi}_f+\pmb{\Pi}_b)^{-1}+e_b\pmb{\Pi}_b(\pmb{\Pi}_f+\pmb{\Pi}_b)^{-1}
    \\&
    =(e_f+e_b)\pmb{\Pi}_b(\pmb{\Pi}_f+\pmb{\Pi}_b)^{-1} = -\pmb{\Pi}_b(\pmb{\Pi}_f+\pmb{\Pi}_b)^{-1}, \nonumber
\end{align}
where the extra minus sign for $\pmb \Pi_b$ comes because $f$ and $b$ are oppositely charged under the emergent gauge field, and all polarization bubbles are evaluated at $(\omega,q)$. Switching $f\leftrightarrow b$ will give the boson result. Therefore, the charge renormalizes to become a tensor. It is worth noting that $\pmb \Pi_b$, $\pmb \Pi_f$, and $\pmb \Pi_b+\pmb \Pi_f$ are $2\times 2$ antisymmetric matrices and therefore commute with each other. When we compute the total current-current correlator due to the $f$ and $b$ sub-systems after renormalizing the currents using the respective charge renormalizations. We find, since there are no current cross-correlations, as discussed in the main text,
\begin{align}
      \pmb{\Pi}_{\text{tot}} &= \pmb{\Pi}_b\pmb{\Pi}^2_f[(\pmb{\Pi}_b+\pmb{\Pi}_f)^{-1}]^2+\pmb{\Pi}_f\pmb{\Pi}_b^2[(\pmb{\Pi}_b+\pmb{\Pi}_f)^{-1}]^2 
      \nonumber \\
      &= (\pmb{\Pi}_b^{-1}+\pmb{\Pi}_f^{-1})^{-1}, 
\end{align}
which implies that the $f$ and $b$ resistivity are added in series.
  
One important point that is glossed over in the above is that the electric and magnetic field are renormalized differently, and $\pmb{\Pi}_b$ and $\pmb{\Pi}_f$ are evaluated for different effective magnetic fields. In our notation, $\pmb \Pi(\omega,q) \approx -i \pmb{\sigma} \omega +\pmb{\chi} q^2$, so the renormalization changes depending on whether the vertex is magnetic $A^\mu(\omega=0,q\to0)$, or electric, $A^\mu(\omega\to0,q=0)$. We find, for instance for the $f$ fermions
\begin{equation}
    E_\text{eff}^f = \pmb{\sigma}_b(\pmb{\sigma}_f+\pmb{\sigma}_b)^{-1} E,\qquad B_\text{eff}^f = \frac{\chi_b}{\chi_b+\chi_f} B,
\end{equation}
for a weak magnetic field $B$. In the magnetic field case, we additionally average over $q$, which replaces $\pmb{\chi}$ with half its trace $\chi = (\chi_{xx}+\chi_{yy})/2$. 

\begin{widetext}
In our derivation, we have neglected contributions to $\pmb{\sigma}$ and $\pmb{\chi}$ from potential cross-correlations $\mathbf{\Pi}_{fb}\sim \langle J_f J_b \rangle$. Doing so is valid, as Model I's site-uncorrelated $g^r_{ijk}$ render them of the form 
\begin{align}
&\mathbf{\Pi}_{fb}(i\omega,q) \sim \int d^2k d^2k' d\Omega d\Omega' v_{f,k}v_{b,k} G_{f,k+q/2}(i\Omega+i\omega/2)G_{f,k-q/2}(i\Omega-i\omega/2) K_{fb}(i\Omega,i\Omega',\omega,q) \nonumber \\
&\times G_{b,k'+q/2}(i\Omega'+i\omega/2)G_{b,k'-q/2}(i\Omega'-i\omega/2),
\label{eq:crossdeath}
\end{align}
where $v_{x,k}=\nabla_k \epsilon_{x,k}$. Since $G_{x,k}=G_{x,-k}$, $G_{x,k+q/2}(i\Omega+i\omega/2)G_{x,k-q/2}(i\Omega-i\omega/2)=G_{x,k}(i\Omega+i\omega/2)G_{x,k}(i\Omega-i\omega/2)+\Xi_{x,k}(i\Omega,i\omega)|q|^2$, with $\Xi_{x,k}=\Xi_{x,-k}$, and $v_{x,k}=-v_{x,-k}$, the $\mathcal{O}(\omega)$ and $\mathcal{O}(q^2)$ terms in the expansion of $\mathbf{\Pi}_{fb}(\omega,q)$ vanish and we can thus neglect these cross-correlations.    
  
\section{Diamagnetic susceptibilities in Model I}\label{app:diamag}

Because of the renormalization of the magnetic field from the internal gauge field, we must find expressions for $\chi_f$ and $\chi_b$. To find them, we evaluate $\chi_\lambda q^2 = \Pi_\lambda(\omega=0,q\to 0)-\Pi_\lambda(\omega=0,q=0)$. We average the two possible directions. Then, we have the bubble contributions (vertex corrections vanish for the same reason as (\ref{eq:crossdeath}) does) 
\begin{equation}
\begin{aligned}
    &\Pi_\lambda(q\to 0)=\frac{\Pi_{\lambda,xx}+\Pi_{\lambda,yy}}{2} = -\eta \frac{1}{V}\sum_k \frac{k^2}{2 m_\lambda^2}T\sum_{i\nu} \left(G_\lambda(k-q/2,i\nu)G_\lambda(k+q/2,i\nu)\right)\\
    &=-2\eta T\int_0^{\tilde{k}_\mathrm{max}} \frac{d\tilde k}{(2\pi)^2} \int_0^{2\pi}d\theta \tilde k^3 \left(\sum_{i\nu}
    \frac{1}{(i\nu/T-\tilde k^2+\tilde k\tilde q\cos(\theta)-\tilde q^2/4+\mu_\lambda/T-\Sigma_\lambda/T)}\frac{1}{(\tilde q\to -\tilde q)}\right)\\
    &\chi_\lambda =-\eta \frac{1}{2m_\lambda} \int_0^{k_\mathrm{max}} k^3 \frac{dk}{2\pi}\left(\sum_{i\nu}\frac{(i\nu_\lambda/T +\mu/T-\Sigma_\lambda/T) }{(i\nu/T-k^2+\mu_\lambda/T-\Sigma_\lambda/T)^4}\right),
    \end{aligned}
\label{eq:chisotropic}    
\end{equation}
where in the second line of the above, we re-scaled the momenta by a factor of $\tilde k = k/\sqrt{2m_\lambda T}$, and we relabeled $\tilde k \to k$ in line 3. 

We can do the Matsubara sums exactly in the bosonic case since $\Sigma_b(i\omega) = -\gamma|\omega|$. We carry them out to find ($z=-\mu/T=\Delta_b/T$):
\begin{equation}\label{eq:diamag1}
\begin{aligned}
     \chi_b&=-\frac{1}{2m_b} \int_0^{\sqrt{\Lambda/T}} \frac{dk}{2\pi}
     k^3\left(\frac{z}{(k^2+z)^4}+\text{Re}\left[\frac{\psi_2\left(\frac{k^2+z}{2 \pi \gamma - 2\pi i}\right)}{(2\pi \gamma - 2\pi i)^3}+\frac{k^2}{3}\frac{\psi_3\left(\frac{k^2+z}{2 \pi \gamma - 2\pi i}\right)}{(2\pi \gamma - 2\pi i)^4}\right]\right),
\end{aligned}
\end{equation}
with $\psi_n(z)$ the polygamma function and $\Lambda$ is the boson bandwidth. This expression diverges as $\chi_b\sim (1/m_b)\ln(\Lambda/\Delta_b)$ when $\Delta_b\rightarrow0$.  

For the $f$ fermions, we can transform (\ref{eq:chisotropic}) to 
\begin{equation}
\chi_f = \frac{1}{2m_f}\int_{-\mu_f}^{\Lambda_f} \frac{d\epsilon}{2\pi}(\epsilon+\mu_f)T\sum_{i\nu}\left(\frac{(i\nu +\mu_f-\Sigma_f(i\nu))}{(i\nu-\epsilon-\Sigma_f(i\nu))^4}\right) = T\sum_{i\nu}\frac{(\Lambda_f+\mu_f)^2(\Lambda_f-2\mu_f+3\Sigma_f(i\nu)-3i\nu)}{24\pi m_f(\Lambda_f+\Sigma_f(i\nu)-i\nu)^3 (\mu_f -\Sigma_f(i\nu)+i\nu)}.
\label{eq:chifsotropic}
\end{equation}
\end{widetext}
We note that $T$ and $|\Sigma_f(i\nu)|$ are always much smaller than the $f$ bandwidth $\Lambda_f$ and Fermi energy $\mu_f$, for any value of $\nu$, since $|\Sigma_f(i\nu)|$ is bounded by a scale controlled by the boson bandwidth $\Lambda\ll \Lambda_f,\mu_f$. Therefore we can expand the summand of (\ref{eq:chifsotropic}) in powers of $\Sigma_f$ and take the $T\rightarrow0$ limit. It may then be seen that the sum of the absolute values of the contributions from all these terms in the expansion is bounded by a quantity that vanishes in the limit of $\Lambda_f,\mu_f\rightarrow\infty$, leaving $\chi_f$ to take its free fermion value of $1/(24\pi m_f)$, which can be easily verified by inserting the result for $\Sigma_f(i\nu)$ and then numerically integrating over $\nu$ in this limit. 

\section{inter-layer instabilities in Model I}\label{app:instabilities}

\begin{figure}[]
 	\centering
 	\includegraphics[width=0.5\textwidth]{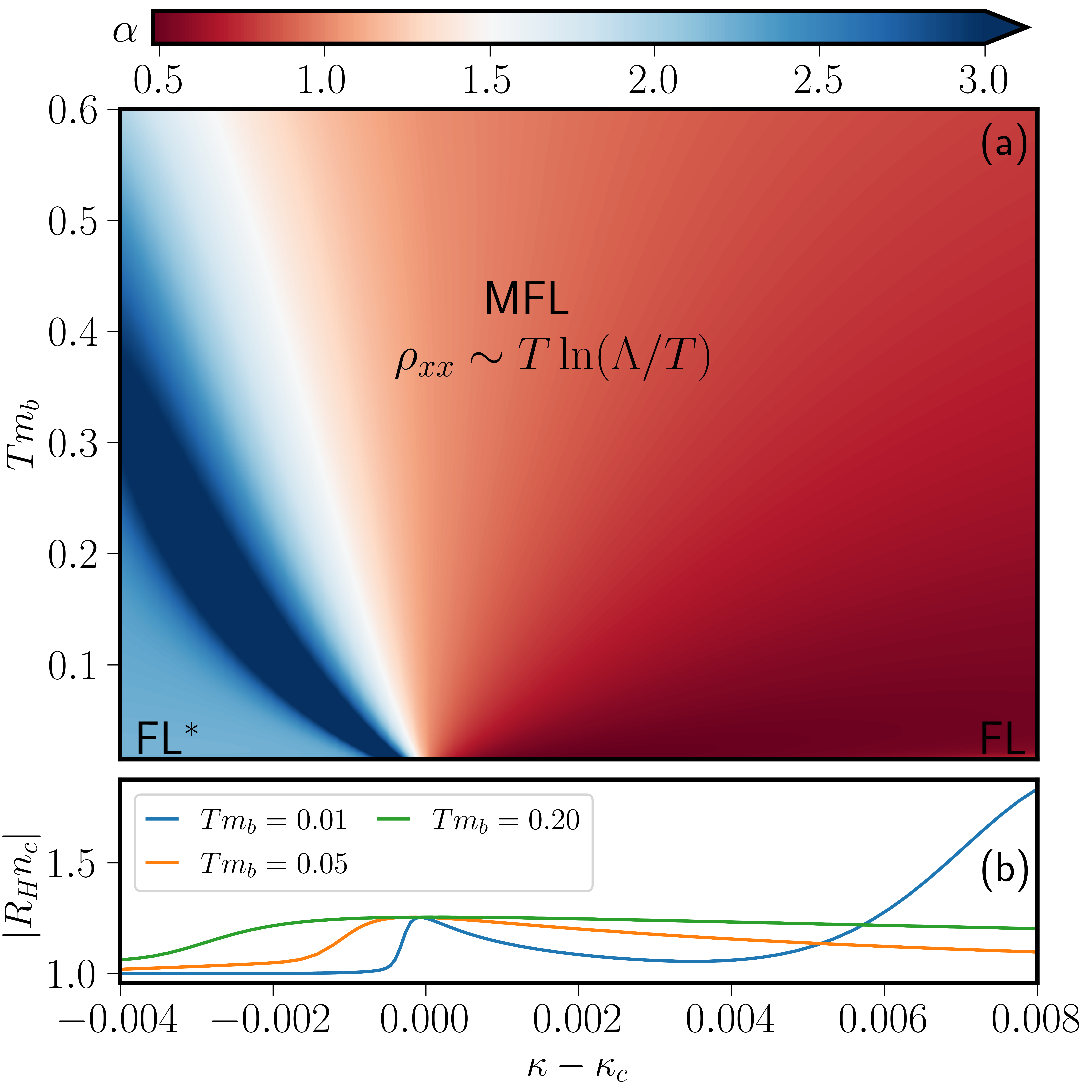}
     	\caption{(a) The phase diagram for Model I in 2D with no inter-layer instabilities. The resistivity is given by $\rho_{xx}-\rho_{xx}(T=0) \sim T^\alpha \ln(\Lambda/T)$, and the color indicates the value of $\alpha=d\ln(\rho_{xx}/\ln(\Lambda/T))/d\ln(T)$.  (b) The plot of the weak-field $R_H$ vs. $\kappa-\kappa_c$.  $R_H$ has a peak near the crossover from Fermi-liquid behavior to $T$-linear resistivity and approaches a constant to either side signaling the change in carrier density. The large peak of $R_H$ seen at low temperatures is occurring as the boson is condensing, as discussed in the main text. The same parameters are used as in Fig.~\ref{fig:phase_diagram} in the main text.}
 	\label{fig:phase_diagram2D}
 \end{figure}

Using the expressions from the previous sections, we can find $\rho_{xx}$ and $R_H$ exactly for a 2D version of Model I without inter-layer couplings, for all values of parameters at small $B$. For the same parameters used in the main text, we plot $R_H$ and $\rho_{xx}$ in Fig.~\ref{fig:phase_diagram2D} while ignoring inter-layer couplings, which should be compared with Fig.~\ref{fig:phase_diagram} in the main text that takes inter-layer couplings into account. Note the large enhancement of $R_H$ seen at low temperatures when $\kappa>\kappa_c$, as also seen in Fig.~\ref{fig:Rhrhoxx_cuts} in the main text.

Despite the exact solvability of Model I in its 2D version described here, to make physical predictions we must analyze possible instabilities that will take us away from our solution. In the 2D version of Model I, the only possible instabilities at large $N$ are BCS-like fermion pairing instabilities, induced by adding weak attractive interactions, which occur at exponentially small energy scales and which we hence ignore. However, the physical version of Model I includes a third spatial dimension, and we should therefore ask what relevant inter-layer interactions are allowed and what their impact on the physics will be.  

An important feature of the physical version of Model I is that the $b$ and $f$ partons are deconfined in a stack of independent 2D layers. We can therefore write down the following large $N$, instability inducing \cite{PatelKim2018}, local, gauge-invariant, quartic interactions between adjacent layers $l$ and $l'$, where $r$ denotes the 2D coordinate of a site within a layer:
\begin{align}
H_{bb} &= -\frac{J_b}{N}\sum_{i,j=1}^N\sum_{r}b^\dagger_{r(l),i}b_{r(l'),i}b^\dagger_{r(l'),j}b_{r(l),j},
\nonumber \\
H'_{bb} &= -\frac{J'_b}{N}\sum_{i,j=1}^N\sum_{r}b^\dagger_{r(l),i}b^\dagger_{r(l'),i}b_{r(l'),j}b_{r(l),j},
\end{align}
\begin{align}
H_{ff} &= -\frac{J_f}{N}\sum_{i,j=1}^N\sum_{\substack{r,\sigma,\sigma',\\\tau,\tau'}}f^\dagger_{r(l),i,\sigma}f^\dagger_{r(l'),i,\sigma'}f_{r(l'),j,\tau}f_{r(l),j,\tau'}, \nonumber \\
H_{bf} &= -\frac{J_{bf}}{N}\sum_{i,j=1}^N\sum_{r,\sigma}\left[b^\dagger_{r(l),i}b_{r(l'),i}f^\dagger_{r(l),j,\sigma}f_{r(l'),j,\sigma} + \mathrm{H.c}\right]. \nonumber
\end{align}

None of these terms contribute directly to the parton self-energies or transport at large $N$. $H_{ff}$ induces BCS-like inter-layer $f$ fermion pairing instabilities, which occur at exponentially small energy scales, and are therefore not of concern to us. The terms in $H_{bb}$ create inter-layer boson instabilities driven by susceptibilities that scale as $\sim m_b J_b' \ln(\Lambda/\Delta_b)$. In the gapped phase of the boson, and in the quantum critical region, these susceptibilities are thus small at the temperature scales of interest, hence we ignore them. However, for $\kappa>\kappa_c$, $\Delta_b(T)$ starts decreasing rapidly below some temperature scale (Fig.~\ref{fig:boson_Mass}), which makes these susceptibilities large, causing the onset of instabilities that lead to the condensation of inter-layer boson bilinears in the gray region of Fig.~\ref{fig:phase_diagram} in the main text. The resulting 3D boson phase will then further have single-boson condensation as temperature is lowered \cite{SachdevQPT}, entering the region below the gray wedge. Once this happens, both the partons will have 3D dispersions as these boson interaction terms will appear like inter-layer hoppings, $b^{\dag}_lb^{\dag}_{l'}b_lb_{l'}\sim c_b b^{\dag}_lb_{l'}$, and $H_{bf}$ will similarly generate inter-layer hopping for the $f$ fermions \footnote{$H'_{bb}$ will also generate inter-layer boson pairing terms $\sim c'_b b^\dagger_l b^\dagger_{l'}$, but the Hugenholtz-Pines theorem \cite{HP1959} nevertheless ensures a 3D gapless boson phase, with the same effects on the fermions.}. This leads to two important changes to the model; first the partons develop an anisotropic dispersion with hopping proportional to the single-boson condensate strength at temperatures well below the gray wedge, second the fermions now scatter off both the $N-1$ critical bosons $b_{2,..,N}$ as well as the condensed mode $\langle b_1\rangle$.

To model these effects, the dispersion of the partons is changed to be
\begin{equation}
    \epsilon_{b/f,k}=\frac{1}{2m_{b/f}}(k_x^2+k_y^2+Y_{b/f} k_z^2), \quad Y_{b/f}= 4\pi^2 J_{b/bf} r_0^2,
\end{equation}
where $r_0$ is the size of the condensate. For Fig.~\ref{fig:phase_diagram} in the main text we take $J_b=1$. Rewriting the SD equations within the condensed phase, the only changes are to the fermion self-energy and the constraint. The constraint equation becomes
\begin{equation}
    \kappa-\kappa_c=r_0^2+(n_b-n_b^c),
\end{equation}
where, in this equation, $n_b$ is the number of bosons not participating in the condensate with $\Delta_b=0$, and using the self-consistently determined dispersion. The self-energy expression is changed to be
\begin{align}
    \text{Im}[\Sigma_{c,R}]  
     &=-r_0^2 g^2 \frac{\nu_{f}}{2} 
     \\
     &- g^2 \frac{\nu_f}{4\pi} \int_{-\infty}^\infty d \epsilon \bar A_b(\epsilon)(n_B(\epsilon) + n_F(\epsilon - \omega)), \nonumber
\end{align}
with $\tilde A_b$ the spectral function of the uncondensed modes.

To keep the number of $f$ fermions fixed, as the dispersion changes, the Fermi energy shifts which in turn modifies the density of states. In order to connect with the 2D model, we introduce a maximum momentum in the $z$ direction, $K$. The spinless density of states is then given by
\begin{equation}
\begin{aligned}
    \nu_f = 
    \begin{cases}
    \frac{K m_f}{\pi} & \epsilon_F^0 > \frac{Y_fK^2}{3m_f} \\
    \frac{m_f}{\pi} \left(\frac{3\epsilon_F^0K m_f}{c_f}\right)^{1/3}& \epsilon_F^0 < \frac{Y_f K^2}{3m_f}
    \end{cases},
\end{aligned}
\label{eq:cond_DOS}
\end{equation}
where $\epsilon_{F,0}$ is the Fermi energy with $Y_f=0$. Note that we take $K=\pi$ so the density of states in the small condensate regime is $\nu_f=m_f$, the same as in the purely 2D case. We will work in the regime where the second condition of \eqref{eq:cond_DOS} is never reached, this is achieved by taking $J_{bf}$ sufficiently small. If the second condition was achieved, $\gamma=g^2\nu_c\nu_f/(2\pi)$ would change. 

The spectral function for the uncondensed modes can be evaluated utilizing the 2D results by replacing $\Delta_b \to \Delta_b + Y_bk_z^2/(2m_{b})$ in \eqref{eq:bos_spec} to find 
\begin{align}
    &\frac{2\pi}{m_b} \tilde A_{b}(\omega,0) =  K\text{sgn}(\omega)-\frac{2K}{\pi} \tan^{-1}\left(\frac{Y_bK^2}{4 \gamma m_{b} \omega} -\frac{1}{\gamma}\right) 
    \\
    &-\frac{4}{\pi Y_b}\text{Im}\left[\sqrt{2(1+i\gamma)\omega m_{b}}\tanh^{-1}\left(\frac{Y_bK}{\sqrt{2m_{b}(1+i\gamma)\omega}}\right)\right] \nonumber.
\end{align}
Unlike the $O(N)$ rotor model, the dispersion is also modified as the condensate grows. This changing dispersion results in a different temperature dependence when $T m_b \gg \kappa-\kappa_c$ and also results in multiple self-consistent values of the condensate size $r_0$ at fixed $\kappa$ and $T$. If we assume interactions which generate a 3D instability at $T=0$, the physical solution for $r_0$ is the one that approaches a non-zero constant at low temperatures, which is the one we use in our numerical calculations. 

Deep in the condensed phase at low temperatures, $r_0$ will be roughly constant and large. In this regime, the frequency dependence of the spectral function for the uncondensed boson modes then goes as as $\sqrt{\omega}$, leading directly to $\mathrm{Im}[\Sigma_{c,R}(\omega=0,T)] \sim T^{3/2} + \mathrm{const.}$ behavior.

\begin{widetext}
\section{Self-energies and critical transport in Model II}\label{app:SDM2}

To evaluate the boson self-energy we start with the individual patch contribution (\ref{eq:bpse}). Integration over $q_\perp$ yields (using $v_{c,f,F}=k_F/m_{c,f}$, where $k_F$ is the Fermi momentum)
\begin{align}
&\Sigma_b^p(i\omega,k) = -ig^2T\sum_{i\nu}\int\frac{d^2q_\parallel}{(2\pi)^2}\left(\mathrm{sgn}(\nu+\omega)-\mathrm{sgn}(\nu)\right) \nonumber \\
&\times \left[v_{f,F}\left(i(\nu+\omega)-\Sigma_c(i\nu+i\omega)-v_{c,F} k_\perp-\frac{k_\parallel q_\parallel \cos(\theta)}{2m_c}-\frac{k_\parallel^2}{2m_c}\right) - v_{c,F}\left(i\nu-\Sigma_f(i\nu)\right)\right]^{-1},
\end{align}
where $\theta$ is the angle between $q_\parallel$ and $k_\parallel$. This further reduces upon integration over $\theta$ to
\begin{align}
&\Sigma^p_b(i\omega,k) = -\frac{ig^2}{2\pi}T\sum_{i\nu}\int_0^{q_{\mathrm{max}}}\frac{q_\parallel dq_\parallel~(\mathrm{sgn}(\nu+\omega)-\mathrm{sgn}(\nu))}{v_{f,F}(i(\nu+\omega)-\Sigma_c(i\nu+i\omega))-v_{c,F}(i\nu-\Sigma_f(i\nu))-\frac{v_{f,F}k_\parallel q_\parallel}{2m_c}-v_{c,F}v_{f,F}k_\perp-\frac{v_{f,F}k_\parallel^2}{2m_c}} \nonumber \\
&\times \frac{1}{\sqrt{1+\frac{\frac{v_{f,F}k_\parallel q_\parallel}{m}}{v_{f,F}(i(\nu+\omega)-\Sigma_c(i\nu+i\omega))-v_{c,F}(i\nu-\Sigma_f(i\nu))-\frac{v_{f,F}k_\parallel q_\parallel}{2m}-v_{c,F}v_{f,F} k_\perp-\frac{v_F k_\parallel^2}{2m}}}} \approx \frac{g^2q_{\mathrm{max}}m_c}{\pi^2v_{f,F}}\frac{|\omega|}{k_\parallel} = \frac{g^2q_{\mathrm{max}}m_cm_f}{\pi^2 k_F}\frac{|\omega|}{k_\parallel}.
\end{align}
Here the cutoff $q_{\mathrm{max}}\sim k_F~d\Omega$, where $d\Omega$ is the solid angle subtended by the patch, is the cutoff on the patch size. If we now average over all patches, we obtain
\begin{equation}
\Sigma_b(i\omega,k) \approx -\int_0^{2\pi}d\phi \int_0^\pi \sin\theta d\theta ~\frac{g^2m_cm_f}{\pi^2}\frac{|\omega|}{k\sin\theta} = -2 g^2 m_c m_f \frac{|\omega|}{k}\equiv -\gamma_2\frac{|\omega|}{k}.
\end{equation}

We now discuss the fermion self-energies (\ref{eq:scfm2}) at criticality. There, the $c,f$ self-energies are expected to show MFL frequency dependence because of the log divergence of the momentum integral over $q_\parallel$. As mentioned at the end of Sec. \ref{sec:M2T} in the main text, in this Appendix, we are interested in the higher temperature regime where the boson is not that strongly damped, so we do {\it not} ignore the $i\nu$ term in the boson propagator in (\ref{eq:scfm2}) while computing the fermion self-energies. The Matsubara frequency sum can then be separated into a UV divergent piece, that is a constant and which may be absorbed by a chemical potential shift, and a UV finite piece, which may be computed analytically. Then we can compute the momentum integral numerically with a UV cutoff $\sim\sqrt{2m_b\Lambda}$ to obtain
\begin{equation}
\Sigma_{c,f}(i\omega) \approx \mathrm{const.} -\frac{i\gamma_2m_b}{m_{c,f}k_F}T\tilde{\varphi}\left(\frac{\omega}{T},\frac{\Lambda}{T},\frac{\Lambda}{\Delta_b(T)}\right).
\end{equation}
where the function $\tilde{\varphi}$ is no longer symmetric between $\pm\omega$ in the higher energy regime $\Lambda\gg \omega,T\gg \gamma_2^2/m_b$, where the $i\nu$ term in the boson propagator in (\ref{eq:scfm2}) is dominant.

In this higher temperature (energy) regime, the small wavevectors in the boson propagator are cut off by temperature as $q_\parallel^2 \sim m_b T$ (by comparison of $q_\parallel^2/(2m_b)$ to the $i\nu$ term), and the boson self-energy $\gamma_2|\nu|/q_\parallel$ (which we now treat as a perturbation), may therefore be approximated to be $\sim \gamma_2 |\nu|/\sqrt{m_b T}$ in (\ref{eq:scfm2}). Model II then behaves similarly to Model I at small $\gamma$, with $\gamma\sim \gamma_2/\sqrt{m_bT}$, from the point of view of the fermions. Then, by virtue of (\ref{eq:rxxc}, \ref{eq:RHc}), we have $\rho_{xx}(T)\sim \sqrt{T}$ (up to log corrections), and $|R_H n_c| \rightarrow 4/3$. 

In Fig.~\ref{fig:Model2Crossover} we show the crossover between the strongly damped low-temperature regime and the weakly damped higher temperature regime over the QCP, by exact numerical calculation of the conduction electron bubble diagram contribution to the conductivity tensor. As we also argued for the case of Model I, this bubble diagram is still the dominant contribution at criticality. Indeed, the contribution of the $f$ fermions and the bosons are still suppressed even in the higher energy regime of Model II (that is similar to the $\gamma\ll 1$ regime of Model I), due to the relatively low conductivity of the bosons. 

\begin{figure*}[ht]
 	\centering
 	\includegraphics[width=0.98\textwidth]{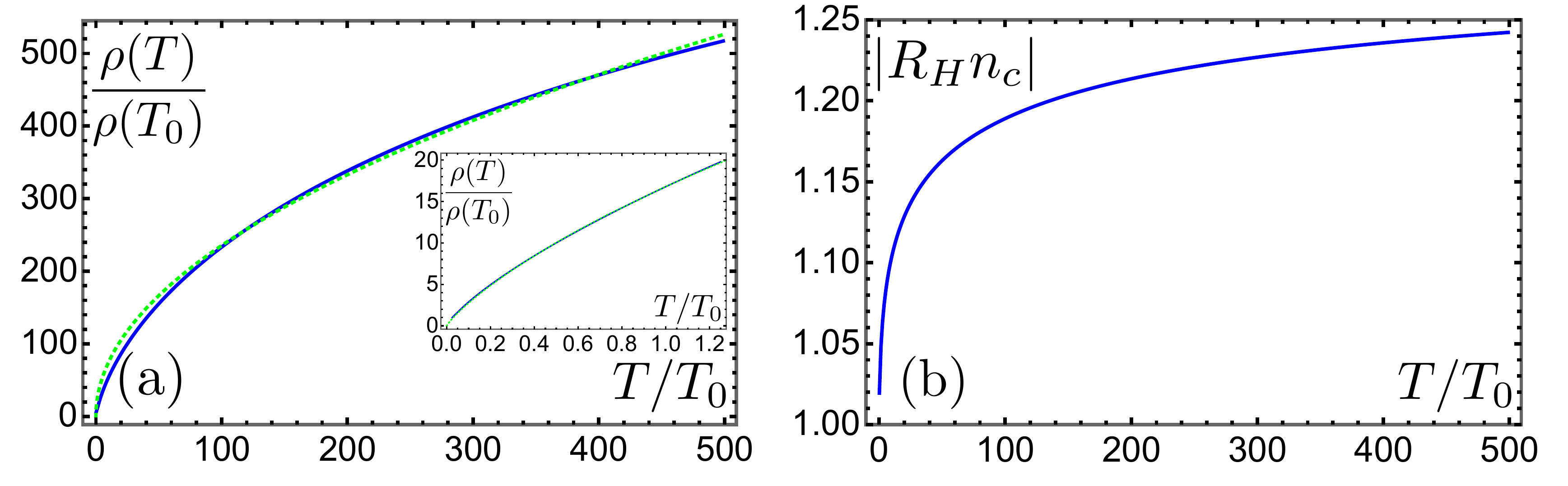}
     	\caption{(a) Temperature dependence of $\rho_{xx}$ over the QCP in Model II. The dashed green lines indicate fits to $\rho(T)/\rho(T_0)=a_1\sqrt{T/T_0}$ and $\rho(T)/\rho(T_0)=a_2(T/T_0)\ln(a_3 T_0/T)$ in the main and inset plots respectively (b) Temperature dependence of $R_H$ in Model II. We use $\gamma_2 = 0.02$, $m_b = 1.0$, the crossover scale $T_0 = \gamma_2^2/m_b = 4\times 10^{-4}$, and the boson bandwidth $\Lambda = \pi^2/2 \approx 1.23\times 10^4~T_0$. The bandwidths of the conduction electrons and $f$ fermions are assumed to be very large.}
 	\label{fig:Model2Crossover}
 \end{figure*}

\end{widetext}

\end{document}